\documentclass[onecolumn, prd, aps, tightenlines, preprintnumbers, showpacs, nofootinbib, superscriptaddress, notitlepage]{revtex4-2}
\pdfoutput=1

\usepackage{tikz}

\usepackage{enumitem}
\usepackage[T1]{fontenc} 
\usepackage{floatrow}
\usepackage{appendix}
\usepackage{braket}
\usepackage{slashed}
\usepackage{pifont}
\usepackage{tabularx}
\usepackage[normalem]{ulem}
\usepackage{xcolor}

\usepackage{hyperref}
\usepackage{epsf}
\usepackage{amsmath}
\usepackage{empheq}
\usepackage[theorems,skins]{tcolorbox}
\usepackage{amsfonts}
\usepackage{amssymb}
\usepackage{psfrag,epsfig,graphicx,graphics}
\usepackage{cancel}

\newcommand\numberthis[1][]{%
    \refstepcounter{equation}%
    \ifx#1\empty\else\label{eq:#1}\fi%
    \tag{\theequation}%
}

\usepackage{xargs} 
\usepackage[colorinlistoftodos,prependcaption,textsize=tiny]{todonotes}
\providecommand{\U}[1]{\protect\rule{.1in}{.1in}}

\newcommand{\cmark}{\ding{51}}
\newcommand{\xmark}{\ding{55}}

\def\slashchar#1{\setbox0=\hbox{$#1$}
   \dimen0=\wd0
   \setbox1=\hbox{/} \dimen1=\wd1
   \ifdim\dimen0>\dimen1
      \rlap{\hbox to \dimen0{\hfil/\hfil}}
      #1
   \else
      \rlap{\hbox to \dimen1{\hfil$#1$\hfil}}
      /
   \fi}

\def\bei{\begin{itemize}}
\def\ei{\end{itemize}}

\def\beeq{\begin{eqnarray}} 
\def\beqa{\begin{eqnarray}}
\def\bea{\begin{eqnarray}}

\def\eea{\end{eqnarray}}
\def\eqa{\end{eqnarray}}
\def\eeeq{\end{eqnarray}}

\def\eqar{\end{array}}
\def\beqar{\begin{array}}

\def\beas{\begin{eqnarray*}}
\def\beqas{\begin{eqnarray*}}

\def\eqas{\end{eqnarray*}}
\def\eeas{\end{eqnarray*}}

\def\beq{\begin{equation}} 
\def\be{\begin{equation}}

\def\ee{\end{equation}}
\def\eq{\end{equation}}
\def\eeq{\end{equation}}

\def\beqd{\begin{displaymath}}
\def\eeqd{\end{displaymath}}
\def\eqd{\end{displaymath}}

\def\beeq{\begin{eqnarray}} \def\eeeq{\end{eqnarray}}

\newcommand{\fin}{\end{document}}

\newcommandx{\MF}[2][1=]{\todo[linecolor=blue,backgroundcolor=blue!25,bordercolor=blue,#1]{#2}}

\newcommandx{\JF}[2][1=]{\todo[linecolor=black,backgroundcolor=white!25,bordercolor=black,#1]{#2}}

\newcommandx{\TA}[2][1=]{\todo[linecolor=red,backgroundcolor=red!25,bordercolor=red,#1]{#2}}

\newcommandx{\GB}[2][1=]{\todo[linecolor=violet,backgroundcolor=violet!25,bordercolor=violet,#1]{#2}}

\begin{document}

\title{\boldmath Reggeization of quarks from next-to-eikonal high-energy QCD}

\author{Tolga Altinoluk}
\affiliation{Theoretical Physics Division, National Centre for Nuclear Research, Pasteura 7, Warsaw, 02-093, Poland}
 
\author{Guillaume Beuf}
\affiliation{Theoretical Physics Division, National Centre for Nuclear Research, Pasteura 7, Warsaw, 02-093, Poland}

\author{Jules Favrel}
\affiliation{Theoretical Physics Division, National Centre for Nuclear Research, Pasteura 7, Warsaw, 02-093, Poland}

\author{Michael Fucilla}
\affiliation{Theoretical Physics Division, National Centre for Nuclear Research, Pasteura 7, Warsaw, 02-093, Poland}









\begin{abstract}
We develop a comprehensive Wilson-line formulation of quark Reggeization in QCD. Starting from a next-to-eikonal operator built from a semi-infinite Wilson line and a background-quark insertion, we identify an interpolating operator for the Reggeized quark and derive its nonlinear rapidity evolution using the background-field method. In the dilute regime, its positive-signature component exhibits Regge-pole evolution governed by the quark Regge trajectory, whereas the negative-signature sector displays mixing between quark and gluon degrees of freedom, in accordance with its known Regge-cut structure. In the planar limit, this mixing is suppressed, and Reggeization emerges without an explicit signature projection, recovering signature degeneracy. We illustrate the universality of the construction by extracting the same Reggeized-quark operator from a more general Wilson-line operator describing a gluon-to-quark transition. Furthermore, we extend the formalism to massive quarks, deriving a coordinate-space evolution kernel whose Fourier transform reproduces the massive-quark Regge trajectory. 
Finally, from the leading operator-mixing structure of the evolution equation and signature arguments, we show that the Reggeized-quark interpolating operator is expected to remain an eigenstate of the rapidity evolution up to next-to-leading logarithmic accuracy.
\end{abstract}

\maketitle

\tableofcontents

\section{Introduction}
\label{sec:intro}

The high-energy limit of scattering amplitudes has long provided a natural setting in which to uncover universal structures of gauge theories~\cite{Grisaru:1973ku,Grisaru:1973wbb,Grisaru:1973vw}. In this regime, Regge theory describes the asymptotic behavior of scattering amplitudes in terms of singularities in the complex angular-momentum plane, leading to characteristic power-law dependence on the center-of-mass energy. In perturbative QCD, this picture finds a remarkably successful realization through the emergence of Reggeized degrees of freedom and the resummation of logarithmically enhanced corrections. Its best-known manifestation is gluon Reggeization~\cite{Lipatov:1976zz} and the associated Balitsky--Fadin--Kuraev--Lipatov (BFKL) equation~\cite{Fadin:1975cb,Kuraev:1976ge,Kuraev:1977fs,Balitsky:1978ic}, which governs the high-energy evolution of scattering amplitudes and constitutes a cornerstone of our understanding of perturbative QCD in the Regge limit. \\

While Reggeization and the BFKL formalism successfully describe high-energy scattering in the dilute regime, the resulting linear evolution predicts a rapid growth of hadronic cross sections with energy that cannot persist asymptotically without conflicting with unitarity~\cite{Gribov:1983ivg}. This observation motivated the development of a nonlinear description of dense QCD, in which large parton occupation numbers induce collective effects and modify the high-energy evolution. In this regime, scattering is naturally formulated within the shockwave formalism~\cite{McLerran:1994vd,Balitsky:1995ub}, where energetic partons are represented by Wilson lines propagating through a strong color background, the QCD shockwave. The energy dependence is then encoded in the nonlinear rapidity evolution of Wilson-line operators, governed by the Balitsky--Kovchegov/Jalilian-Marian--Iancu--McLerran--Weigert--Leonidov--Kovner (BK/JIMWLK) equations~\cite{Balitsky:1995ub,Kovchegov:1999yj,Kovchegov:1999ua,Jalilian-Marian:1996mkd,Jalilian-Marian:1997qno,Jalilian-Marian:1997jhx,Jalilian-Marian:1997ubg,Kovner:2000pt,Weigert:2000gi,Iancu:2000hn,Iancu:2001ad,Ferreiro:2001qy}. \\

Historically, the Reggeization and shockwave approaches developed largely independently, providing complementary descriptions of high-energy QCD. The shockwave formalism became the natural framework for gluon saturation in dense systems~\cite{Gelis:2010nm}, whereas the language of parton Reggeization remained central to the investigation of the all-order structure of QCD amplitudes in the Regge limit~\cite{DelDuca:2013ara,DelDuca:2014cya,DelDuca:2017twk}. A major conceptual advance was achieved in Ref.~\cite{Caron-Huot:2013fea}, where gluon Reggeization was shown to emerge directly from the rapidity evolution of Wilson lines upon expansion around the dilute regime. This correspondence provides an operator description of gluon Reggeization at leading and next-to-leading logarithmic accuracy (LLA and NLLA) and, beyond the Regge-pole approximation, a framework for studying multi-Reggeon exchanges and Regge cuts. Indeed, at NNLLA, the simple Regge-pole structure of amplitudes with gluon quantum-number exchange is known to break down~\cite{DelDuca:2001gu} through Regge-cut contributions generated by multi-Reggeon exchange. These contributions represent one of the principal obstacles to extending the BFKL framework beyond NLLA~\cite{Falcioni:2020lvv,DelDuca:2021vjq,Caola:2021izf,Falcioni:2021buo,Falcioni:2021dgr,Byrne:2022wzk,Fadin:2023roz,Buccioni:2024gzo,Abreu:2024xoh}, and their complete structure in QCD remains an open problem~\cite{Fadin:2024eyf}. Significant progress has been achieved within the Wilson-line formalism~\cite{Caron-Huot:2013fea,Caron-Huot:2017fxr,Caron-Huot:2017zfo,Caron-Huot:2020grv}, while an alternative strategy based on the direct analysis of high-order Feynman diagrams has been developed in Refs.~\cite{Fadin:2017nka,Fadin:2023aen}. The derivation of Regge trajectories from the Lipatov effective action approach~\cite{Lipatov:1991nf,Kirschner:1994gd,Kirschner:1994xi,Lipatov:2000se,Nefedov:2019mrg,Hentschinski:2011xg} is limited to NLLA for the gluon~\cite{Chachamis:2012gh,Chachamis:2013hma} and LLA for the quark~\cite{Nefedov:2017qzc}, with no attempt to tackle Regge cuts so far. A related direction is represented by Glauber Soft-Collinear Effective Theory (SCET)~\cite{Rothstein:2016bsq,Moult:2022lfy,Gao:2024fyz,Gao:2024qsg,Moult:2017xpp,Moult:2019vou}, where the effective degrees of freedom are Glauber rather than Reggeized gluons, although the two descriptions are expected to be closely connected. Despite these developments, the Wilson-line operator description of Ref.~\cite{Caron-Huot:2013fea} has so far been available only for amplitudes carrying $t$-channel gluon quantum numbers. Its extension to $t$-channel quark exchange has remained an open problem. \\

The origin of this asymmetry is rooted in high-energy power counting. Whereas gluon exchange contributes at eikonal accuracy and is naturally encoded by conventional infinite Wilson lines, amplitudes mediated by $t$-channel quark exchange are suppressed by one power of the center-of-mass energy and belong to the subeikonal sector. A Wilson-line realization of quark Reggeization therefore requires extending the shockwave formalism beyond the eikonal approximation. Such extensions have been explored extensively in different contexts: i.) the spin and orbital-angular-momentum structure of hadrons~\cite{Kovchegov:2015pbl,Kovchegov:2016zex,Kovchegov:2016weo,Kovchegov:2017jxc,Kovchegov:2017lsr,Kovchegov:2018znm,Kovchegov:2018zeq,Kovchegov:2020kxg,Kovchegov:2020hgb,Adamiak:2021ppq,Kovchegov:2021lvz,Kovchegov:2021iyc,Cougoulic:2022gbk,Kovchegov:2022kyy,Borden:2023ugd,Kovchegov:2024aus,Borden:2024bxa,Adamiak:2025dpw,Kovchegov:2025gcg,Borden:2025ehe,Hatta:2016aoc,Kovchegov:2019rrz,Boussarie:2019icw,Kovchegov:2023yzd,Kovchegov:2024wjs,Cougoulic:2019aja,Cougoulic:2020tbc}, ii.) the precision studies of gluon saturation at the Electron-Ion Collider~\cite{Chirilli:2018kkw,Chirilli:2021lif,Altinoluk:2014oxa,Altinoluk:2015gia,Altinoluk:2015xuy,Agostini:2019avp,Agostini:2019hkj,Altinoluk:2020oyd,Agostini:2022ctk,Agostini:2022oge,Altinoluk:2021lvu,Altinoluk:2023qfr,Agostini:2023cvc,Altinoluk:2022jkk,Altinoluk:2024zom,Altinoluk:2024dba,Altinoluk:2024tyx,Agostini:2024xqs,Altinoluk:2025ang,Agostini:2025vvx,Altinoluk:2025ivn,Li:2023tlw,Li:2024fdb,Li:2024xra,Li:2026azt,Jalilian-Marian:2017ttv,Jalilian-Marian:2018iui,Jalilian-Marian:2019kaf}, and iii.) the connection between the small- and moderate-$x$ regimes of QCD~\cite{Balitsky:2015qba,Balitsky:2016dgz,Balitsky:2017flc,Boussarie:2020fpb,Boussarie:2021wkn,Boussarie:2023xun,Mukherjee:2026cte,Kar:2026vzk,Mukherjee:2026six}. Here, we show that the same next-to-eikonal (NEik) framework provides the natural operator language to understand quark Reggeization. \\

In this work, we establish a Wilson-line formulation of quark Reggeization for both massless and massive quarks. Starting from next-to-eikonal operators built from semi-infinite Wilson lines and a quark background-field insertion, we derive their nonlinear one-loop rapidity evolution using the background-field method. In the dilute regime, projection onto the positive-signature sector identifies a Reggeized-quark interpolating operator whose evolution reproduces the one-loop quark Regge trajectory. The negative-signature sector instead exhibits a coupled quark--gluon evolution, providing an operator realization of its known non-Regge-pole structure. In the planar limit, the mixing is suppressed and Reggeization emerges without an explicit signature projection, recovering the expected signature degeneracy. We further illustrate the universality of the construction by extracting the same positive-signature fundamental operator from both photon- and gluon-initiated transitions. Extending the formalism to finite quark mass, we derive a mass-dependent coordinate-space kernel whose Fourier transform reproduces the known massive-quark Regge trajectory. Finally, the perturbative structure of the allowed operator mixing provides strong evidence that the positive-signature Regge-pole eigenvalue remains unaffected by multi-Reggeon-operator contributions through NLLA. \\

The remainder of the paper is organized as follows. In Sec.~\ref{sec:Parton_Reggeization}, we review the essential features of gluon and quark Reggeization, with particular emphasis on signature. Section~\ref{sec:intro_Shockwave} introduces the shockwave formalism, rapidity evolution, and the NEik propagators required to describe $t$-channel quark exchange. In Sec.~\ref{sec:Gluon_Reggeization_Wilson_line}, we review the Wilson-line realization of the Reggeized gluon, which provides the blueprint for the quark construction. Our main results are presented in Sec.~\ref{sec:QuarkReggeShockwaveApproach}, where we construct the Reggeized-quark interpolating operator, derive its rapidity evolution, and analyze its signature structure. Section~\ref{sec:UniversGluon} establishes the universality of this operator construction using the gluon-initiated channel. In Sec.~\ref{sec:Massive_quarks}, we extend the construction to massive quarks and recover the corresponding Regge trajectory. We summarize our results and discuss future directions in Sec.~\ref{sec:Summary}.

\section{Review of parton Reggeization}
\label{sec:Parton_Reggeization}

In this section, we review the main results on parton Reggeization in perturbative QCD that will serve as benchmarks for the Wilson-line construction developed below. We first summarize the standard Regge-pole structure associated with gluon and quark exchange, emphasizing the role of color and signature. We then discuss the one-loop Regge trajectory of a massive quark, providing an expression exact in both the quark mass and the dimensional-regularization parameter.

\subsection{Summary of gluon Reggeization}
\label{sec:GluonRegge}

We begin with gluon Reggeization, which provides the reference framework for the quark construction developed in the remainder of this paper. In the high-energy limit,
\begin{equation}
s \simeq -u \gg |t| ,
\end{equation}
scattering amplitudes $A+B\to A'+B'$ carrying color-octet, negative-signature gluon quantum numbers in the $t$-channel assume the factorized Regge-pole form
\begin{equation}
{\cal A}_{AB\rightarrow A'B'}^{(8,-)}
=
\Gamma_{A'A}^{c}\,
\frac{s}{t}
\left[
\left(\frac{s}{-t}\right)^{\omega(t)}
+
\left(\frac{-s}{-t}\right)^{\omega(t)}
\right]
\Gamma_{B'B}^{c} .
\label{Pole:Eq:ReggeForm}
\end{equation}
Here, $1+\omega(t)$ denotes the gluon Regge trajectory, while $\Gamma_{A'A}^{c}$ and $\Gamma_{B'B}^{c}$ are particle--Reggeon--particle vertices describing the coupling of the external states to the exchanged Reggeized gluon. All dependence on the center-of-mass energy is carried by the Regge factors in Eq.~\eqref{Pole:Eq:ReggeForm}, whereas the effective vertices depend on $t$ but not on $s$. This factorized form has been established to all orders in perturbation theory at both leading and next-to-leading logarithmic accuracy.

At one loop and in $D=2+d=4-2\epsilon$ space-time dimensions, the gluon Regge trajectory is given by~\cite{Lipatov:1976zz}
\begin{align}
\omega^{(1)}(t)
&=
\frac{g^2 t}{(2\pi)^{d+1}}\,
\frac{N_c}{2}\,
(\mu^2)^{1-d/2}
\int
\frac{d^d\boldsymbol{k}}
{\boldsymbol{k}^2(\boldsymbol{p}-\boldsymbol{k})^2}
\nonumber\\
&=
-
\frac{g^2N_c\,\Gamma(1+\epsilon)}
{(4\pi)^{2-\epsilon}}
\left(\frac{\boldsymbol{p}^2}{\mu^2}\right)^{-\epsilon}
\frac{\Gamma^2(-\epsilon)}{\Gamma(-2\epsilon)} ,
\label{Eq:GluonReggeTraj}
\end{align}
where
\begin{equation}
t=p^2\simeq-\boldsymbol{p}^2 .
\end{equation}
At Born level, the quark--Reggeon--quark and gluon--Reggeon--gluon effective vertices take the common form
\begin{equation}
\Gamma_{A'A}^{c(0)}
=
g\,
\delta_{\lambda_{A'}\lambda_A}\,
\braket{A'|T^c|A},
\label{Eq:LOpRp}
\end{equation}
where $\lambda_A$ and $\lambda_{A'}$ denote the helicities of the incoming and outgoing partons, respectively, and $T^c$ is the color generator in the corresponding representation.

The Regge trajectory and the particle--Reggeon--particle vertices can be extracted by comparing the high-energy expansion of fixed-order amplitudes with the Regge-pole ansatz in Eq.~\eqref{Pole:Eq:ReggeForm}. Consider, for example, elastic quark--quark scattering projected onto color-octet exchange with negative signature:
\begin{align}
{\cal A}_{qq\rightarrow qq}^{(8,-)}
&=
\Gamma_{qq}^{c}\,
\frac{s}{t}
\left[
\left(\frac{s}{-t}\right)^{\omega(t)}
+
\left(\frac{-s}{-t}\right)^{\omega(t)}
\right]
\Gamma_{qq}^{c}
\nonumber\\
&\simeq
\Gamma_{qq}^{c(0)}
\frac{2s}{t}
\Gamma_{qq}^{c(0)}
\nonumber\\
&\quad
+
\Gamma_{qq}^{c(0)}
\frac{s}{t}\,
\omega^{(1)}(t)
\left[
\ln\left(\frac{s}{-t}\right)
+
\ln\left(\frac{-s}{-t}\right)
\right]
\Gamma_{qq}^{c(0)}
+
\Gamma_{qq}^{c(1)}
\frac{2s}{t}
\Gamma_{qq}^{c(0)}
+
\Gamma_{qq}^{c(0)}
\frac{2s}{t}
\Gamma_{qq}^{c(1)}
+
\mathcal{O}(g^6) .
\label{Pole:Eq:ReggeFormEx1}
\end{align}
Comparison with the corresponding one-loop amplitude determines both the trajectory $\omega^{(1)}(t)$ and the one-loop correction $\Gamma_{qq}^{c(1)}$ to the quark--Reggeon--quark vertex. Similarly, the gluon--Reggeon--gluon vertex $\Gamma_{gg}^{c(1)}$ can be extracted from elastic gluon--gluon scattering~\cite{Fadin:2001dc}. The universality implied by gluon Reggeization can then be tested in the mixed quark--gluon channel: the high-energy limit of the elastic quark--gluon amplitude must be reproduced using the same trajectory $\omega^{(1)}(t)$ together with the independently extracted vertices $\Gamma_{qq}^{c(1)}$ and $\Gamma_{gg}^{c(1)}$. More generally, the consistency of the Regge-pole ansatz with $s$-channel unitarity is encoded in the bootstrap conditions~\cite{Braun:1999uz,Fadin:2000ww,Fadin:2002hz}, whose fulfillment provides the basis for gluon Reggeization and for the BFKL construction at LLA and NLLA.

\subsection{Summary of quark Reggeization}
\label{sec:QuarkRegge}

The Reggeization of the gluon has a direct analogue for amplitudes carrying quark quantum numbers in the $t$-channel~\cite{Fadin:1976nw,Fadin:1977jr}, which led to considerable progress in the formulation of BFKL for quark~\cite{Kotsky:2002aq,Bogdan:2004cg,Bogdan:2006wq,Bogdan:2007qj}. In this case, Regge-pole behavior occurs in the positive-signature color-triplet channel. The corresponding amplitudes can be written as~\cite{Fadin:2001dc}
\begin{equation}
{\cal A}_{AB\rightarrow A'B'}^{(3,+)}
=
\Gamma_{A'A}\,
\frac{\sqrt{s}}{m-\slashed{p}_{\perp}}\,
\frac{1}{2}
\left[
\left(\frac{-s}{-t}\right)^{\delta(\slashed{p}_{\perp},m)}
+
\left(\frac{s}{-t}\right)^{\delta(\slashed{p}_{\perp},m)}
\right]
\Gamma_{B'B},
\label{Pole:Eq:QuarkReggeForm}
\end{equation}
where $m$ is the quark mass and $\Gamma_{A'A}$ and $\Gamma_{B'B}$ denote the particle--Reggeon transition vertices coupling the external states to the exchanged Reggeized quark. The quantity $\delta(\slashed{p}_{\perp},m)$ defines the quark Regge trajectory. Because the massive trajectory contains nontrivial Dirac structures, it is understood as a function of the transverse Dirac matrix $\slashed{p}_{\perp}$. \\

At one loop, the quark Regge trajectory is given by~\cite{Fadin:1977jr}
\begin{equation}
\delta^{(1)}(\slashed{p}_{\perp},m)
=
\frac{g^2C_F}{(2\pi)^{d+1}}\,
(\slashed{p}_{\perp}-m)\,
(\mu^2)^{1-d/2}
\int
\frac{d^d\boldsymbol{k}}
{(\slashed{k}_{\perp}-m)(\boldsymbol{p}-\boldsymbol{k})^2}.
\label{Eq:QuarkTraj}
\end{equation}
As in the gluon case, the Regge-pole form in Eq.~\eqref{Pole:Eq:QuarkReggeForm} is accompanied by process-dependent transition vertices, while the trajectory is universal. It is useful to first consider the massless limit of Eq.~\eqref{Eq:QuarkTraj}. Setting $m=0$ gives
\begin{align}
\delta^{(1)}(\slashed{p}_{\perp})
&=
\frac{g^2C_F}{(2\pi)^{d+1}}\,
\slashed{p}_{\perp}\,
(\mu^2)^{1-d/2}
\int
\frac{d^d\boldsymbol{k}}
{\boldsymbol{k}^2(\boldsymbol{p}-\boldsymbol{k})^2}\,
\slashed{k}_{\perp}
\nonumber\\
&=
\frac{g^2t}{(2\pi)^{d+1}}\,
\frac{C_F}{2}\,
(\mu^2)^{1-d/2}
\int
\frac{d^d\boldsymbol{k}}
{\boldsymbol{k}^2(\boldsymbol{p}-\boldsymbol{k})^2}.
\label{Eq:MasslessQuarkTrajectoryIntegral}
\end{align}
Comparison with the one-loop gluon trajectory in Eq.~\eqref{Eq:GluonReggeTraj} immediately yields
\begin{equation}
\delta^{(1)}(\slashed{p}_{\perp})
=
\delta^{(1)}(t)
=
\frac{C_F}{N_c}\,
\omega^{(1)}(t).
\label{Eq:QuarkReggeTraj}
\end{equation}
This one-loop Casimir-scaling relation between the quark and gluon trajectories will provide an important benchmark for the Wilson-line derivation presented below. For later comparison with the coordinate-space calculation, it is also convenient to derive an explicit expression for the massive trajectory that is exact in the dimensional-regularization parameter. Introducing a Feynman parameter in Eq.~\eqref{Eq:QuarkTraj}, one obtains
\begin{align}
\delta^{(1)}(\slashed{p}_{\perp},m)
&=
\frac{g^2\Gamma(1+\epsilon)}
{(4\pi)^{2-\epsilon}}\,
2C_F\,
(\slashed{p}_{\perp}-m)
(\mu^2)^\epsilon
\int_0^1 dx\,
\frac{x\slashed{p}_{\perp}+m}
{\left[(1-x)(x\boldsymbol{p}^2+m^2)\right]^{1+\epsilon}}
\nonumber\\
&=
-
\frac{g^2\Gamma(1+\epsilon)}
{(4\pi)^{2-\epsilon}}\,
2C_F
\left(\frac{m^2}{\mu^2}\right)^{-\epsilon}
\Bigg\{
\left(1+\frac{\boldsymbol{p}^2}{m^2}\right)
\int_0^1 dx\,
x(1-x)^{-1-\epsilon}
\left(1+x\frac{\boldsymbol{p}^2}{m^2}\right)^{-1-\epsilon}
\nonumber\\
&\hspace{4.1cm}
+
\left(1-\frac{\slashed{p}_{\perp}}{m}\right)
\int_0^1 dx\,
(1-x)^{-\epsilon}
\left(1+x\frac{\boldsymbol{p}^2}{m^2}\right)^{-1-\epsilon}
\Bigg\}.
\label{Eq:QuarkTrajectoryFeynmanParameter}
\end{align}
The two parameter integrals can be expressed in terms of Gauss hypergeometric functions, yielding
\begin{align}
\delta^{(1)}(\slashed{p}_{\perp},m)
&=
-
\frac{g^2\Gamma(1+\epsilon)}
{(4\pi)^{2-\epsilon}}\,
2C_F
\left(\frac{m^2}{\mu^2}\right)^{-\epsilon}
\Bigg\{
\left(1+\frac{\boldsymbol{p}^2}{m^2}\right)
\frac{\Gamma(-\epsilon)}{\Gamma(2-\epsilon)}
\nonumber\\
&\hspace{2.2cm}\times
{}_2F_1\left(
1+\epsilon,2;2-\epsilon;
-\frac{\boldsymbol{p}^2}{m^2}
\right)
\nonumber\\
&\hspace{1.2cm}
+
\left(1-\frac{\slashed{p}_{\perp}}{m}\right)
\frac{\Gamma(1-\epsilon)}{\Gamma(2-\epsilon)}
{}_2F_1\left(
1+\epsilon,1;2-\epsilon;
-\frac{\boldsymbol{p}^2}{m^2}
\right)
\Bigg\}.
\label{Eq:QuarkReggeTrajEpsilonExact}
\end{align}
Equation~\eqref{Eq:QuarkReggeTrajEpsilonExact} is exact in both $m$ and $\epsilon$. Its massless limit can be obtained using the large-argument expansion
\begin{align}
{}_2F_1(a,b;c;z)
\underset{|z|\to\infty}{\sim}
&\,
\frac{\Gamma(c)\Gamma(b-a)}
{\Gamma(b)\Gamma(c-a)}
(-z)^{-a}
+
\frac{\Gamma(c)\Gamma(a-b)}
{\Gamma(a)\Gamma(c-b)}
(-z)^{-b},
\label{Eq:Exp2F1}
\end{align}
which reproduces Eq.~\eqref{Eq:QuarkReggeTraj} in the limit $\boldsymbol{p}^2/m^2\to\infty$. \\

Alternatively, expanding the hypergeometric functions around $\epsilon=0$ gives
\begin{align}
{}_2F_1(1+\epsilon,2;2-\epsilon;z)
&=
\frac{1}{1-z}
-
\epsilon
\left[
\frac{1}{1-z}
+
\frac{1+z}{z}\ln(1-z)
\right]
+
\mathcal{O}(\epsilon^2),
\label{Eq:HypergeometricExpansionOne}
\\
{}_2F_1(1+\epsilon,1;2-\epsilon;z)
&=
-\frac{\ln(1-z)}{z}
+
\mathcal{O}(\epsilon).
\label{Eq:HypergeometricExpansionTwo}
\end{align}
Substituting these expansions into Eq.~\eqref{Eq:QuarkReggeTrajEpsilonExact}, one finds
\begin{align}
\delta^{(1)}(\slashed{p}_{\perp},m)
&=
\frac{g^2\,2C_F\,\Gamma(1+\epsilon)}
{(4\pi)^{2-\epsilon}}
\left(\frac{m^2}{\mu^2}\right)^{-\epsilon}
\Bigg[
\frac{1}{\epsilon}
-
\ln\left(1+\frac{\boldsymbol{p}^2}{m^2}\right)
+
\frac{m\slashed{p}_{\perp}}{\boldsymbol{p}^2}
\ln\left(1+\frac{\boldsymbol{p}^2}{m^2}\right)
\Bigg]
+
\mathcal{O}(\epsilon).
\label{Eq:QuarkReggeTrajEpsilonExp}
\end{align}
This expression gives the massive one-loop quark trajectory through finite order in the dimensional-regularization parameter.

\subsection{More on the signature of the Reggeized quark}
\label{sec:QuarkSignature}

In this subsection, we discuss in more detail the role of signature in amplitudes mediated by Reggeon exchange. This quantum number is central to the analysis developed in the remainder of the paper. We first recall its definition and then summarize the current status of Reggeization for amplitudes carrying quark quantum numbers in the $t$-channel. Besides providing useful benchmarks for our Wilson-line construction, this discussion will clarify the different logarithmic counting of the positive- and negative-signature sectors.

As a reference process, consider quark--photon Compton scattering in the high-energy limit,
\begin{equation}
s\simeq -u\gg |t|.
\end{equation}
The components with definite signature are obtained by symmetrizing or antisymmetrizing the amplitude under the crossing transformation $s\leftrightarrow u$:
\begin{equation}
\mathcal{M}^{(\pm)}_{\gamma q\rightarrow q\gamma}
=
\frac{1}{2}
\left[
\mathcal{M}_{\gamma q\rightarrow q\gamma}
\pm
\left.
\mathcal{M}_{\gamma q\rightarrow q\gamma}
\right|_{s\leftrightarrow u}
\right].
\label{Eq:QuarkSignatureProjection}
\end{equation}
Accordingly, the full amplitude can be decomposed as
\begin{equation}
\mathcal{M}_{\gamma q\rightarrow q\gamma}
=
\mathcal{M}^{(+)}_{\gamma q\rightarrow q\gamma}
+
\mathcal{M}^{(-)}_{\gamma q\rightarrow q\gamma}.
\label{Eq:QuarkSignatureDecomposition}
\end{equation}
More precisely, for amplitudes involving external fermions, signaturization is naturally defined for truncated amplitudes, with the external wave functions removed, together with the appropriate analytic continuation between the crossed channels. In the Regge limit, this procedure reduces to the transformation in Eq.~\eqref{Eq:QuarkSignatureProjection} within the conventions adopted here.

The Reggeization of amplitudes carrying positive-signature quark exchange was first proposed in Refs.~\cite{Fadin:1976nw,Fadin:1977jr} and subsequently established to all orders at leading logarithmic accuracy through the bootstrap approach~\cite{Bogdan:2006af}. Moreover, the high-energy limit of the two-loop quark--gluon scattering amplitude was found to be compatible with the Regge-pole form at next-to-leading logarithmic accuracy~\cite{Bogdan:2002sr}, allowing the extraction of the two-loop quark Regge trajectory. Although this fixed-order result does not constitute an all-order proof at NLLA, it provides strong evidence that the positive-signature sector continues to Reggeize at this accuracy.

The situation is qualitatively different in the negative-signature sector. At finite $N_c$, negative-signature amplitudes are not described by a single Regge pole~\cite{Fadin:1976nw,Fadin:1977jr}. Their partial waves contain branch-point singularities associated with the compound exchange of a Reggeized quark and a Reggeized gluon. \\

An important point concerns the perturbative order at which the two signature components first appear. The positive-signature amplitude contains a Born contribution, whereas the negative-signature amplitude starts only at one loop:
\begin{equation}
\mathcal{M}^{(+)}_{\gamma q\rightarrow q\gamma}
\sim \mathcal{O}(\alpha_s^0),
\qquad
\mathcal{M}^{(-)}_{\gamma q\rightarrow q\gamma}
\sim \mathcal{O}(\alpha_s),
\label{Eq:SignaturePerturbativeCounting}
\end{equation}
where the common couplings associated with the Born process are understood to have been factored out. Their leading logarithmic towers consequently have the schematic form
\begin{align}
\mathcal{M}^{(+)}_{\gamma q\rightarrow q\gamma}
&\sim
\mathcal{M}_{\rm Born}
\sum_{n=0}^{\infty}
c_n\left(\alpha_s\ln s\right)^n,
\nonumber\\
\mathcal{M}^{(-)}_{\gamma q\rightarrow q\gamma}
&\sim
\alpha_s\,\mathcal{M}_{\rm Born}
\sum_{n=0}^{\infty}
d_n\left(\alpha_s\ln s\right)^n.
\label{Eq:SignatureLogarithmicCounting}
\end{align}
Thus, the leading-logarithmic series of the negative-signature component is suppressed by one power of $\alpha_s$ relative to the corresponding positive-signature series. Equivalently, it contributes only at NLLA to the full, non-signaturized amplitude. This distinction between the logarithmic accuracy intrinsic to a definite-signature sector and that of the full amplitude will be important below.

It follows that the full amplitude Reggeizes at LLA: at this accuracy, only the positive-signature component contributes. At NLLA, however, the leading contribution from the negative-signature sector must also be included. Since this contribution contains a Regge-cut component at finite $N_c$, the full non-signaturized amplitude can no longer be represented by a single Reggeized-quark pole, even though its positive-signature projection remains compatible with Reggeization.

The structure simplifies considerably in the planar limit. Fixed-order results and the large-$N_c$ color structure indicate that the non-Regge-pole contribution in the negative-signature channel is suppressed, so that this sector is governed by the same trajectory as its positive-signature counterpart at its leading logarithmic order. This phenomenon is commonly referred to as \emph{signature degeneracy}. Consequently, in the large-$N_c$ limit, the full amplitude is expected to recover Regge-pole behavior through NLLA without requiring an explicit signature projection.

Notice that establishing the full amplitude at NLLA requires the positive-signature component at NLLA but the negative-signature component only at its own LLA. The NLLA evolution intrinsic to the negative-signature sector would first contribut at the NNLLA to the full amplitude and remains unknown. The status of the different signature sectors is summarized in Table~\ref{tab:QuarkReggeizationStatus}.

\begin{table}[t]
\centering
\renewcommand{\arraystretch}{1.25}
\begin{tabular}{|c|cc|cc|}
\hline
&
\multicolumn{2}{c|}{Finite $N_c$}
&
\multicolumn{2}{c|}{Large-$N_c$ limit}
\\
\cline{2-5}
&
LLA & NLLA & LLA & NLLA
\\
\hline
Positive signature
&
{\color{green!95!black}\cmark}
&
\cmark
&
{\color{green!95!black}\cmark}
&
\cmark
\\
\hline
Negative signature
&
{\color{red}\xmark}
&
{\color{red}\xmark}
&
\cmark
&
\textbf{?}
\\
\hline
Full amplitude
&
{\color{green!95!black}\cmark}
&
{\color{red}\xmark}
&
{\color{green!95!black}\cmark}
&
\cmark
\\
\hline
\end{tabular}
\caption{Status of quark Reggeization in the different signature sectors and logarithmic approximations. Green symbols denote statements established to all orders at the indicated accuracy, while black symbols denote expectations supported by fixed-order results. A cross indicates the presence of contributions that prevent a description in terms of a single Reggeized-quark pole. The question mark denotes a logarithmic sector that remains undetermined.}
\label{tab:QuarkReggeizationStatus}
\end{table}
\section{Shockwave formalism and its NEik extension}
\label{sec:intro_Shockwave}

For the convenience of the reader, we briefly review the semi-classical effective approach to QCD at small-$x$, within which scattering amplitudes are naturally described in terms of eikonal Wilson lines. We will use what in the literature is generically referred to as the shockwave formalism~\cite{McLerran:1994vd,Balitsky:1995ub}. \\

\subsection{Kinematics and notation}

We introduce a light-cone basis composed of $n_1$ and $n_2$, with $n_1 \cdot n_2 = 1$ defining the $+/-$ direction. In this basis, any vector $k$ admits the Sudakov decomposition
\begin{equation}
k^\mu = k^+ n_1^\mu + k^- n_2^\mu + k_\perp^\mu
\end{equation}
and the scalar product of two such vectors takes the form
\begin{equation}
    k \cdot q = k^+ q^- + k^- q^+ + k_\perp \cdot q_\perp  
    = k^+ q^- + k^- q^+ -\boldsymbol{k} \cdot \boldsymbol{q}\, ,
\end{equation}
where, throughout this work, transverse momenta written in bold denote Euclidean vectors, while the same quantities carrying a $\perp$ index are understood in the Minkowski sense. Throughout, we adopt what we refer to as the projectile frame: a reference frame in which the target moves ultrarelativistically towards the projectile. In this frame, particles associated with the projectile travel predominantly along $n_1$ (the $+$ direction), while those on the target side carry a large component along $n_2$ (the $-$ direction).

\subsection{Effective Lagrangian and Wilson lines}

The shockwave approach proceeds by separating the QCD gluon field $\mathbf{A}$ into an external background component $\mathcal{A}$ and an internal component $A$, depending on whether the corresponding $+$-momentum lies below or above an arbitrary rapidity cutoff $e^\eta p_{p}^+$, with $\eta<0$. Once boosted from the target rest frame into the projectile frame, this background field becomes highly localized in $z^+$ and effectively takes the form
\begin{equation}
    \mathcal{A}^\mu (z) = \mathcal{A}^-(z^+, z^- = 0, z_\perp) n_2^\mu \sim \delta (z^+) \mathcal{A} (\boldsymbol{z})  n_2^\mu \,,
\end{equation}
independent of $z^-$ and sharply peaked around $z^+=0$ — the configuration commonly known as the (eikonal) shockwave approximation. Our starting point is the QCD Lagrangian, which we split into free and interacting parts,
\begin{gather} 
\mathcal{L} =-\frac{1}{4} F_{a \mu \nu} F^{a \mu \nu} + \bar{\boldsymbol{\psi}} (i\slashed{D}-m) \boldsymbol{\psi}  =\mathcal{L}_{\text {free }}+\mathcal{L}_{\text {int }} \nonumber \; , 
\end{gather}
\begin{gather}
 \mathcal{L}_{\text {int }}  =-g f^{a b c} {\bf A}_\mu^b {\bf A}_\nu^c \partial^\mu {\bf A}^{\nu a} - \frac{1}{4} g^2 f^{a b c} f^{a d e} {\bf A}_\mu^b {\bf A}_\nu^c {\bf A}^{\mu d} {\bf A}^{\nu e}+i \bar{ \boldsymbol{\psi}} \left(-i g t^a \slashed{\boldsymbol{A}^a} \right) \boldsymbol{\psi} \; .
\end{gather}
We then implement the field split announced above,
\begin{equation}
    {\bf A}_{\mu}^a (k) = A_\mu^a ( k^+ > e^\eta p_{p}^+, k^-, k_{\perp}) + \mathcal{A}^a ( k^+ < e^\eta p_{p}^+, k^-, k_{\perp}) \; ,
\end{equation}
and choose the light-cone gauge $n_2 \cdot A = 0$. Together with the fact that $\mathcal{A}^2 \propto n_2^2 = 0$, this choice considerably simplifies the resulting effective field theory, so that the interacting Lagrangian reorganizes into
\begin{gather}
\mathcal{L}_{\text {int }}=-g f^{a b c} A_\mu^b\left(\partial^\mu A^{\nu a}\right) A_\nu^c-\frac{1}{4} g^2 f^{a b c} f^{a d e} A_\mu^b A^{\mu d} A^{\nu e} A_\nu^c+\bar{ \boldsymbol{\psi}}\left(g t^a \slashed{A}^a\right) \boldsymbol{\psi} \nonumber \\
-g f^{a b c} \mathcal{A}_\mu^b\left(\partial^\mu A^{\nu a}\right) A_\nu^c+\bar{ \boldsymbol{\psi}}\left(g t^a \slashed{\mathcal{A}}^a\right) {\boldsymbol{\psi}} \; .
\end{gather}
The first line reproduces the standard QCD interaction vertices. The second line instead contains the couplings between the fast projectile fields and the shockwave background and can be written in the compact form
\begin{gather}
\mathcal{L}_{\text {int }}^S =-g f^{a c b} \mathcal{A}^{-c} g^{\alpha \beta}\left(\frac{\partial A_\beta^a}{\partial x^{-}}\right) A_\alpha^b+\bar{ \boldsymbol{\psi}}\left(g t^a \slashed{\mathcal{A}}^a\right) \boldsymbol{\psi}  = - g f^{a c b} \mathcal{A}^{-c} g^{\alpha \beta}\left(\frac{\partial A_\beta^a}{\partial x^{-}}\right) A_\alpha^b+\bar{ \boldsymbol{\psi}} g \mathcal{A}^{-} \gamma^{+} \boldsymbol{\psi} \nonumber \\ = g f^{c a b} \mathcal{A}^{-c} g_{\perp}^{\alpha \beta}\left(\frac{\partial A_\beta^a}{\partial x^{-}}\right) A_\alpha^b+\bar{ \boldsymbol{\psi}} g \mathcal{A}^{-} \gamma^{+} \boldsymbol{\psi} =i g T_{a b}^c \mathcal{A}^{-c} g_{\perp}^{\alpha \beta}\left(\frac{\partial A_\beta^a}{\partial x^{-}}\right) A_\alpha^b+ g \bar{\boldsymbol{\psi}}  \mathcal{A}^{-} \gamma^{+} \boldsymbol{\psi} \; ,
\label{Eq:Effective_Lagran_inter_Shockwave}
\end{gather}
where we have set $\mathcal{A}^{-}(x) = t^a \mathcal{A}^{-a} (x)$ and $T_{ab}^{c} = - i f^{abc}$. \\

Resumming this interaction to all orders in the background field, within the leading eikonal approximation, gives rise to a Wilson line describing propagation through the shockwave localized around $z^+=0$:
\begin{equation}
   \mathcal{U}_R^{\eta} (\boldsymbol{z} ) \equiv \mathcal{U}_R (\boldsymbol{z} )  = \mathcal{P} \exp \left(i g \int d z^+ \mathcal{A}^-(z^+, \boldsymbol{z}) \cdot T_R \right) \, ,
   \label{Eq:WilsonLineDef}
\end{equation}
with $\mathcal{P}$ the path-ordering operator along the $+$ direction. From this all-order resummation, one can derive the effective Feynman rules governing the interaction of a fast-moving quantum projectile with the classical shockwave background — in particular the shockwave propagators\footnote{As shown in Ref.~\cite{Boussarie:2024bdo,Boussarie:2024pax}, this construction can be further generalized by defining more general effective background-field operators.} that capture arbitrarily many interactions with the background field (see Ref.~\cite{Li:2023ihv} for the complete list at eikonal accuracy). \\

The scattering amplitude then follows the familiar small-$x$ factorized structure: the convolution of the projectile impact factor with the non-perturbative matrix element of Wilson-line operators evaluated between target states. The operators are functional of the Wilson lines, $\mathcal{O}[\mathcal{U}]$, and, as such, they implicitly depend on the rapidity cut-off $\eta$. 

\subsection{Rapidity renormalization group evolution in the shockwave formalism}

In this formalism, the small-$x$ evolution of such operators is determined by renormalization-group equation in $\eta$. The methodology employed in this work to derive the evolution equations for shockwave operators is based on the high-energy operator product expansion (OPE) introduced by Balitsky~\cite{Balitsky:1995ub} in the leading eikonal approximation and more recently extended to next-to-eikonal accuracy in Refs.~\cite{Chirilli:2018kkw,Chirilli:2021lif}. \\

For illustration, we briefly review the procedure at leading eikonal accuracy, following closely~\cite{Grabovsky:2013mba}. Within the shockwave formalism, a rapidity cut-off $\eta$ is introduced in order to separate slow (classical) and fast (quantum) degrees of freedom. As in standard renormalization-group approaches, the dependence of the relevant operators on this cut-off determines their evolution, which in the present case corresponds to the small-$x$ evolution. \\

To derive this dependence, one shifts the rapidity cut-off from $\eta$ to $\eta+\Delta\eta$ and decomposes the background field as
\begin{gather}
\mathcal{A}^{-}_{\eta+\Delta\eta}(z)
=
\mathcal{A}^{-}_{\eta}(z)
+
\mathcal{A}^{-}_{\Delta\eta}(z),
\end{gather}
where $\mathcal A^-_{\Delta\eta}$ denotes the semi-fast fluctuation field contained in the rapidity slice, \(
e^\eta p_p^+ < k^+ < e^{\eta+\Delta\eta} p_p^+
\), which is integrated out in evolving from $\eta$ to $\eta+\Delta\eta$, namely
\begin{gather}
\mathcal{A}^{-}_{\Delta\eta}(z)
=
\int
\frac{d^Dk}{(2\pi)^D}
e^{-ik\cdot z}
\mathcal{A}^{-}_{\Delta\eta}(k)
\,
\theta(k^+-e^\eta p_p^+)
\theta(e^{\eta+\Delta\eta}p_p^+-k^+).
\end{gather}
Substituting this decomposition into the Wilson line and expanding in the fluctuation field yields
\begin{gather}
    \mathcal{U}_F^{\eta+ \Delta \eta} (x^+, y^+, \boldsymbol{z}) = \mathcal{U}_F^{\eta} (x^+, y^+, \boldsymbol{z}) + ig \int_{y^+}^{x^+} d z_1^+ \; \mathcal{U}_F^{\eta} (x^+, z_1^+, \boldsymbol{z}) \mathcal{A}^{-}_{\Delta \eta} (z_1^+, \boldsymbol{z} )  \mathcal{U}_F^{\eta} (z_1^+, y^+, \boldsymbol{z}) \nonumber \\ + (ig)^2 \int_{y^+}^{x^+} d z_1^+ \; \int_{y^+}^{z_1^+} d z_2^+ \; \mathcal{U}_F^{\eta} (x^+, z_1^+, \boldsymbol{z}) \mathcal{A}^{-}_{\Delta \eta} (z_1^+, \boldsymbol{z} )  \mathcal{U}_F^{\eta} (z_1^+, z_2^+, \boldsymbol{z}) \mathcal{A}^{-}_{\Delta \eta} (z_2^+, \boldsymbol{z} ) \mathcal{U}_F^{\eta} (z_2^+, y^+, \boldsymbol{z}) \nonumber \\ +\mathcal O((\mathcal A^-_{\Delta\eta})^3) \; .
    \label{Eq:ExpansionFluct}
\end{gather}
where the omitted terms correspond to insertions of more quantum fields $\mathcal{A}_{\Delta\eta}^-$ along the eikonal trajectory. The evolution equation is then obtained by treating the modes with
$ \displaystyle \ln \left( \frac{k^+}{p_p^+} \right) <\eta$ as a classical external background and integrating over the quantum fluctuations in the rapidity slice
\[
\eta<\ln \left( \frac{k^+}{p_p^+} \right) <\eta+\Delta\eta.
\]
Accordingly, one computes the expectation value of a given operator $\mathcal{O}$ built from the Wilson lines
\begin{equation}
\mathcal{O}^{\eta+\Delta\eta}[\mathcal U]= \frac{ \langle0| T\left\{ \mathcal{O}^{\eta}[\mathcal U] e^{i\int dz\,\mathcal L(z)}
\right\} |0\rangle}{\langle0| T\left( e^{i\int dz\,\mathcal L(z)}\right) |0\rangle} \; ,
\end{equation}
where the average is taken over the free quantum fields restricted to the rapidity interval $\Delta\eta$. \\

In practice, the effect of quantum fluctuations is evaluated through Feynman diagrams involving propagators in the shockwave background. One obtains
\begin{gather}
\mathcal{O}^{\eta+\Delta\eta}[\mathcal U]
=
\mathcal{O}^{\eta}[\mathcal U]
+
\Delta\eta\,\mathcal K[\mathcal O^\eta]
+
\mathcal O(\Delta\eta^2) \; ,
\end{gather}
where
\begin{gather}
\mathcal K[\mathcal O^\eta]\equiv \frac{I_R+I_V}{\Delta \eta}
\label{Eq:KernelDiagram}
\end{gather}
is the one-loop evolution kernel. Here, $I_R$ denotes the "real" contributions, corresponding to diagrams in which the emitted gluon crosses the shockwave, while $I_V$ denotes the "virtual" contributions, corresponding to diagrams without shockwave crossing. Taking the infinitesimal limit in rapidity, one obtains
\begin{equation}
\frac{\partial}{\partial\eta}
\mathcal O^\eta[\mathcal U]
=
\lim_{\Delta\eta\to0}
\frac{
\mathcal O^{\eta+\Delta\eta}[\mathcal U]
-
\mathcal O^\eta[\mathcal U]
}{\Delta\eta}
=
\mathcal K[\mathcal O^\eta],
\end{equation}
which determines the one-loop evolution equation for the operator $\mathcal O^\eta[\mathcal U]$. \\

In order to determine the diagrams through the shockwave, $I_R$ and $I_V$, it is useful to know the effective propagators through the shockwave. At eikonal accuracy, we will need the gluon propagator, which is known since the early years of gluon saturation~\cite{McLerran:1994vd}. We present it in mixed momentum-coordinate space representation (see e.g. Eq.~(2.18) of ref.~\cite{Boussarie:2024bdo}) 
\begin{align}
    G_{\mu \nu}^{ab} (z_1, z_2)  & = - \frac{(-i)^d}{2 (2 \pi)^{1+d} }  \int d^d \boldsymbol{z}_3 [ z_2^+ g_{\perp \mu}^{\alpha} - z_{23 \perp}^{\alpha} n_{2 \mu}  ] \mathcal{U}_A^{ab} (\boldsymbol{z}_3)  [ -z_1^+ g_{\perp \alpha \nu} - z_{31 \perp \alpha} n_{2 \nu}  ] 
    \nonumber \\ 
    & \times \int_0^{\infty^+} d k^+ \frac{ (k^+)^{d-1} }{(-z_1^+ z_2^+)^{1+d/2}}  e^{i k^+ \left( - z_{21}^{-} + \frac{ \boldsymbol{z}_{23}^2 }{2 z_2^+} - \frac{ \boldsymbol{z}_{31}^2 }{2 z_1^+} +  i \varepsilon\right)} \; .
\end{align}
To derive evolution equations, we only need propagators constructed from $\mathcal{A}_{\Delta\eta}^-$, whose interaction with the fast-moving projectile remains purely eikonal, namely
\begin{align}
   G_{\Delta \eta}^{--, ab} (z_1, z_2)  & = n_{1 \mu} n_{1 \nu} G_{\Delta \eta}^{\mu \nu, ab}  =   \frac{(-i)^d}{2 (2 \pi)^{1+d} }  \int d^d \boldsymbol{z}_3 \; \mathcal{U}_A^{ab} (\boldsymbol{z}_3) \; \boldsymbol{z}_{23} \cdot \boldsymbol{z}_{31}    
   \nonumber \\ 
   & \times \int_{e^{\eta} p_p^+}^{e^{\eta + \Delta \eta} p_p^+} d k^+ \frac{ (k^+)^{d-1} }{(-z_1^+ z_2^+)^{1+d/2}}  e^{i k^+ \left( - z_{21}^{-} + \frac{ \boldsymbol{z}_{23}^2 }{2 z_2^+} - \frac{ \boldsymbol{z}_{31}^2 }{2 z_1^+} +  i \varepsilon\right)} \; .
   \label{Eq:GluonPropEvoLim}
\end{align}

\subsection{Beyond the eikonal approximation: $t$-channel quark exchange and NEik propagators}

The discussion presented so far relies entirely on the eikonal approximation, in which the interaction of a fast-moving projectile with the shockwave is encoded in the eikonal Wilson lines. In this limit, the interaction is insensitive to the quark background field inside the shockwave, and the incoming parton preserves its identity upon traversing the target. Consequently, the eikonal formalism naturally describes high-energy processes mediated by $t$-channel gluon exchange. By construction, interactions that change the identity of the propagating parton are absent in this limit. Extending the shockwave formalism beyond the eikonal approximation has therefore attracted considerable attention in recent years. In particular, NEik corrections to effective parton propagators in a pure gluon background have been systematically derived and applied to various high-energy observables \cite{Chirilli:2021lif,Altinoluk:2014oxa,Altinoluk:2022jkk}. It has long been recognized that interactions with the quark background field through $t$-channel quark exchange also constitute a NEik effect, requiring a corresponding extension of the shockwave formalism \cite{Kovchegov:2015pbl,Altinoluk:2023qfr,Chirilli:2026pkv,Mukherjee:2026six}. Since the present work is concerned with the Wilson-line realization of quark Reggeization, we restrict our attention to this particular class of NEik interactions. Before presenting the effective parton propagators relevant for this construction, it is useful to briefly discuss their power counting. 

The total quark field can be split into fast (quantum) and slow (classical) modes, in complete analogy with the gluon field, such that 
 \begin{equation}
    \boldsymbol{\psi} (k) = \psi ( k^+ > e^\eta p_{p}^+, k^-, k_{\perp}) + \Psi ( k^+ < e^\eta p_{p}^+, k^-, k_{\perp}) \; ,
\end{equation}
where the scattering is described by the classical component $\Psi$ of the background field in the high-energy limit which can be isolated by a longitudinal boost $\gamma_t$ along the $x^-$ direction. The background quark field $\Psi(z)$ associated with the target can be conveniently decomposed into its “good” and “bad” components
\begin{align}
\label{Eq:GoodComp}
\Psi^{(-)}( z )&=\frac{\gamma^+\gamma^-}{2} \Psi( z ) \, ,
\end{align}
and
\begin{align}
\label{Eq:BadComp}
\Psi^{(+)}( z )&=\frac{\gamma^-\gamma^+}{2}\Psi( z ) \, ,
\end{align}
respectively. Under the longitudinal boost $\gamma_t$ of the target, the two components scale as 
\begin{align}
\label{eq:good-bad_scaling}
&\Psi^{(-)}( z ) \propto (\gamma_t)^{1/2} \, ,\\
&\Psi^{(+)}( z ) \propto (\gamma_t)^{-1/2} \, . 
\end{align}
The “bad” component is power suppressed with respect to the “good” component under the longitudinal boost $\gamma_t$ and therefore does not contribute at the accuracy considered in this work. Consequently, only the “good” component  $\Psi^{(-)}$ of the quark background field contributes to the present analysis. 

For comparison, the dominant component of the background gluon field scales as ${\cal A^-}\propto (\gamma_t)$, giving rise to the familiar eikonal interaction. While the “good” component $\Psi^{(-)}$ is enhanced with respect to the “bad” component under the same boost, it still scales only as $(\gamma_t)^{1/2}$ and is therefore suppressed by one power of $(\gamma_t)^{1/2}$ with respect to the eikonal gluon background. Consequently, interactions with the background quark field first contribute at NEik accuracy in observables whose leading contribution is eikonal.\footnote{The above statement refers to observables whose leading contribution is eikonal. For observables that are intrinsically sub-eikonal, for example spin-dependent observables, couplings to the background quark field can already appear at leading order in the appropriate power counting.}. This simple power-counting argument is therefore not accidental, but follows directly from the Lorentz transformation properties of the fields. The weaker boost enhancement of spin-$1/2$ fields compared to spin-1 fields explains why quark-induced interactions are intrinsically sub-eikonal and why the Wilson-line realization of quark Reggeization naturally belongs to the NEik sector of the shockwave formalism.

Besides the power-counting discussed above, as in the eikonal treatment of the background gluon field, the quark background field is considered as localized in the longitudinal direction such that 
\begin{gather}
\label{eq:Psi_loc}
   \Psi^{(-)}( z ) \longrightarrow  \Psi^{(-)} (z^+, \boldsymbol{z}) \sim \delta(z^+) \; . 
\end{gather}

Unlike the eikonal interactions discussed in the previous subsections, NEik interactions involving a single $t$-channel quark exchange necessarily change the identity of the propagating parton. The corresponding corresponding operators therefore constitute the natural generalization of the infinite eikonal Wilson lines describing identity-preserving interactions at eikonal accuracy. For example, the transition of a photon into a quark through a $t$-channel quark exchange is described by  
\begin{gather*}
 \int_{-\frac{L^+}{2}}^{ \frac{L^+}{2}} d z^+ \,\mathcal{U}_F (\infty^+, z^+, \boldsymbol{z}) \,\Psi^{(-)} (z^+, \boldsymbol{z}) \; .
\end{gather*}
Similarly, the transition of a gluon into a quark through a $t$-channel quark exchange is described by 
\begin{gather*}
    \int_{-\frac{L^+}{2}}^{ \frac{L^+}{2}} d z^+ \,\mathcal{U}_F (\infty^+, z^+, \boldsymbol{z}) t^b \,\Psi^{(-)} (z^+, \boldsymbol{z}) \mathcal{U}_A^{ba} (z^+, -\infty^+, \boldsymbol{z})
\end{gather*}
The operators introduced above constitute the fundamental building blocks for the Wilson-line description of quark exchange developed in the following sections. 

The longitudinal support of $z^+$ integral is identified by the longitudinal extend of the target from $-L^+/2$ to $L^+/2$. Since the quark field is localized in a longitudinal interval around $z^+=0$ (see \eqref{eq:Psi_loc}), one can take safely $L^+\to\infty$. However, when deriving the evolution, considering a finite longitudinal extent $L^+$ is useful for providing an intuitive light-cone time picture of the interaction and for the graphical representation of the corresponding evolution diagrams. 

The identity-changing operators introduced above require a corresponding set of effective propagators describing the interaction of the parton with the quark background field. At NEik accuracy, these propagators constitute natural extension of the familiar eikonal propagators built from Wilson lines. For the purposes of the present, only a subset of these propagators is required. We derive them below in a coordinate-space representation particularly suited for the calculation of their rapidity evolution.  In particular, we require propagators describing antiquark-to-gluon and gluon-to-quark transitions. Starting from the building block propagators derived in Ref.~\cite{Altinoluk:2024dba}, we express them in a coordinate-space representation that is convenient for the calculation of the evolution equation. Please note that our convention for the Wilson lines differs by an overall sign from that of Ref.~\cite{Altinoluk:2024dba}. \\

We begin by deriving the effective antiquark-to-gluon propagator. As discussed above, these transition propagators are constructed by combining the eikonal before-to-inside and inside-to-after shockwave propagators derived in Ref.~\cite{Altinoluk:2024dba}. The two building blocks are connected through an insertion of localized background quark field, which mediates the transition between the two partonic degrees of freedom. The effective antiquark-to-gluon propagator is therefore given by
\begin{gather}
    [S^{ \bar{\psi} \rightarrow A } (z_3, z_2) ]_{\mu, l}^{b} =  i g \int_{-\frac{L}{2}^+}^{\frac{L}{2}^+} d^D z_1 [ S^{b.i.}_{\bar{\psi}} (z_3, z_1)  \gamma^{\sigma} t^c \Psi^{(-)} (z_1^+, \boldsymbol{z}_1)]_l  [ S^{i.a.}_{A} (z_1, z_2) ]_{\mu \sigma}^{cb} \; ,
    \label{Eq:AntiQuarkGluonPropStart}
\end{gather}
where the interaction with the background quark field takes place at the intermediate point $z_1$. The first building block is the before-to-inside antiquark propagator, $S^{b.i.}_{\bar{\psi}}(z_3,z_1)$, describing the propagation of an incoming antiquark from before the shockwave ($z_3^+<-L^+/2$) to the interaction point inside the shockwave ($-L^+/2<z^+_1<L^+/2$); see the left panel of Fig.~\ref{Fig:QIOAtOneLoop}). It reads  
\begin{align}
    S^{b.i.}_{\bar{\psi}} (z_3, z_1) & = (-1) \int \frac{d k_1^+}{(2 \pi)} \frac{d^{d} \boldsymbol{k}_1}{(2 \pi)^d} \frac{\theta (-k_1^+)}{2 k_1^+} e^{-i z_3 \cdot  \check{k}_1} (\check{\slashed{k}}_1 + m)  
    \nonumber \\ 
    & \, \times \mathcal{U}_F^{\dagger} ( z_1^+, z_3^+, \boldsymbol{z}_1) \left[ 1 - \frac{\gamma^+ \gamma^i}{2k_1^+} i \overleftarrow{D}_{F, \boldsymbol{z}^i} \right] e^{i z_1^- k_1^+ - i \boldsymbol{z}_1 \cdot \boldsymbol{k}_1 } \; ,
\end{align}
where we have introduced the vector $\check{k}_1$ representing the on-shell counterpart of $k_1$, i.e.
\begin{gather}
    k_1^{\mu} = \check{k}_1^{\mu} + \frac{k_1^2}{2 k_1^+} n_2^{\mu} \; ,
\end{gather}
and the covariant derivative
\begin{gather}
    \overleftarrow{D_{\boldsymbol{z}_1^i}^R} = \overleftarrow{\partial}_{\boldsymbol{z}_1^i} + i g \; T_R \cdot A_i(z_1^+, \boldsymbol{z}_1) \; .
\end{gather}
The second building block is the inside-to-after gluon propagator, $ S^{i.a.}_{A} (z_1, z_2) $, describing the propagation of the gluon from the interaction point inside the target ($-L^+/2<z_1^+<L^+/2$) to an end point outside the shockwave at positive light-cone time ($z_2^+ > L^+/2 $). It reads 
\begin{gather}
    [S^{i.a.}_{A} (z_1, z_2)]_{\mu \sigma} = \int \frac{d k_2^+}{(2 \pi)} \frac{d^{d} \boldsymbol{k}_2}{(2 \pi)^d} \frac{\theta (k_2^+)}{2 k_2^+} e^{ - i z_2 \cdot \check{k}_2 } \mathcal{U}^{bc}_A (z_2^+, z_1^+, \boldsymbol{z}_1) \left[ g_{\perp \mu}^{j} - \frac{n_{2 \mu} \boldsymbol{k}_2^j }{ k_2^+ } \right] \nonumber \\ \times  \left[ g_{\perp \sigma}^{j } - \frac{n_{2 \sigma}}{k_2^+} ( i \overleftarrow{D}_{\boldsymbol{z}_1^j}^A + \boldsymbol{k}_2^j ) \right] e^{- i \boldsymbol{k}_2 \cdot \boldsymbol{z}_1 + i k_2^+ z_1^- } \; ,
\end{gather}
The expression in Eq.~\eqref{Eq:AntiQuarkGluonPropStart} simplifies considerably once the projection properties of the ``good'' component of the background quark field are taken into account. The term proportional to $n_{2\sigma}$ in the gluon propagator generates the Dirac structure $\gamma^+ \Psi^{(-)}$, which vanishes identically because of the projector defining the good component. Consequently, $\gamma^{\sigma}$ becomes purely transverse. The same argument then eliminates the term proportional to $\gamma^+$ in the antiquark propagator after commuting it through the transverse Dirac matrix. As a result, all contributions containing covariant derivatives vanish, leaving
\begin{align}
    [S^{ \bar{\psi} \rightarrow A } (z_3, z_2) ]_{\mu, l}^{b} & =  i g \int \frac{d k^+}{(2 \pi)} \frac{d^{d} \boldsymbol{k}_1}{(2 \pi)^d}  \frac{d^{d} \boldsymbol{k}_2}{(2 \pi)^d} \frac{\theta (k^+)}{4 (k^+)^2}  \int_{-\frac{L}{2}^+}^{\frac{L}{2}^+} d z_1^+ d^d \boldsymbol{z}_1 
    \nonumber \\ 
     & \times \,  e^{-i z_3 \cdot  \check{k}_1 - i z_2 \cdot \check{k}_2 - i \boldsymbol{z}_1 \cdot ( \boldsymbol{k}_1 + \boldsymbol{k}_2) } (\check{\slashed{k}}_1 + m) \mathcal{U}_F ( z_3^+, z_1^+, \boldsymbol{z}_1) \gamma^{\sigma} t^c \Psi^{(-)} (z_1^+, \boldsymbol{z}_1) 
    \nonumber \\ 
    & \times \,  \mathcal{U}_A^{bc} (z_2^+, z_1^+, \boldsymbol{z}_1) \left[ g_{\perp \mu \sigma} + \frac{n_{2 \mu} k_{2 \perp \sigma} }{ k^+ } \right] , 
    \label{Eq:AntiQuarkGluonPropMiddle}
\end{align}
For the purposes of the present work, the propagator can be simplified further. Since we are interested in the leading logarithmic evolution, only the rapidity-singular contribution needs to be retained. This is achieved by contracting the gluon propagator with $n_{2\mu}$, thereby projecting onto its eikonal coupling to a fast-moving projectile along the $k^+$. Within this projection, the $g_{\perp \mu \sigma}$ term in Eq.~\eqref{Eq:AntiQuarkGluonPropMiddle} and the $k^+ \gamma^-$ contribution in $\slashed{\check{k}}_1$ are subleading and can be neglected. The propagator therefore reduces to 
\begin{align}
    [S_{\Delta \eta}^{ \bar{\psi} \rightarrow A } (z_3, z_2) ]_{l}^{-,b} & =   i g  \int_{e^{\eta}p_p^+}^{e^{\eta+ \Delta \eta} p_p^+}  \frac{d k^+}{(2 \pi)} \frac{d^{d} \boldsymbol{k}_1}{(2 \pi)^d}  \frac{d^{d} \boldsymbol{k}_2}{(2 \pi)^d} \frac{\theta (k^+)}{4 (k^+)^3 }  \int_{-\frac{L}{2}^+}^{\frac{L}{2}^+} d z_1^+ d^d \boldsymbol{z}_1 
    \nonumber \\ 
    & \, \times e^{-i z_3 \cdot  \check{k}_1 - i z_2 \cdot \check{k}_2 - i \boldsymbol{z}_1 \cdot ( \boldsymbol{k}_1 + \boldsymbol{k}_2) }  (\slashed{k}_{1\perp}+m) \slashed{k}_{2 \perp} [\mathcal{U}_F ( z_3^+, z_1^+, \boldsymbol{z}_1)  t^c \Psi^{(-)} (z_1^+, \boldsymbol{z}_1 )]_l \mathcal{U}_A^{bc} (z_2^+, z_1^+, \boldsymbol{z}_1) ,
    \label{Eq:AntiQuarkGluonPropMiddle2}
\end{align}
Throughout the present work, we only encounter the configuration $\boldsymbol{z}_2 = \boldsymbol{z}_3$, which will be assumed in the following derivation. After performing the shifts
\begin{gather*}
    \boldsymbol{k}_1 \longrightarrow \boldsymbol{k}_1 + \frac{k^+}{z_3^+}  \boldsymbol{z}_{12} \; , \hspace{1 cm }  \boldsymbol{k}_2 \longrightarrow \boldsymbol{k}_2 - \frac{k^+}{z_2^+}  \boldsymbol{z}_{12} \; ,
\end{gather*}
the transverse momentum integrals become Gaussian and can be carried out in a straightforward manner. The result reads
\begin{gather}
    [S_{\Delta \eta}^{ \bar{\psi} \rightarrow A } (z_3, z_2) ]_{l}^{-,b} = \frac{g}{4 (2 \pi)^{1+d}} \frac{-(-i)^{1+d}}{(-z_2^+ z_3^+)^{1+d/2}} \int d^d \boldsymbol{z}_1  \int d z_1^+ [\mathcal{U}_F ( - \infty^+, z_1^+, \boldsymbol{z}_1)  t^c \Psi^{(-)} (z_1^+, \boldsymbol{z}_1 )]_l \nonumber \\ \times \mathcal{U}_A^{bc} (\infty^+, z_1^+, \boldsymbol{z}_1) 
    \int_{e^{\eta} p_p^+}^{e^{\eta + \Delta \eta} p_p^+} d k^+ (k^+)^{d-1} e^{-i k^+ \left( z_{23}^{-} - \frac{ \boldsymbol{z}_{12}^2 }{2 z_2^+} + \frac{ \boldsymbol{z}_{12}^2 }{2 z_3^+} +  i \varepsilon\right) + i \frac{z_3^+}{2 k^+} m^2 } \left( \slashed{z}_{12 \perp} + \frac{z_3^+}{k^+} m \right) \slashed{z}_{12 \perp}  \; ,
    \label{Eq:AntiquarkToGluonPropNEik}
\end{gather}
where the end points of the adjoint Wilson lines are extended to $z_2^+\to+\infty$ and $z_3^+\to-\infty$ since the background gauge fields vanish outside the extent of the target (see \cite{Altinoluk:2024zom} for a more detailed discussion). Moreover, thanks to the shockwave approximation, the integration over $z_1^+$ has been extended to the entire light-cone time.   

The second propagator required for the evolution is the crossed gluon-to-quark propagator, obtained by reversing the direction of the light-cone time. It is constructed from the before-to-inside gluon propagator and the inside-to-after quark propagator connected through an insertion of the background quark field. Since the derivation follows exactly the same steps as those presented above for the antiquark-to-gluon propagator, we do not repeat it here and simply present the reduced expression relevant for the leading-logarithmic evolution which reads 

\begin{gather}
    [S_{\Delta \eta}^{A \rightarrow \psi } (z_3, z_2) ]_{l}^{-,b}   = \frac{g}{4 (2 \pi)^{1+d}} \frac{ (-i)^{1+d}}{(-z_2^+ z_3^+)^{1+d/2}} \int d^d \boldsymbol{z}_1  \int d z_1^+ [\mathcal{U}_F (  \infty^+, z_1^+, \boldsymbol{z}_1)  t^c \Psi^{(-)} (z_1^+, \boldsymbol{z}_1 )]_l 
    \nonumber \\ 
     \times \mathcal{U}_A^{bc} (-\infty^+, z_1^+, \boldsymbol{z}_1) \,  
\int_{e^{\eta} p_p^+}^{e^{\eta + \Delta \eta} p_p^+} d k^+ (k^+)^{d-1} e^{-i k^+ \left( z_{23}^{-} - \frac{ \boldsymbol{z}_{12}^2 }{2 z_2^+} + \frac{ \boldsymbol{z}_{12}^2 }{2 z_3^+} +  i \varepsilon\right) - i \frac{z_2^+}{2 k^+} m^2 } \left( \slashed{z}_{12 \perp} - \frac{z_2^+}{k^+} m \right) \slashed{z}_{12 \perp} \; .
    \label{Eq:GluonToQuarkProp}
\end{gather}

\section{The Reggeized gluon from eikonal Wilson lines}
\label{sec:Gluon_Reggeization_Wilson_line}

Ref.~\cite{Caron-Huot:2013fea} established a deep connection between the Reggeized gluon and eikonal Wilson lines appearing in the semi-classical theory of small $x$ QCD. This correspondence provides the conceptual foundation for the Wilson-line description of high-energy gluon exchange and serves as the blueprint for the construction developed in the present work. Since the Wilson-line realization of quark Reggeization closely parallels the gluon case, we briefly review the essential ingredients of this framework. 

\subsection{A Reggeized gluon interpolating operator and the dilute expansion}

The Reggeized gluon interpolating  (RGI) operator is defined as the logarithm of an infinite eikonal Wilson line in the adjoint representation~\cite{Caron-Huot:2013fea}. In the original construction, the Wilson line extends over the entire light-cone time. Since the Wilson line realization of the quark Reggeization developed in the following section naturally involves semi-infinite Wilson lines, we adopt a slightly more general definition, which reads  
\begin{align}
 R^a (z_1^+, z_2^+, \boldsymbol{z}) &= \frac{f^{abc}}{g N_c} [\ln \mathcal{U}_A (z_1^+, z_2^+, \boldsymbol{z})]^{bc} \nonumber \\ 
 &= \frac{f^{abc}}{g N_c} [ \mathcal{U}_A (z_1^+, z_2^+, \boldsymbol{z})- 1]^{bc} - \frac{1}{2} \frac{f^{abc}}{g N_c} [ \mathcal{U}_A (z_1^+, z_2^+, \boldsymbol{z})- 1]^{bd} [ \mathcal{U}_A (z_1^+, z_2^+, \boldsymbol{z})- 1]^{dc} + ... \nonumber \\ 
 &= \int_{z_2^+}^{z_1^+} d x^+ \mathcal{A}^a (x^+, \boldsymbol{z})  - \frac{g f^{abc}}{2}  \int_{z_2^+}^{z_1^+} d x_1^+ \int^{z_1^+}_{z_2^+} d x_2^+ \mathcal{A}^b (x_2^+, \boldsymbol{z})  \mathcal{A}^c (x_1^+, \boldsymbol{z}) \theta (x_2^+ - x_1^+) + ...
 \label{Eq:ReggeizedGluon}
\end{align}
The expansion in Eq. \eqref{Eq:ReggeizedGluon} shows that the operator $R^a (z_1^+, z_2^+, \boldsymbol{z})$ starts at order $g^0$ in perturbation theory, where it interpolates a single free gluon. 
Higher order terms describe a light-cone time $x^+$ ordered cascade of gluons generated through the logarithmic expansion of the Wilson line. The operator $R^a$ possesses three important properties that play a central role in the discussion of the subsequent subsections: 
\begin{itemize}
    \item[(i.)] The Reggeized gluon field \(R^a\) can be interpreted as the components, in a basis of color generators, of the Lie-algebra element whose exponentiation yields the Wilson line~\cite{Caron-Huot:2013fea}\footnote{Interestingly, the use of Lie-algebra-valued variables associated with Wilson lines predates the Reggeon-operator formulation and was already explored in the derivation of a small-\(x\) effective action in Ref.~\cite{Jalilian-Marian:2000pwi}.}. Indeed, a Wilson line in an arbitrary representation \(R\) is a group element and can be parametrized as
\begin{gather}
\mathcal{U}_R(z_1^+,z_2^+,\boldsymbol{z})
=
\exp \left \{ ig T_R^a R^a(z_1^+,z_2^+,\boldsymbol{z}) \right \},
\label{Eq:UniversExp}
\end{gather}
where \(T_R^a\) are the generators in the representation \(R\), while \(R^a\) are the components of the corresponding Lie-algebra-valued field \(T_R^aR^a\).
 Expanding the group element around the identity gives
\begin{gather}
    \mathcal{U}_R (z_1^+, z_2^+, \boldsymbol{z})
    =
    1
    + i g T_R^a R^a (z_1^+, z_2^+, \boldsymbol{z})
    - \frac{g^2}{2} T_R^a T_R^b
    R^a (z_1^+, z_2^+, \boldsymbol{z})
    R^b (z_1^+, z_2^+, \boldsymbol{z})
    + \ldots .
\end{gather}
In the dilute regime, $gR\ll 1$, the successive terms in the expansion are parametrically suppressed, and the Wilson line can be approximated by truncating the series at a finite order\footnote{This statement should not be interpreted as a prescription to discard all higher-order terms in $gR$, but rather as a power-counting rule. The expansion in $gR$ organizes contributions according to Reggeon multiplicity and identifies the logarithmic accuracy at which each multi-Reggeon sector can contribute.}. Depending on the desired accuracy, this allows one to describe the interaction in terms of a finite number of Reggeized gluon exchanges. Conversely, in the non-linear regime, $gR\sim 1$, the full group element must be retained.

    \item[(ii.)] The Reggeized gluon is odd under path reversal,
    \begin{gather}
         R^{a} (z_2^+, z_1^+, \boldsymbol{z}) = - R^{a} (z_1^+, z_2^+, \boldsymbol{z}) \; .   \label{Eq:Signature_ReggeQuark}
    \end{gather}
    a property that reflects its negative signature. In terms of scattering amplitudes, this corresponds to odd behavior under the interchange of the Mandelstam invariants (e.g. $s\leftrightarrow u \approx -s$). At the operator level, the relation follows from
    \begin{gather*}
       \mathcal{U}_A^{bc} (z_1^+, z_2^+, \boldsymbol{z}) \rightarrow \mathcal{U}_A^{cb} (z_1^+, z_2^+, \boldsymbol{z}) = [\mathcal{U}_A^{bc} (z_1^+, z_2^+, \boldsymbol{z})]^{\dagger} = \mathcal{U}_A^{bc} (z_2^+, z_1^+, \boldsymbol{z}) \; ,
    \end{gather*}
namely from exchanging the initial and final states.
    \item[(iii.)]  It is Hermitian, i.e.
     \begin{gather}
         R^{a \dagger} (z_1^+, z_2^+, \boldsymbol{z}) = R^{a} (z_1^+, z_2^+, \boldsymbol{z}) \; .
    \end{gather}
\end{itemize}
The fundamental feature of $R^{a}$ is its rapidity renormalization group equation in the cut-off $\eta$, which can be determined from the corresponding one of a single Wilson line~\cite{Balitsky:1995ub,Balitsky:2013fea}. As a preliminary step towards the derivation of the evolution equation for the subeikonal operators relevant to quark Reggeization, we provide a pedagogical rederivation of the evolution equation for a single Wilson line.

\subsection{Evolution equation of a single Wilson line}

\begin{figure}
    \centering
    \includegraphics[width=0.5\linewidth]{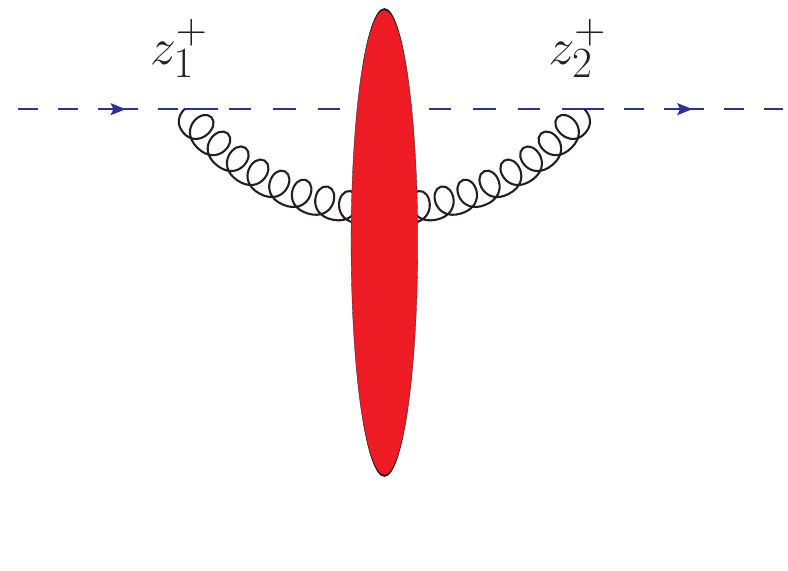}
    \caption{Diagram contributing to the one-loop evolution of a Wilson line}
    \label{Fig:WilsonLineEvo}
\end{figure}

Adopting the choice $\epsilon \equiv \epsilon_{\rm UV} = \epsilon_{\rm IR}$ to remove scaleless integrals, the expansion \eqref{Eq:ExpansionFluct} receives a contribution from a single non-vanishing diagram shown in Fig.~\ref{Fig:WilsonLineEvo}. Consequently, the evolution kernel $\mathcal{K}[\mathcal{U}_F (\boldsymbol{z}_1)]$ is determined from
\begin{align}
I_{R} = (\mu^2)^{1-d/2} (i g)^2 \int_{-\infty}^0 d z_1^+ \int_0^{\infty} d z_2^+ [ t^a    \mathcal{U}_F (z_2^+, z_1^+, \boldsymbol{z}_1)  t^b  ]_{ij} n_{1 \mu} n_{1 \nu} G_{\Delta \eta}^{\mu \nu, ab} (z_2^+, z_1^+, \boldsymbol{z}_1) \; .  
\end{align}
The remaining integrations can now be performed straightforwardly. As discussed earlier, the end points of the Wilson lines (including those appearing in the effective propagators) may be extended to $z_1^+\rightarrow-\infty^+$ and $z_2^+\rightarrow\infty^+$. Moreover, the propagator $G_{\Delta \eta}^{- -, ab} (z_2^+, z_1^+, \boldsymbol{z}_1)$ is obtained from Eq.~\eqref{Eq:GluonPropEvoLim} by setting $\boldsymbol{z}_1 = \boldsymbol{z}_2$. The longitudinal integrals over $z_1^+$ and $z_2^+$ can be conveniently performed using the Schwinger formula
\begin{align}
\int_{0}^{\infty} d \sigma \; \sigma^{d/2 -1} e^{i \sigma A } = \frac{i^{d/2} \Gamma[ \frac{d}{2}]}{A^{d/2}} \; ,
\label{Eq:SchwingerPar}
\end{align}
while the integration over $k^+$ gives
\begin{align}
    \int_{e^{\eta} p_p^+}^{e^{\eta + \Delta \eta} p_p^+} \frac{d k^+}{k^+} e^{-i k^+ ( z_{21}^- - i \varepsilon ) } = \Delta \eta + \mathcal{O} (\Delta \eta^2) \; .
    \label{Eq:KplusInt}
\end{align}
Combining these integrations yields
\begin{align}
I_{R,g} = \frac{g^2 (\mu^2)^{1-d/2} [ \Gamma \left( \frac{d}{2} \right)]^2 }{4 \pi^{1+d}} \Delta \eta \int d^d \boldsymbol{z}_3 \frac{1}{(\boldsymbol{z}_{13}^2)^{d-1}} [ t^a \mathcal{U}_F (\boldsymbol{z}_1) t^b ]_{ij} \mathcal{U}_A^{ab} (\boldsymbol{z}_3) \; . 
\end{align}
The color structure can be simplified using the relation
\begin{align}
\mathcal{U}_A^{ab} (\boldsymbol{z}_3) t^b = \mathcal{U}_F^{\dagger} (\boldsymbol{z}_3) t^a \mathcal{U}_F (\boldsymbol{z}_3),
\label{Eq:FundAdjoi}
\end{align}
together with the Fierz identity
\begin{align}
t^a_{ik} t^a_{lm} = \frac{1}{2} \left[ \delta_{im} \delta_{lk} - \frac{1}{N_c} \delta_{ik} \delta_{lm} \right] ,
\label{Eq:FierzIdentity}
\end{align}
which gives 
\begin{align}
\frac{\partial}{ \partial \eta } [\mathcal{U}_{F} (\boldsymbol{z}_1)]_{ij} =   a_s \int d^d \boldsymbol{z}_{3}  \frac{1}{(\boldsymbol{z}_{13}^2)^{d-1}}  \left \{  {\rm Tr_c} [\mathcal{U}_{F} (\boldsymbol{z}_1) \mathcal{U}_{F}^{\dagger} (\boldsymbol{z}_3) ] \; [\mathcal{U}_{F} (\boldsymbol{z}_3)]_{ij} - \frac{1}{N_c} [\mathcal{U}_{F} (\boldsymbol{z}_1)]_{ij} \right \} .
    \label{Eq:SingleWilsonEvoAlm}
\end{align}
Here we introduced
\begin{gather}
\label{def:a_s}
    a_s = \frac{g^2 }{8 \pi^{1+d}} \left[\Gamma \left( \frac{d}{2} \right) \right]^2 (\mu^2)^{1-d/2} \; .
\end{gather}
Within our regularization scheme, the second contribution in the curly brackets produces only scaleless integrals and therefore vanishes. Therefore, the evolution equation for a single Wilson line reduces to  

\begin{gather}
    \frac{\partial}{ \partial \eta } [\mathcal{U}_{F} (\boldsymbol{z}_1)]_{ij} = a_s \int d^d \boldsymbol{z}_{2} \frac{1}{(\boldsymbol{z}_{12}^2)^{d-1}} {\rm Tr_c} [\mathcal{U}_{F} (\boldsymbol{z}_1) \mathcal{U}_{F}^{\dagger} (\boldsymbol{z}_2) ] \; [\mathcal{U}_{F} (\boldsymbol{z}_2)]_{ij} \; .
    \label{Eq:SingleWilsonLineEvo}
\end{gather}

\subsection{Linearization and the gluon Regge trajectory}

The evolution equation in Eq.~(\ref{Eq:SingleWilsonLineEvo}) provides the starting point for recovering the well-known phenomenon of gluon Reggeization. This is achieved by expanding the Wilson line in the dilute regime, where the Reggeized gluon field is weak, ($gR \ll 1$). Starting from Eq.~\eqref{Eq:SingleWilsonLineEvo}, inserting the expansion of the Wilson line given in Eq.~\eqref{Eq:ReggeizedGluon}, and projecting onto the color-octet channel, one obtains a linear evolution equation for the Reggeized gluon field ($R^a$), valid up to corrections of $\mathcal{O}(g^4R^3)$. In coordinate space, the resulting equation reads~\cite{Caron-Huot:2013fea}
\begin{gather}
    \frac{\partial}{ \partial \eta } R^a (\boldsymbol{z}_1) =  N_c \; \mathcal{K}_{\rm RP} (\boldsymbol{z}_{1}, \boldsymbol{z}_{2}) \otimes_{\boldsymbol{z}_2} R^a (\boldsymbol{z}_2) + \mathcal{O} (g^4 R^3) \; ,
    \label{Eq:ReggeGluonEvoCoord}
\end{gather}
where
\begin{gather}
    \mathcal{K}_{\rm RP} (\boldsymbol{z}_{1}, \boldsymbol{z}_{2}) =  \frac{a_s  }{(\boldsymbol{z}_{12}^2)^{d-1}} \; ,
    \label{Eq:KRegge}
\end{gather}
is the Regge-pole (RP) coordinate space kernel with $a_s$ is defined in Eq. \eqref{def:a_s}. For later convenience, we introduced the compact notation
\begin{gather}
    f(\boldsymbol{z}) \otimes_{z} g(\boldsymbol{z}) =\int d^d z \;f(\boldsymbol{z}) g(\boldsymbol{z}) \; .
\end{gather}
The equation can be diagonalized in momentum space,
\begin{gather}
    \frac{\partial}{ \partial \eta } R^a (\boldsymbol{p}) = N_c \; a_s \int d^d \boldsymbol{z}_1 \int d^d \boldsymbol{z}_2 \frac{e^{-i \boldsymbol{p} \cdot \boldsymbol{z}_1 }}{(\boldsymbol{z}_{12}^2)^{d-1}}  R^a (\boldsymbol{z}_2) \; .
\end{gather}
After shifting the integration variable ($\boldsymbol{z}_1 \rightarrow \boldsymbol{z}_1 + \boldsymbol{z}_2$) and using the standard Fourier transform of the coordinate-space kernel, 
\begin{gather}
    \int d^d \boldsymbol{z}_1 \frac{e^{-i \boldsymbol{p} \cdot \boldsymbol{z}_1 }}{(\boldsymbol{z}_{1}^2)^{d-1}} = \frac{4^{1-d/2} \pi^{d/2} \Gamma (1-d/2)}{\Gamma (d-1) } (\boldsymbol{p}^2)^{d/2-1} \; ,
\end{gather}
the convolution reduces to a multiplicative factor,
\begin{gather}
    \frac{\partial}{ \partial \eta } R^a (\boldsymbol{p}) =  \omega^{(1)} (- \boldsymbol{p}^2) R^a (\boldsymbol{p}) + \mathcal{O} \left( g^4 R^3 \right) \; .
\end{gather}

The evolution equation therefore takes the form of a first-order differential equation, whose eigenvalue is precisely the one-loop gluon Regge trajectory in Eq.~(\ref{Eq:GluonReggeTraj}), which explicitly shows that the operator $R^a$ is the interpolating operator for the Reggeized gluon. The phenomenon of \textit{gluon Reggeization} in QCD thus possesses an intriguing connection with the evolution structure of the Wilson lines, which becomes manifest upon linearizing Eq.~(\ref{Eq:ReggeizedGluon}) in the dilute limit. \\

Equation~(\ref{Eq:ReggeGluonEvoCoord}) is consistent with the evolution structure schematized in Fig.~3 of Ref.~\cite{Caron-Huot:2013fea}, where the rapidity evolution derived from the B-JIMWLK equation takes the form illustrated there, with mixing with three-Reggeon operators first appearing at order $g^4$. Although these contributions are clearly negligible in the LL analysis, the fact that they start at order $g^4$ is an essential ingredient for gluon Reggeization at the NLLA. In Ref.~\cite{Caron-Huot:2013fea}, the rapidity evolution was shown to take the schematic matrix form
\begin{equation}
 \frac{\partial}{\partial \eta}
\begin{pmatrix}
R^1 \\
R^2 \\
R^3 \\
R^4 \\
R^5 \\
...
\end{pmatrix}
=
\begin{pmatrix}
  g^2 H_{11} & 0 &   g^4 H_{31} & 0 &   g^6 H_{51} & ... \\
0 & g^2 H_{22} & 0 & g^4 H_{42} & 0 & ... \\
g^4 H_{13} & 0 & g^2 H_{33} & 0 & g^4 H_{53} & ... \\
0 & g^4 H_{24} & 0 & g^2 H_{44} & 0 & ... \\
g^6 H_{15} & 0 & g^4 H_{35} & 0 & g^2 H_{55} & ... \\
... & ... & ... & ... & ... & ...
\end{pmatrix}
\begin{pmatrix}
R^1 \\
R^2 \\
R^3 \\
R^4 \\
R^5 \\
...
\end{pmatrix} \; ,
\label{Eq:MatrixEvo}
\end{equation}
where each $H_{ij}$ admits a perturbative expansion starting at order $g^0$ and the terms on the diagonal are non-degenerate. Here, $R^n$ denotes schematically the space of operators containing $n$ Reggeized-gluon fields, with their transverse coordinates and color structures left implicit. Consequently, at order $g^2$ the evolution matrix is purely diagonal, immediately implying gluon Reggeization at the LLA. Signature conservation implies that operators containing an odd number of Reggeized gluons can mix only with odd-Reggeon operators, and similarly for the even sector. The first off-diagonal transitions therefore connect sectors whose Reggeon numbers differ by two~\cite{Caron-Huot:2013fea}. Restricting ourselves to NLLA, all contributions of order $g^{4+n}$ with $n>0$ can be neglected. The relevant evolution therefore reduces to~\cite{Caron-Huot:2013fea}
\begin{equation}
 \frac{\partial}{\partial \eta}
\begin{pmatrix}
R^1 \\
R^3 
\end{pmatrix}
=
\begin{pmatrix}
  g^2 H_{11} & g^4  H_{31} \\
 g^4 H_{13}  &  g^2 H_{33}  
\end{pmatrix}
\begin{pmatrix}
R^1 \\
R^3 
\end{pmatrix} \; ,
\label{Eq:ReggeizedGluonMixingNLLA}
\end{equation}
One eigenvalue of the evolution matrix is identified with the Regge trajectory,
\begin{equation}
\lambda_{\rm RP} = \frac{g^2}{2} \left( H_{11} + H_{33} + \sqrt{ (H_{11} - H_{33} )^2 +  4 \; g^4  H_{13} H_{31} } \right).
\label{Eq:ReggePoleEigen}
\end{equation}
Expanding in powers of the coupling yields
\begin{equation}
\lambda_{\rm RP} = g^2 H_{11} + g^6 \frac{H_{13} H_{31}}{ \sqrt{ (H_{11} - H_{33} )^2 } } + \mathcal{O}(g^8).
\label{Eq:ReggeTrajEigen}
\end{equation}
Although the off-diagonal kernels start at order $g^4$, their contribution to the eigenvalue is second order in the mixing and is divided by $g^2$. It therefore scales as $g^4g^4/g^2=g^6$.
It follows that, although mixing with three-Reggeon operators first appears at two loops (order $g^4$), it modifies the Regge trajectory only at three-loop order (order $g^6$). At the amplitude level, three-Reggeon cuts first appear at two loops, where they contribute only to energy-independent terms, without any accompanying logarithms of \(s\). They therefore enter at next-to-next-to-leading logarithmic accuracy (NNLLA)~\cite{DelDuca:2001gu}. We emphasize that, had the mixing instead started at order $g^3$, for instance through two-Reggeon operators (whose mixing is forbidden by signature), gluon Reggeization would already break down at NLLA. Therefore, up to two loops, the Regge trajectory is entirely determined by the second term in the perturbative expansion of $H_{11}$,
\begin{equation}
\lambda_{\rm RP} = g^2 H_{11} + \mathcal{O}(g^6)
= g^2 H_{11}^{\rm 1-loop} +  g^4 H_{11}^{\rm 2-loop} + \mathcal{O}(g^6).
\end{equation}
This observation will provide the relevant benchmark for the quark case. There, the logarithmic accuracy at which Regge-pole evolution can be affected depends not only on the perturbative order of the off-diagonal kernels, but also on the different lowest-order scaling of the operators belonging to the coupled sectors. 
\section{A Reggeized quark interpolating operator}
\label{sec:QuarkReggeShockwaveApproach}

We now turn to the central result of this work. First focusing on the massless quarks, we study the one-loop rapidity evolution of the simplest Wilson-line operator carrying the quantum numbers of a $t$-channel Reggeized quark. 
In the dilute limit, the nonlinear evolution equation can be linearized, allowing us to identify the Reggeized Quark Interpolating (RQI) operator by recovering the one-loop quark Regge trajectory.   

\begin{figure}
    \centering
    \includegraphics[width=0.5\linewidth]{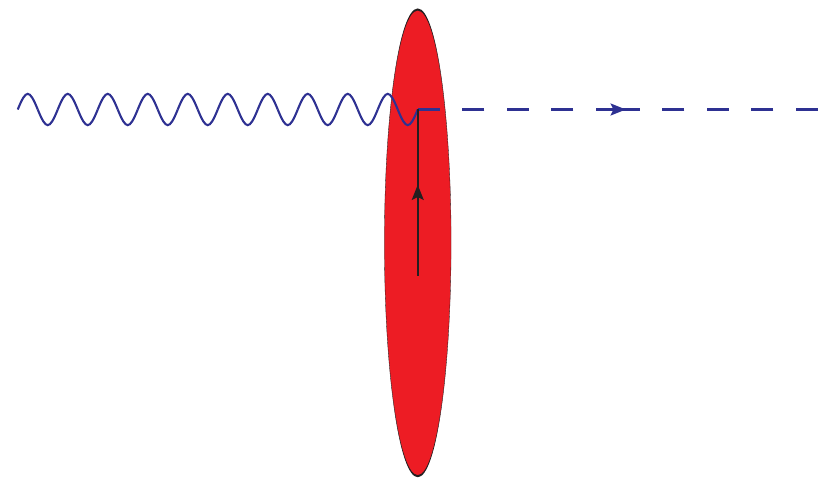}
    \caption{A fast-moving photon interacting with the quark background field from the target and changing its nature to a fast-moving quark, which is then dressed by gluon shockwave interactions (the dashed line denotes a Wilson line in the fundamental representation).}
    \label{fig:QIOLeading}
\end{figure}
%

\subsection{Constructing the Reggeized quark from semi-infinite Wilson lines}

Unlike the gluon exchange, the exchange of a quark in the $t$-channel is a genuinely subeikonal effect in the high-energy expansion. Nevertheless, the corresponding operator must still isolate the leading energy dependence of the scattering amplitude. The simplest realization is provided by the Wilson-line operator,
\begin{gather}
  Q_i (\boldsymbol{z}) \equiv Q_i ( \infty^+, -\infty^+, \boldsymbol{z})
    = \int d z^+ \,[\mathcal{U}_F (\infty^+, z^+, \boldsymbol{z})]_{ij} \,\Psi_j^{(-)} (z^+, \boldsymbol{z}) \; ,
\label{Eq:AlmostReggeizedQuarkInter}
\end{gather}
which enters the description of the process shown in Fig.~\ref{fig:QIOLeading}. In this process, an incoming photon annihilates with a quark field of the target, producing a fast-moving quark that is subsequently dressed by eikonal interactions with the target gluon background field. The operator consists of a Wilson line in the fundamental representation extending from $z^+$, where the good component of the $t$-channel quark field is inserted,  up to $\infty^+$. The Wilson line resums multiple soft gluon interactions between the fast-moving quark and the target color field along its eikonal trajectory, while the photon propagates freely from $-\infty^+$ to $z^+$, without interacting with the QCD background.  \\

The generalization to the gluon-initiated channel is intuitive and requires the inclusion of a Wilson line in the adjoint representation that resums the soft interactions of the gluon with the target from $-\infty^+$ to $z^+$. For clarity, we restrict ourselves to the photon-initiated process, which is sufficient to isolate the color-triplet exchange and allows for a particularly transparent derivation of the evolution equation. The gluon-initiated  channel will be discussed in Sec.~\ref{sec:UniversGluon}. \\

The Wilson-line operator $Q_i (\boldsymbol{z})$ can be linearized by expanding Eq. \eqref{Eq:AlmostReggeizedQuarkInter} to leading order in RGI operator and it yields 
\begin{gather}
   Q_i (\boldsymbol{z})
    =  \int d z^+ \, \left[ \delta_{ij} + ig  t^a_{ij} R^a (\infty^+, z^+ \boldsymbol{z}) \right] \Psi_j^{(-)} (z^+, \boldsymbol{z}) + \mathcal{O} (g^2 R^2) \equiv R_{Q, i}(\boldsymbol{z}) + \mathcal{O} (g^2 R^2)
   \label{Eq:AlmostReggeizedQuarkInterLinear}
\end{gather}
where the operator $R_{Q, i}$ denotes the linearized quark operator. Its interpretation follows directly from that of the Reggeized gluon in Eq.~(\ref{Eq:AlmostReggeizedQuarkInter}). At leading order in the coupling $g$, it reduces to a source of a free quark field, while higher-order terms encode the emission of an arbitrary number of soft gluons through the coupling to the Reggeized background field $R^a$, thereby resumming eikonal interactions in the gluon background. \\

As we will show in the following, the operator in Eq.~(\ref{Eq:AlmostReggeizedQuarkInterLinear}) does not yet correspond to a genuine Reggeized-quark exchange. To isolate the Regge pole, it is necessary to project onto operators with definite signature. For this purpose, we introduce the signature transformation $\mathcal{S}$, defined as the exchange $\infty^+ \leftrightarrow -\infty^+$ in the Wilson lines. At the amplitude level, this operation implements the crossing transformation $s \leftrightarrow u$. In contrast to the RGI operator given in Eq.~(\ref{Eq:ReggeizedGluon}), which is odd under $\mathcal{S}$,
\begin{align}
\mathcal{S}\, R^a(\boldsymbol{z})=-R^a(\boldsymbol{z})\,,
\end{align}
the operator obtained from Eq.~(\ref{Eq:AlmostReggeizedQuarkInterLinear}) is not an eigenstate of the signature transformation. Indeed, acting with $\mathcal{S}$ on its Wilson-line realization, $Q_i(\boldsymbol{z})$, maps it into
\begin{gather}
    \mathcal{S} \{ Q_i (\boldsymbol{z}) \}  = \bar{Q}_i (\boldsymbol{z}) = \int d z^+ \,[\mathcal{U}_F^{\dagger} ( z^+, -\infty^+, \boldsymbol{z})]_{ij} \,\Psi_j^{(-)} (z^+, \boldsymbol{z}) \; , \label{SignatureOperation}
\end{gather}
where $\bar{Q}_i (\boldsymbol{z})$ is the operator associated with the crossed process $\bar{q} q \rightarrow \gamma $. Its expansion in the dilute limit reads 
\begin{gather}
    \bar{Q}_i (\boldsymbol{z})  = \int d z^+ \, \left[ \delta_{ij} - ig  t^a_{ij} R^a (z^+, -\infty^+, \boldsymbol{z}) \right] \Psi_j^{(-)} (z^+, \boldsymbol{z}) + \mathcal{O} (g^2 R^2) = \bar{R}_{Q,i} (\boldsymbol{z}) + \mathcal{O} (g^2 R^2) \; .
    \label{Eq:CrossedOperatorQuarkLO}
\end{gather}
%

\subsection{Evolution equation at one-loop level}

\begin{figure}
    \centering
    \includegraphics[width=0.4\linewidth]{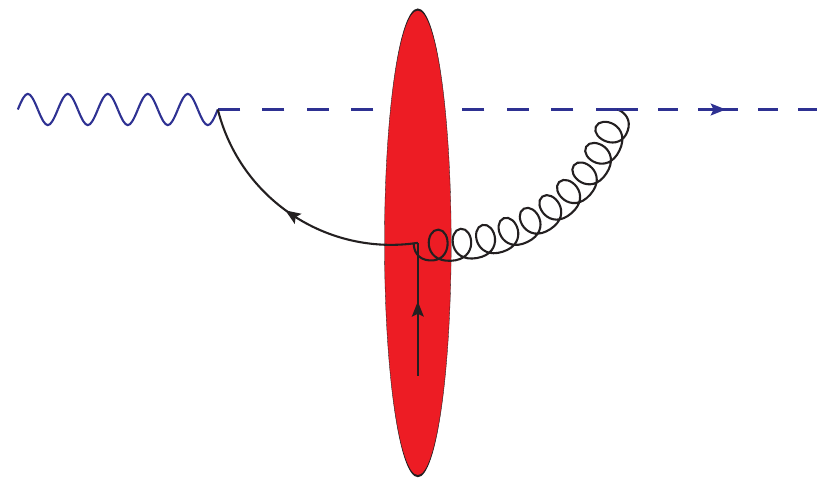} \hspace{2 cm}
    \includegraphics[width=0.4\linewidth]{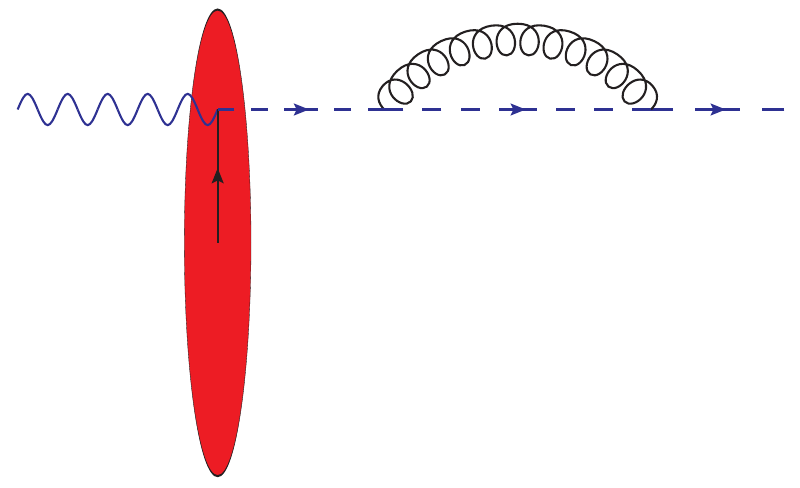}
    \caption{Diagrams contributing to the one-loop evolution of the operator $Q_i (\boldsymbol{z})$. For massless fermions, the right diagram leads to a scaleless integral.}
    \label{Fig:QIOAtOneLoop}
\end{figure}

We now derive the one-loop evolution of the operator $Q_i (\boldsymbol{z})$. The evolution is determined by the left diagram in Fig.~\ref{Fig:QIOAtOneLoop}.\footnote{The right diagram in Fig.~\ref{Fig:QIOAtOneLoop} leads to a scaleless integral. Again, we choose $\epsilon \equiv \epsilon_{\rm UV} = \epsilon_{\rm IR}$, thereby discarding contributions proportional to scaleless integrals.} The corresponding contribution is
\begin{gather}
    I_{R, Q} = (\mu^2)^{1-d/2} ( i g ) \int_{-\infty}^0 d z_3^+ \int_0^{\infty} d z_2^+ [ t^b    \mathcal{U}_F ( \boldsymbol{z}_2)  ]_{il} [S_{\Delta \eta}^{ \bar{\psi} \rightarrow A } (z_3, z_2) ]_{l}^{-, b} \; .  
\end{gather}

Substituting the expression for the antiquark-to-gluon effective propagator given in Eq.~(\ref{Eq:AntiquarkToGluonPropNEik}) for the massless case, the shockwave approximation allows us to set the end points of the Wilson lines $z_3^+$ and $z_2^+$ to $-\infty^+$ and $+\infty^+$. The remaining longitudinal integrations over  $z_3^+$ and $z_2^+$ are then performed using Eq.~(\ref{Eq:SchwingerPar}), while the $k^+$ integration is evaluated with Eq.~(\ref{Eq:KplusInt}). Expressing the result in terms of the evolution kernel (see Eq. \eqref{Eq:KernelDiagram}), one obtains  
%
%
\begin{gather}
   \mathcal{K}[Q_i(\boldsymbol{z}_2)] = a_s \int d^d \boldsymbol{z}_1 \frac{1}{(\boldsymbol{z}_{12}^2)^{d-1}}  \int_{-\frac{L}{2}^+}^{\frac{L}{2}^+} d z_1^+ [ t^b    \mathcal{U}_F ( \boldsymbol{z}_2) \mathcal{U}_F (-\infty^+, z_1^+, \boldsymbol{z}_1) t^c \Psi^{(-)} (z_1^+, \boldsymbol{z}_1) ]_{i} \;\mathcal{U}_A^{cb} (z_1^+, \infty^+, \boldsymbol{z}_1) \; .
\end{gather}

Using the identity given in Eq.~(\ref{Eq:FundAdjoi}) together with a relabeling of the transverse integration variable as $\boldsymbol{z}_2$ and the external integration coordinate as $\boldsymbol{z_1}$, the kernel reduces to a simpler form, yielding the one-loop evolution equation for $Q_i (\boldsymbol{z}_1)$,
\begin{gather}
    \frac{\partial Q_i (\boldsymbol{z}_1)}{ \partial \eta} = a_s \int d^d \boldsymbol{z}_2 \frac{1}{(\boldsymbol{z}_{12}^2)^{d-1}}  \frac{1}{2} \bigg \{  \left( {\rm Tr}[ \mathcal{U}_F (\boldsymbol{z}_1 ) \mathcal{U}_F^{\dagger} (\boldsymbol{z}_2 ) ] - \frac{1}{N_c} \right) \delta_{ij} - \frac{1}{N_c} \left( [ \mathcal{U}_F (\boldsymbol{z_1} ) \mathcal{U}_F^{\dagger} (\boldsymbol{z}_2 ) ]_{ij} - \delta_{ij} \right)  \bigg \} Q_j (\boldsymbol{z}_2) \; .
    \label{Eq:AlmostReggeEvoCoord}
\end{gather} 

The evolution for the ``crossed'' operator $\bar{Q}_i (\boldsymbol{z}_1)$ is obtained analogously and reads 
%
%
\begin{gather}
    \frac{\partial \bar{Q}_i (\boldsymbol{z}_1)}{ \partial \eta} = a_s \int d^d \boldsymbol{z}_2 \frac{1}{(\boldsymbol{z}_{12}^2)^{d-1}}  \frac{1}{2} \bigg \{  \left( {\rm Tr}[ \mathcal{U}_F^{\dagger} (\boldsymbol{z}_1 ) \mathcal{U}_F (\boldsymbol{z}_2 ) ] - \frac{1}{N_c} \right) \delta_{ij} - \frac{1}{N_c} \left( [ \mathcal{U}_F^{\dagger} (\boldsymbol{z}_1 ) \mathcal{U}_F (\boldsymbol{z}_2 ) ]_{ij} - \delta_{ij} \right)  \bigg \} \bar{Q}_j (\boldsymbol{z}_2) \; .
    \label{Eq:EvoOfcrossedOperator}
\end{gather}
Equation~(\ref{Eq:EvoOfcrossedOperator}) may be obtained either by explicitly evaluating the crossed counterpart of the left diagram in Fig.~\ref{Fig:QIOAtOneLoop}, using the effective propagator given in Eq. \eqref{Eq:GluonToQuarkProp}, or equivalently by applying the signature transformation $\mathcal{S}$, corresponding to the exchange $\infty^+ \leftrightarrow -\infty^+$ in the Wilson lines. 
%

\subsection{Linearization and the planar limit}
\label{sec:QuarkLinearization}

We now investigate how quark Reggeization emerges from the nonlinear evolution equation~\eqref{Eq:AlmostReggeEvoCoord}. As a first consistency check, setting all Wilson lines to the identity reduces the color structure in the curly brackets to $C_F\delta_{ij}$. The resulting equation governs the rapidity evolution of the good component of the quark background field,
\begin{equation}
\frac{\partial}{\partial\eta}
\int dz^+\,\Psi_i^{(-)}(z^+,\boldsymbol z_1)
=
C_F\,
\mathcal K_{\rm RP}(\boldsymbol z_1,\boldsymbol z_2)
\otimes_{\boldsymbol z_2}
\int dz^+\,\Psi_i^{(-)}(z^+,\boldsymbol z_2).
\label{Eq:FreeQuarkEvolution}
\end{equation}
After Fourier transformation, the eigenvalue of this equation is the one-loop massless-quark trajectory,
\begin{equation}
\delta^{(1)}(-\boldsymbol p^2)
=
\frac{C_F}{N_c}\,
\omega^{(1)}(-\boldsymbol p^2),
\end{equation}
in agreement with Eq.~\eqref{Eq:QuarkReggeTraj}. This provides a nontrivial one-loop check of Eq.~\eqref{Eq:AlmostReggeEvoCoord}, but does not by itself establish Reggeization beyond this order. \\

The relation between Eq.~\eqref{Eq:AlmostReggeEvoCoord} and Regge-pole evolution becomes particularly transparent in the planar limit. The second color structure in the curly brackets is suppressed by $1/N_c$, while the dipole operator admits the dilute expansion
\begin{equation}
{\rm Tr}\!\left[
\mathcal U_F(\boldsymbol z_1)
\mathcal U_F^\dagger(\boldsymbol z_2)
\right]
=
N_c+\mathcal O(g^2R^2).
\label{Eq:dipoleExp}
\end{equation}
Restricting the evolution to the one-Reggeized-gluon sector then gives
\begin{equation}
\frac{\partial R_{Q,i}(\boldsymbol z_1)}
{\partial\eta}
=
\frac{N_c}{2}\,
\mathcal K_{\rm RP}(\boldsymbol z_1,\boldsymbol z_2)
\otimes_{\boldsymbol z_2}
R_{Q,i}(\boldsymbol z_2)
+
\mathcal O\left(\frac{1}{N_c},g^4R^2\right).
\label{Eq:ReggeQuarkLargeNc}
\end{equation}
Upon Fourier transformation, the evolution is governed by
\begin{equation}
\frac{1}{2}\,
\omega^{(1)}(-\boldsymbol p^2),
\end{equation}
which precisely reproduces the large-$N_c$ limit of the quark trajectory, since $C_F/N_c\to1/2$. This simplification has a natural interpretation in terms of signature. At finite $N_c$, amplitudes carrying fermion quantum numbers contain analytic structures more involved than a single Regge pole, and pure Regge-pole behavior is recovered only in the positive-signature sector~\cite{Fadin:1976nw,Fadin:1977jr}. In the planar limit, the nonplanar color structure responsible for this obstruction is suppressed. Equation~\eqref{Eq:AlmostReggeEvoCoord} provides a direct Wilson-line realization of this mechanism: its leading color structure factorizes into a dipole multiplying the quark operator, and the dipole reduces to the identity in the one-Reggeized-gluon sector. More importantly, the planar evolution contains no contribution of order $g^3R$ coupling $R_Q$ to an additional Reggeized-gluon degree of freedom. The first departures from the one-Reggeized-quark sector arise at order $g^4R^2$. This closely parallels the gluon evolution discussed in Sec.~\ref{sec:Gluon_Reggeization_Wilson_line}, where the absence of an $O(g^3)$-order mixing with $R^2$, due to signature conservation, ensures that the Regge-pole eigenvalue remains unaffected through NLLA. By the same power-counting argument, Eq.~\eqref{Eq:ReggeQuarkLargeNc} strongly suggests that, in the planar limit, the non-signaturized Reggeized-quark degree of freedom remains a Regge-pole eigenstate through NLLA. We will return to this point in Sec.~\ref{sec:QuarkBeyondLLA}.  \\

\begin{figure}
    \centering
    \includegraphics[width=0.54\linewidth]{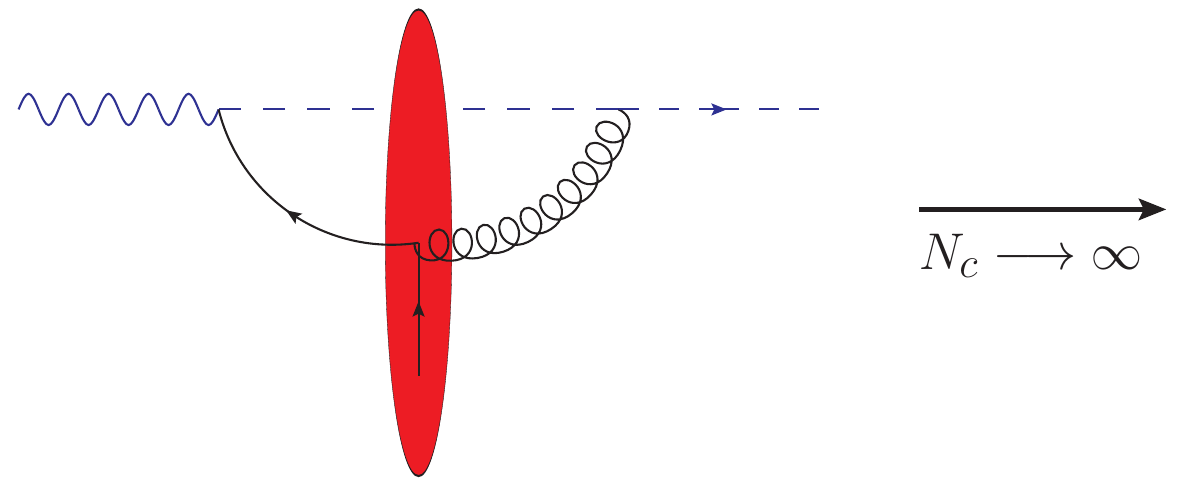}
    \hspace{0.3 cm}
    \includegraphics[width=0.37\linewidth]{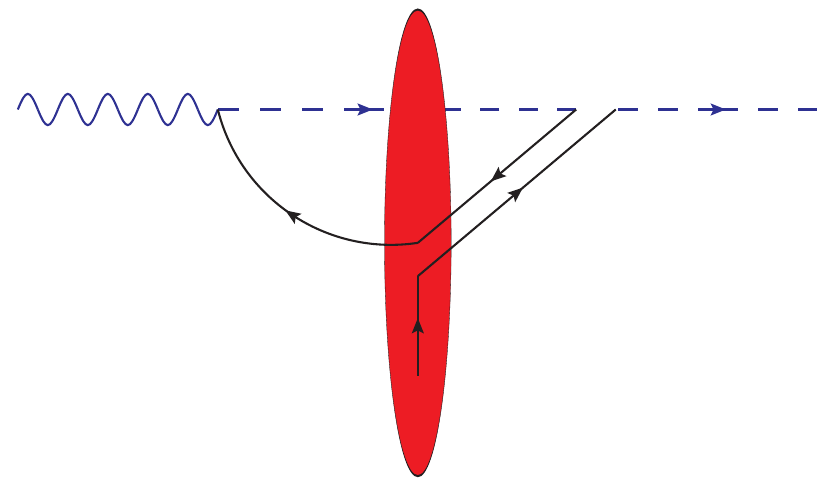}
    \caption{Simplification of the one-loop evolution in the planar limit.}
    \label{Fig:QIOAtOneLoopPlanar}
\end{figure}

At finite $N_c$, the second color structure in Eq.~\eqref{Eq:AlmostReggeEvoCoord} survives. Expanding the Wilson lines in the dilute regime and retaining the contribution linear in the Reggeized-gluon field gives
\begin{align}
\frac{\partial R_{Q,i}(\boldsymbol z_1)}
{\partial\eta}
&=
C_F\,
\mathcal K_{\rm RP}(\boldsymbol z_1,\boldsymbol z_2)
\otimes_{\boldsymbol z_2}
R_{Q,i}(\boldsymbol z_2)
\nonumber\\
&\quad
+
\frac{1}{2N_c}
[\mathcal K_{\rm RPB}(\boldsymbol z_1,\boldsymbol z_2)]_{ij}
\otimes_{\boldsymbol z_2}
\left[
R^a(\boldsymbol z_2)-R^a(\boldsymbol z_1)
\right]
\bar R_{Q,j}(\boldsymbol z_2)
+
\mathcal O(g^4R^2),
\label{Eq:ReggePlusReggeBreak1}
\end{align}
where
\begin{equation}
[\mathcal K_{\rm RPB}(\boldsymbol z_1,\boldsymbol z_2)]_{ij}
\equiv
ig\,t^a_{ij}\,
\frac{a_s}{(\boldsymbol z_{12}^2)^{d-1}}
=
ig\,t^a_{ij}\,
\mathcal K_{\rm RP}(\boldsymbol z_1,\boldsymbol z_2).
\label{Eq:KNonRegge}
\end{equation}
We refer to this contribution as the Regge-pole-breaking (RPB) term. \\

The first term in Eq.~\eqref{Eq:ReggePlusReggeBreak1} closes on the one-Reggeized-quark sector and reproduces the one-loop quark Regge trajectory at LLA. The RPB term has a qualitatively different operator structure: it couples $R_{Q,i}$ to the composite $R \Psi$ operator
\begin{equation}
\mathcal O_{R \Psi,j}^a (\boldsymbol z_1,\boldsymbol z_2)
\equiv
\left[
R^a(\boldsymbol z_2)-R^a(\boldsymbol z_1)
\right]
\bar R_{Q,j}(\boldsymbol z_2).
\label{Eq:qRCompositeOperator}
\end{equation}
Its kernel starts at order $g^3$, one power of the coupling higher than the diagonal Regge-pole kernel, while the operators $R_{Q,i}$ and $O_{R \Psi,i}$ starts both at $\mathcal{O}(g^0)$. Consequently, this term does not affect the LLA evolution of the full amplitude, but it enters at the first subleading logarithmic order and prevents the non-signaturized operator $R_Q$ from evolving autonomously within the NLLA. The structure of the RPB contribution strongly suggests that the obstruction to Regge-pole evolution is tied to the absence of definite signature. First, $R_Q$ is not an eigenstate of the signature transformation, which maps it into the crossed operator $\bar R_Q$. This latter explicitly couples to $R_Q$ under evolution. Second, under the signature transformation, the diagonal Regge-pole contribution is left invariant, whereas the RPB term changes sign as a consequence of the negative signature of the Reggeized-gluon field. The crossed evolution equation therefore contains the same non-diagonal operator structure with the opposite sign. These properties indicate that the RPB contribution should be analyzed in a basis of definite-signature operators.

\subsection{Definite-signature operators at finite $N_c$}
\label{sec:QuarkDefiniteSignature}

The signature projection is naturally implemented after expanding the Wilson-line operators in the dilute regime. We define
\begin{equation}
R_{Q,i}^{(\pm)}(\boldsymbol z)
=
\frac{1}{2}
\left[
R_{Q,i}(\boldsymbol z)
\pm
\bar R_{Q,i}(\boldsymbol z)
\right].
\label{Eq:SignProjections}
\end{equation}
These combinations should be understood as projections of the linearized Reggeized operators. 

\subsubsection{Positive-signature sector}

Using Eqs.~\eqref{Eq:AlmostReggeizedQuarkInterLinear} and \eqref{Eq:CrossedOperatorQuarkLO}, the positive-signature operator reads
\begin{equation}
R_{Q,i}^{(+)}(\boldsymbol z)
=
\int dz^+\,
\left[
\delta_{ij}
+
ig\,t^a_{ij}
\frac{
R^a(\infty^+,z^+,\boldsymbol z)
-
R^a(z^+,-\infty^+,\boldsymbol z)
}{2}
\right]
\Psi_j^{(-)}(z^+,\boldsymbol z).
\label{Eq:PositiveQuarkOperatorR}
\end{equation}
The difference of the two semi-infinite Reggeized-gluon operators can be expressed by means of the Baker--Campbell--Hausdorff formula,
\begin{align}
\ln X+\ln Y
&=
\ln(XY)
-\frac{1}{2}[\ln X,\ln Y]
-\frac{1}{12}[\ln X,[\ln X,\ln Y]]
-\frac{1}{12}[\ln Y,[\ln Y,\ln X]]
+\cdots.
\label{Eq:BCHQuark}
\end{align}
Within the one-Reggeized-gluon approximation, one obtains
\begin{align}
&R^a(\infty^+,z^+,\boldsymbol z)
-
R^a(z^+,-\infty^+,\boldsymbol z)
=
\frac{f^{abc}}{gN_c}
\left[
\ln\left(
\mathcal U_A(\infty^+,z^+,\boldsymbol z)
\mathcal U_A^\dagger(z^+,-\infty^+,\boldsymbol z)
\right)
\right]^{bc}
+
\mathcal O(gR^2).
\label{Eq:PositiveBCH}
\end{align}
The positive-signature operator can therefore be represented as
\begin{align}
R_{Q,i}^{(+)}(\boldsymbol z)
&=
\int dz^+\,
\Bigg\{
\delta_{ij}
+
\frac{if^{abc}t^a_{ij}}{2N_c}
\left[
\ln\left(
\mathcal U_A(\infty^+,z^+,\boldsymbol z)
\mathcal U_A^\dagger(z^+,-\infty^+,\boldsymbol z)
\right)
\right]^{bc}
\Bigg\}
\Psi_j^{(-)}(z^+,\boldsymbol z)
+
\mathcal O(g^2R^2).
\label{Eq:RQIoperator}
\end{align}

To derive its evolution, we combine Eq.~\eqref{Eq:ReggePlusReggeBreak1} with the crossed equation
\begin{align}
\frac{\partial\bar R_{Q,i}(\boldsymbol z_1)}
{\partial\eta}
&=
C_F\,
\mathcal K_{\rm RP}(\boldsymbol z_1,\boldsymbol z_2)
\otimes_{\boldsymbol z_2}
\bar R_{Q,i}(\boldsymbol z_2)
\nonumber\\
&\quad
-
\frac{1}{2N_c}
[\mathcal K_{\rm RPB}(\boldsymbol z_1,\boldsymbol z_2)]_{ij}
\otimes_{\boldsymbol z_2}
\left[
R^a(\boldsymbol z_2)-R^a(\boldsymbol z_1)
\right]
R_{Q,j}(\boldsymbol z_2)
+
\mathcal O(g^4R^2).
\label{Eq:ReggePlusReggeBreak2}
\end{align}
This equation follows either by linearizing Eq.~\eqref{Eq:EvoOfcrossedOperator} or by applying the signature transformation to Eq.~\eqref{Eq:ReggePlusReggeBreak1}. Under this transformation, $\mathcal K_{\rm RP}$ and $\mathcal K_{\rm RPB}$ are invariant, whereas the Reggeized-gluon combination
\begin{equation}
R^a(\boldsymbol z_2)-R^a(\boldsymbol z_1)
\end{equation}
changes sign. Within $\mathcal{O}(g^4 R^2)$, the operators entering the RPB term can be replaced by their common leading component,
\begin{equation}
R_{Q,i}(\boldsymbol z)
=
\int dz^+\,\Psi_i^{(-)}(z^+,\boldsymbol z)
+\mathcal O(gR),
\qquad
\bar R_{Q,i}(\boldsymbol z)
=
\int dz^+\,\Psi_i^{(-)}(z^+,\boldsymbol z)
+\mathcal O(gR).
\label{Eq:SimpReggeQuark}
\end{equation}
The RPB contributions then cancel in the positive-signature combination, leaving
\begin{equation}
\frac{\partial R_{Q,i}^{(+)}(\boldsymbol z_1)}
{\partial\eta}
=
C_F\,
\mathcal K_{\rm RP}(\boldsymbol z_1,\boldsymbol z_2)
\otimes_{\boldsymbol z_2}
R_{Q,i}^{(+)}(\boldsymbol z_2)
+
\mathcal O(g^4R^2).
\label{Eq:FinalReggeization}
\end{equation}
Crucially, Eq.~\eqref{Eq:FinalReggeization} contains no $O(g^3R)$ contribution coupling the positive-signature operator to an additional Reggeized-gluon degree of freedom. The first possible contamination of the one-Reggeized-quark sector starts instead at order $g^4 R^2$. This is the same perturbative pattern encountered in the Reggeized-gluon evolution discussed in Sec.~\ref{sec:Gluon_Reggeization_Wilson_line}, where multi-Reggeon mixing beginning at order $g^4$ does not modify the Regge-pole eigenvalue through NLLA. The structure of Eq.~\eqref{Eq:FinalReggeization} therefore provides strong evidence that the positive-signature Reggeized-quark degree of freedom likewise remains a Regge-pole eigenstate through NLLA. A more detailed discussion of the corresponding mixing structure is presented in Sec.~\ref{sec:QuarkBeyondLLA}. After Fourier transformation, the eigenvalue is precisely the complete one-loop massless-quark trajectory in Eq.~\eqref{Eq:QuarkReggeTraj}. Equation~\eqref{Eq:FinalReggeization} therefore identifies the positive-signature operator in Eq.~\eqref{Eq:RQIoperator} as the Reggeized-quark interpolating operator.

\subsubsection{Negative-signature sector}

The negative-signature operator is perturbatively suppressed with respect to its positive-signature counterpart~\cite{Fadin:1976nw,Fadin:1977jr}. Indeed, from Eq.~\eqref{Eq:SignProjections}, one finds
\begin{equation}
R_{Q,i}^{(-)}(\boldsymbol z)
=
\frac{ig\,t^a_{ij}}{2}
\int dz^+\,
\left[
R^a(\infty^+,z^+,\boldsymbol z)
+
R^a(z^+,-\infty^+,\boldsymbol z)
\right]
\Psi_j^{(-)}(z^+,\boldsymbol z).
\label{Eq:NegativeSignatureBeforeBCH}
\end{equation}
The BCH formula gives
\begin{equation}
R^a(\infty^+,z^+,\boldsymbol z)
+
R^a(z^+,-\infty^+,\boldsymbol z)
=
R^a(\boldsymbol z)
+
\mathcal O(gR^2),
\label{Eq:NegativeBCH}
\end{equation}
and hence
\begin{equation}
R_{Q,i}^{(-)}(\boldsymbol z)
=
\frac{ig\,t^a_{ij}}{2}
R^a(\boldsymbol z)
\int dz^+\,
\Psi_j^{(-)}(z^+,\boldsymbol z)
+
\mathcal O(g^2R^2).
\label{Eq:NegativeSignQuark}
\end{equation}
Thus, the leading field content of \(R_Q^{(-)}\) is a composite \( R \Psi\) configuration: it contains an elementary background-quark insertion multiplied by one Reggeized-gluon field. The quark field appearing in Eq.~\eqref{Eq:NegativeSignQuark} should not itself be identified with the positive-signature Reggeized-quark operator. \\

Combining Eqs.~\eqref{Eq:ReggePlusReggeBreak1} and \eqref{Eq:ReggePlusReggeBreak2}, one obtains
\begin{align}
\frac{\partial R_{Q,i}^{(-)}(\boldsymbol z_1)}
{\partial\eta}
&=
C_F\,
\mathcal K_{\rm RP}(\boldsymbol z_1,\boldsymbol z_2)
\otimes_{\boldsymbol z_2}
R_{Q,i}^{(-)}(\boldsymbol z_2)
\nonumber\\
&\quad
+
\frac{1}{2N_c}
[\mathcal K_{\rm RPB}(\boldsymbol z_1,\boldsymbol z_2)]_{ij}
\otimes_{\boldsymbol z_2}
\left[
R^a(\boldsymbol z_2)-R^a(\boldsymbol z_1)
\right]
R_{Q,j}^{(+)}(\boldsymbol z_2)
+
\mathcal O(g^4R^2).
\label{Eq:NegativeSignatureEvolution}
\end{align}
Both terms on the right-hand side contribute at the leading nonvanishing order of the negative-signature operator. Indeed, $R_Q^{(-)}\sim g R \Psi$, so that
\begin{equation}
K_{\rm RP} R_Q^{(-)} \sim g^2 R_Q^{(-)}
\sim
g^3 R \Psi,
\qquad \qquad
\mathcal K_{\rm RPB} \,R\,R_Q^{(+)}
\sim g^3 R\,R_Q^{(+)} \sim
g^3 R \Psi.
\end{equation}
The negative-signature evolution therefore does not close on \(R_Q^{(-)}\) alone at its own LLA. Instead, it belongs to a coupled Reggeized quark/gluon sector, in agreement with the observation of Ref.~\cite{Fadin:1977jr}, that the non-Regge pole structure of amplitudes with negative-signature exchange could be explained, in LLA, by branch cuts generated by the exchange of one Reggeized quark plus one Reggeized gluon. \\

As expected, rapidity evolution does not mix sectors of different total signature. The operator \(R_Q^{(-)}\) has negative signature, while the composite \(R\,R_Q^{(+)}\) also has negative total signature:
\begin{equation}
{\rm signature}(R\,R_Q^{(+)})
=
{\rm signature}(R)\,
{\rm signature}(R_Q^{(+)})
=
(-1)(+1)
=
-1.
\end{equation}
The evolution in Eq.~\eqref{Eq:NegativeSignatureEvolution} therefore preserves signature, even though its operator structures contain constituents with different individual signatures.

\subsection{Operator mixing and quark Reggeization beyond LLA}
\label{sec:QuarkBeyondLLA}

We finally discuss what the structure derived above implies for quark Reggeization beyond the leading logarithmic approximation. Our aim is not to determine the complete higher-order evolution kernels, but to identify the first multi-Reggeon-operator sectors allowed by signature and their perturbative power counting.

\subsubsection{Positive-signature sector}

Beyond the one-Reggeized-gluon approximation used in deriving Eq.~\eqref{Eq:FinalReggeization}, i.e. retaining $\mathcal{O} (g^4 R^2)$-terms, the positive-signature evolution also contains the contribution
\begin{align}
\frac{\partial R_{Q,i}^{(+)}(\boldsymbol z_1)}
{\partial\eta}
\supset
-&
\frac{1}{2N_c}
[\mathcal K_{\rm RPB}(\boldsymbol z_1,\boldsymbol z_2)]_{ij}
\otimes_{\boldsymbol z_2}
\left[
R^a(\boldsymbol z_2)-R^a(\boldsymbol z_1)
\right]
R_{Q,j}^{(-)}(\boldsymbol z_2).
\label{Eq:PositiveSignatureFirstMixing}
\end{align}
The symbol \(\supset\) emphasizes that Eq.~\eqref{Eq:PositiveSignatureFirstMixing} isolates only the mixing generated by relaxing Eqs.~(\ref{Eq:SimpReggeQuark}), rather than giving the complete evolution at order \(g^4 R^2\). Signature is conserved because the two negative-signature factors combine into a positive-signature composite operator:
\begin{equation}
{\rm signature}(R\,R_Q^{(-)})
=
(-1)(-1)
=
+1.
\end{equation}
Equation~\eqref{Eq:PositiveSignatureFirstMixing} identifies one of the operator structures belonging to the positive-signature composite sector. Additional operators with the same quantum numbers contribute at the same order. Their multiplicity does not alter the coupling power counting below. For simplicity, in the following we retain the operator displayed in Eq.~\eqref{Eq:PositiveSignatureFirstMixing} as a representative element of this sector, sufficient to illustrate the structure and perturbative consequences of the mixing. \\

From Eq.~\eqref{Eq:NegativeSignQuark}, the leading field content of $R_Q^{(-)}$ is \(   R \Psi\), while of $R_Q^{(+)}$ is \( \Psi\). Thus,
\begin{equation}
R_Q^{(+)}\sim g^0,
\qquad
R\,R_Q^{(-)}\sim g.
\end{equation}
After introducing the rescaled operator, $
g^{-1}R\,R_Q^{(-)} $, both components have the same leading perturbative scaling. Their evolution is expected to take the schematic form
\begin{equation}
\frac{\partial}{\partial\eta}
\begin{pmatrix}
R_{Q,i}^{(+)}
\\[0.2cm]
g^{-1}R^aR_{Q,i}^{(-)}
\end{pmatrix}
=
\begin{pmatrix}
g^2H_{R_Q^{(+)}\to R_Q^{(+)}}
&
g^4H_{RR_Q^{(-)}\to R_Q^{(+)}}
\\[0.2cm]
g^4H_{R_Q^{(+)}\to RR_Q^{(-)}}
&
g^2H_{RR_Q^{(-)}\to RR_Q^{(-)}}
\end{pmatrix}
\begin{pmatrix}
R_{Q,i}^{(+)}
\\[0.2cm]
g^{-1}R^aR_{Q,i}^{(-)}
\end{pmatrix}
+\cdots.
\label{Eq:MatrixEvoQuark}
\end{equation}
Here, coordinate, color, and spinor structures are left implicit, and each \(H\) starts at order \(g^0\). The power counting of the first row follows from the explicit one-loop calculation of the present paper. The diagonal evolution of the second sector must start at order \(g^2\), as required by perturbative rapidity evolution. The only nontrivial assumption is the \(g^4\) scaling of the lower-left transition, which has not been derived in the present work. Although the reverse transition has not been computed explicitly, the power counting displayed in Eq.~\eqref{Eq:MatrixEvoQuark} is supported by the expected Hermiticity of the rapidity-evolution Hamiltonian. After rescaling the composite operator $R^aR_Q^{(-)}$ by one inverse power of $g$, both operator sectors start at the same perturbative order. The explicit calculation then shows that the transition from the rescaled composite sector to $R_Q^{(+)}$ starts at order $g^4$. If the evolution Hamiltonian is Hermitian with respect to an appropriate inner product in this normalized Reggeon space, the reverse matrix element must start at the same order. This is analogous to the relation between the one-to-three and three-to-one Reggeon transitions in the gluon sector~\cite{Caron-Huot:2013fea}. In the Wilson-line formulation, scattering amplitudes can be viewed as inner products between projectile and target states represented by functionals of Wilson lines. More precisely, the projectile and target are described by functionals of right ($U$)- and left ($\bar U$)- moving Wilson lines, lying respectively on the null planes $x^-=0$ and $x^+=0$, and their inner product is defined through the vacuum expectation value of their time-ordered product~\cite{Caron-Huot:2013fea},
\begin{equation}
\langle \mathcal O_1,\mathcal O_2\rangle
\equiv
\langle 0|T \{ \mathcal O_1[U],\mathcal O_2[\bar U] \} |0\rangle \; . 
\end{equation}
This correlator provides the corresponding high-energy scattering amplitude (once the appropriate external-state impact factors are specified). In this sense, the inner product gives a direct connection between the Wilson-line operator description and physical scattering amplitudes. A rigorous implementation of this argument would require identifying the inner product appropriate to the NEik operator space, and establishing the corresponding self-adjointness of the rapidity Hamiltonian~\cite{Caron-Huot:2013fea}, which we leave for future work. Under this assumption, the off-diagonal mixing modifies the Regge-pole eigenvalue only at order
\begin{equation}
\Delta\lambda_{\rm mixing}
\sim
\frac{g^4g^4}{g^2}
=
g^6.
\end{equation}
Multi-operator mixing is therefore not expected to affect the positive-signature Regge-pole eigenvalue through NLLA. At order \(g^4\), the two-loop quark trajectory is therefore expected to arise entirely from the next perturbative correction to the diagonal kernel $H_{R_Q^{(+)}\to R_Q^{(+)}}$. This is consistent with the successful fixed-order extraction of the two-loop quark trajectory~\cite{Bogdan:2002sr}. \\

The role of signature in this argument is essential and closely parallels the Reggeized-gluon case discussed in Sec.~\ref{sec:Gluon_Reggeization_Wilson_line}. Since a single Reggeized gluon has negative signature, a two-Reggeized-gluon operator $R^2$ has positive signature and cannot mix with the one-Reggeized-gluon sector. Were such a transition allowed, the lower perturbative suppression of the two-Reggeon operator could generate a correction to the Regge-pole eigenvalue already at NLLA. Signature conservation instead postpones the first allowed mixing to the three-Reggeon sector, $R\leftrightarrow R^3$, whose off-diagonal kernels start at order $g^4$ and modify the eigenvalue only at order $g^6$. An analogous mechanism operates for the Reggeized quark. The operator $R_Q^{(+)}$ has positive signature, whereas the composite $R\,R_Q^{(+)}$ has negative total signature, and is therefore forbidden from mixing with $R_Q^{(+)}$. This exclusion is particularly important because $R_Q^{(+)}$ does not carry the additional perturbative suppression associated with $R_Q^{(-)}$ and could otherwise affect the Regge-pole evolution already at NLLA. The first allowed composite operator is instead $R\,R_Q^{(-)}$, whose total signature is positive. Since $R_Q^{(-)}$ starts one power of $g$ higher than $R_Q^{(+)}$, the allowed composite sector is correspondingly suppressed. After the appropriate perturbative rescaling, its mixing with $R_Q^{(+)}$ starts at order $g^4$ and, under the Hermiticity assumption discussed above, modifies the Regge-pole eigenvalue only at order $g^6$. Signature conservation therefore removes precisely the potentially dangerous lower-order mixing, providing the key structural reason why the positive-signature quark trajectory is expected to remain unaffected by multi-Reggeon-operator contributions through NLLA.

\subsubsection{Negative-signature sector and the full amplitude}

The leading negative-signature operator and the composite operator appearing in its evolution scale as
\begin{equation}
g^{-1}R_Q^{(-)}
\sim
R\,R_Q^{(+)}
\sim
g^0.
\end{equation}
The corresponding coupled evolution is therefore expected to have the schematic form
\begin{equation}
\frac{\partial}{\partial\eta}
\begin{pmatrix}
g^{-1}R_{Q,i}^{(-)}
\\[0.2cm]
R^aR_{Q,i}^{(+)}
\end{pmatrix}
=
\begin{pmatrix}
g^2H_{R_Q^{(-)}\to R_Q^{(-)}}
&
g^2H_{RR_Q^{(+)}\to R_Q^{(-)}}
\\[0.2cm]
g^2H_{R_Q^{(-)}\to RR_Q^{(+)}}
&
g^2H_{RR_Q^{(+)}\to RR_Q^{(+)}}
\end{pmatrix}
\begin{pmatrix}
g^{-1}R_{Q,i}^{(-)}
\\[0.2cm]
R^aR_{Q,i}^{(+)}
\end{pmatrix}
+ \mathcal{O} (g^3 R^2).
\label{Eq:MatrixEvoQuarkNegative}
\end{equation}
All entries may contribute at the leading logarithmic order intrinsic to the negative-signature sector. This prevents its evolution from reducing to an autonomous single-pole equation and provides a Wilson-line realization of the branch-point structure associated with quark--gluon exchange identified in Refs.~\cite{Fadin:1976nw,Fadin:1977jr}. \\

In the planar limit, the nonplanar off-diagonal contributions are suppressed, while the diagonal quark kernel approaches
\begin{equation}
C_F\mathcal K_{\rm RP}
\longrightarrow
\frac{N_c}{2}\mathcal K_{\rm RP}.
\end{equation}
Since the corrections to the equation (\ref{Eq:MatrixEvoQuarkNegative}) are of $\mathcal{O} (g^3 R^2)$, the negative-signature sector recovers the same Regge-pole trajectory as the positive-signature sector at its leading logarithmic order, realizing signature degeneracy. \\

Finally, the logarithmic accuracy intrinsic to the negative-signature sector should be distinguished from that of the full amplitude. Since \(R_Q^{(-)}\) starts one power of the coupling later than \(R_Q^{(+)}\), its leading-logarithmic evolution first contributes at NLLA to the full non-signaturized amplitude. The results obtained above therefore reproduce the pattern summarized in Sec.~\ref{sec:QuarkSignature}: the full amplitude Reggeizes at LLA, while at finite \(N_c\) its pure Regge-pole form is obstructed at NLLA by the leading negative-signature \(R \Psi\) sector. In the positive-signature projection, this obstruction cancels, and the quark Regge-pole eigenvalue is expected to remain unaffected by multi-operator mixing through NLLA.

\section{Universality of quark Reggeization}
\label{sec:UniversGluon}

In this section, we discuss the universality of the construction of the RQI operator developed in the previous section. \\


\begin{figure}
    \centering
    \includegraphics[width=0.5\linewidth]{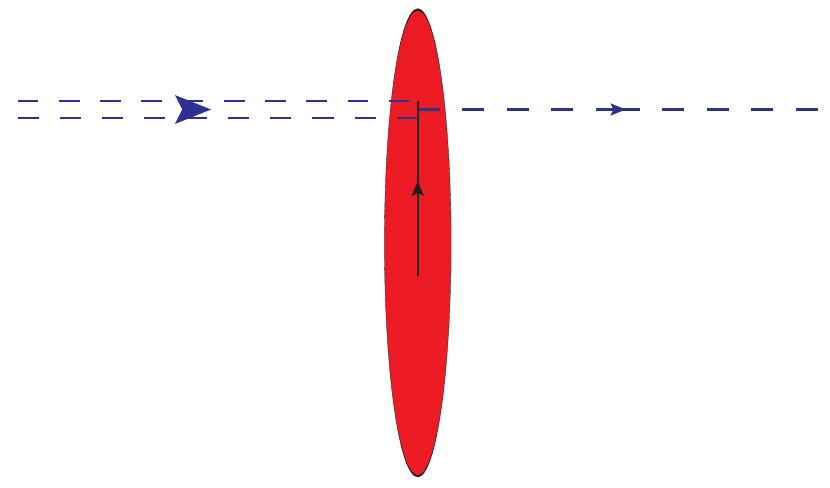}
    \caption{A fast-moving gluon (the double dashed line denotes a Wilson line in the adjoint representation) interacting with the quark background field from the target and changing its nature to a fast-moving quark. Both the quark and the gluon are dressed by gluon shockwave interactions.}
    \label{fig:QIOLeadingGluon}
\end{figure}

In the previous section, we constructed the RQI operator from the simplest identity-changing transition, namely the one associated with the $ \gamma q \longrightarrow q$ transition. However, the Reggeized quark is expected to represent a universal degree of freedom of the high-energy limit. Consequently, the corresponding interpolating operator should be independent of the particular hard process used in its construction. In this subsection, we demonstrate explicitly that same RQI operator indeed emerges from a more general identity-changing transition, thereby illustrating the universality of quark Reggeization within this formulation. \\


In the case of the Reggeized-gluon exchange, this universality property is immediately evident from the fact that the Wilson lines in both fundamental and adjoint representations admit an expansion around the identity in terms of the same $R^a (\boldsymbol{z})$ (see Eq.~(\ref{Eq:UniversExp})). Consequently, in the dilute limit, the high-energy evolution of every scattering amplitude involving identity-preserving transitions in the $s$-channel (quark-to-quark or gluon-to-gluon) is governed by the same Reggeized-gluon degree of freedom. \\

In the processes involving a $t$-channel quark exchange, the most general identity-changing transition in the $s$-channel is the one associated with a gluon transforming into a quark\footnote{All other possible transitions are trivially related to this one.}, Fig. \ref{fig:QIOLeadingGluon}, which reads 
\begin{gather}
Q^a_{j} (\boldsymbol{z}) = \int d z^+ [\mathcal{U}_F ( \infty^{+}, z^+, \boldsymbol{z} ) t^b \Psi^{(-)} (z^+, \boldsymbol{z} )]_j [\mathcal{U}_A ( z^+, - \infty^{+},  \boldsymbol{z} )]^{ba}.
\end{gather}
Unlike the operator associated with the previously considered photon-initiated channel, this operator carries a richer color structure in the $t$-channel. It is also important to stress that, in general, its evolution equation differs from that of the operator $Q_i(\boldsymbol{z})$ defined in Eq. (\ref{Eq:AlmostReggeizedQuarkInter}). For $SU(3)$, the color decomposition of this operator reads
\begin{equation}
\mathbf 3\otimes\mathbf 8
=
\mathbf 3\oplus\bar{\mathbf 6}\oplus\mathbf{15}.
\end{equation}
For generic $N_c$, the corresponding decomposition contains the
fundamental representation and its two higher-dimensional analogues.
In the following, we focus on the fundamental channel, which reduces
to the color triplet for $N_c=3$.

Although the extraction of the evolution equations governing the $\bar{\mathbf{6}}$ and $\mathbf{15}$ exchanges lies beyond the scope of the present work and is left for future studies, we now show that projecting onto the positive-signature color-triplet representation reproduces exactly the same RQI operator within the one-Reggeized-gluon approximation. Using the normalized projection tensor onto the fundamental channel,
\begin{equation}
[\mathcal P_F]^a_{ij}=\frac{t^a_{ij}}{C_F},
\end{equation}
we obtain
\begin{gather}
[\mathcal{P}_3]^a_{ij} Q^a_{j} (\boldsymbol{z})  = \frac{1}{C_F} \int d z^+ [ t^a \mathcal{U}_F ( \infty^{+}, z^+, \boldsymbol{z} ) t^b \Psi^{(-)} (z^+, \boldsymbol{z} )]_i [\mathcal{U}_A ( z^+, - \infty^{+},  \boldsymbol{z} )]^{ba}
\end{gather}
By using the relation (\ref{Eq:FundAdjoi}) to rewrite the adjoint Wilson line in terms of the fundamental ones, this expression becomes 
\begin{gather}
[\mathcal{P}_3]^a_{ij} Q^a_{j} (\boldsymbol{z})  = \frac{1}{C_F} \int d z^+ [ \mathcal{U}_F^{\dagger} ( z^+, - \infty^{+},  \boldsymbol{z} ) t^b \mathcal{U}_F ( \boldsymbol{z} ) t^b \Psi^{(-)} (z^+, \boldsymbol{z} )]_i 
\end{gather}
Applying the Fierz identity (\ref{Eq:FierzIdentity}) yields
\begin{gather}
[\mathcal{P}_3]^a_{ij} Q^a_{j} (\boldsymbol{z})  = \frac{1}{C_F} \int d z^+ \frac{1}{2} \left[ \delta_{nj} \delta_{lk} - \frac{1}{N_c} \delta_{nl} \delta_{kj} \right] [ \mathcal{U}_F^{\dagger} ( z^+, - \infty^{+},  \boldsymbol{z} )]_{in}  [\mathcal{U}_F (  \boldsymbol{z} )]_{lk}  \Psi^{(-)}_j (z^+, \boldsymbol{z} )  \nonumber \\
 = \frac{1}{ C_f} \int d z^+  \frac{1}{2} \left[ {\rm Tr}_c [\mathcal{U}_F (  \boldsymbol{z} )] \; [ \mathcal{U}_F^{\dagger} ( \infty^+, - \infty^{+},  \boldsymbol{z} )]_{in} - \frac{ \delta_{in} }{N_c} \right] [ \mathcal{U}_F ( \infty^{+}, z^+  \boldsymbol{z} )   \Psi^{(-)}  (z^+, \boldsymbol{z} )]_n \; .
\end{gather}
Expanding the above expression within the one-Reggeized-gluon approximation, we obtain
\begin{align}
    [\mathcal{P}_3]^a_{ij} Q^a_{j} (\boldsymbol{z}) & =  \int d z^+  \left[ \delta_{ij} + ig  t^a_{ij} R^a (\infty^+, z^+, \boldsymbol{z}) \right]  \Psi^{(-)}_j  (z^+, \boldsymbol{z} ) \; - \frac{N_c}{2  C_F} \int d z^+ \; i g t^a_{ij} R^a (\boldsymbol{z} )  \Psi^{(-)}_j  (z^+, \boldsymbol{z} ) + \mathcal{O} (g^2R^2) \nonumber \\ &= R_{Q,i} (\boldsymbol{z}) - \frac{N_c}{C_F} R_{Q,i}^{(-)} (\boldsymbol{z}) + \mathcal{O} (g^2R^2) \; ,
 \end{align}
where the second term is manifestly odd under the signature transformation. Taking the positive-signature component of this expression in the common dilute Reggeon space, the second term drops out, and one obtains
\begin{equation}
\left.[\mathcal P_F]^a_{ij}Q_j^a(\boldsymbol z)
\right|_{\text{positive signature}}
=
R_{Q,i}^{(+)}(\boldsymbol z)
+
\mathcal O(g^2R^2).
\end{equation}

We therefore conclude that the positive-signature color-triplet exchange is universally described by the same RQI operator $R^{(+)}_{Q,i}(\boldsymbol{z})$, independently of the identity-changing process used in its construction, thereby establishing the universality of quark Reggeization within this formalism. The negative-signature projection is likewise proportional to
$R_Q^{(-)}$, confirming that the same negative-signature degree of
freedom governs the fundamental channel, up to a process-dependent
normalization.

\section{Massive quark Regge trajectory}
\label{sec:Massive_quarks}

In the last section of the paper, we demonstrate that the framework can accommodate a non-zero quark mass and use it to extract the one-loop Regge trajectory of massive quarks. 

\subsection{Evolution equation of $Q_i (\boldsymbol{z})$ at one-loop accuracy with non-zero quark mass}

In the presence of a non-vanishing quark mass, the calculation requires the full propagator given in Eq.~(\ref{Eq:AntiquarkToGluonPropNEik}). Compared to the massless case, the nontrivial modification arises in the integration over the longitudinal coordinate $z_3^+$. While for $m=0$ the integral over $z_3^+$ reads 
\begin{gather}
   I_{z_3^+} (m=0) = -\boldsymbol{z}_{12}^2 \int_{0}^{\infty} d z_3^+ \; e^{i \frac{k^+}{2 z_3^+} \boldsymbol{z}_{12}^2} (z_{3}^+)^{-1-d/2}   = -(\boldsymbol{z}_{12}^2)^{1-d/2} \frac{ i^{d/2} 2^{d/2}  \Gamma \left( d/2 \right)}{ \left( k^+  \right)^{d/2} } \; ,
\end{gather}
the corresponding integral in the massive case becomes
\begin{gather}
\label{eq:intz3+_m_1}
   I_{z_3^+} (m) = \int_{0}^{\infty} d z_3^+ \; e^{i \frac{k^+}{2 z_3^+} \boldsymbol{z}_{12}^2 -i \frac{z_3^+}{2 k^+} m^2 } \left( \slashed{z}_{12 \perp} - \frac{z_3^+}{k^+} m \right) \slashed{z}_{12 \perp} (z_{3}^+)^{-1-d/2} \; .
\end{gather}
The integration over $z_3^+$ in the case of massive quarks can be performed by employing the integral representation of the modified Bessel function of the second kind which reads  
\begin{gather}
\label{eq:Int_rep_Bessel}
    \int_0^{\infty} x^{\nu-1} \exp \left \{ \frac{i \mu}{2} \left( x - \frac{\beta^2}{x} \right) \right \}  dx = 2 \beta^{\nu} i^{\nu} K_{\nu} (\beta \mu) \; . 
\end{gather}
Using Eq. \eqref{eq:Int_rep_Bessel} to perform the $z_3^+$ integration in Eq. \eqref{eq:intz3+_m_1} yields
\begin{gather}
   I_{z_3^+} (m) = -(\boldsymbol{z}_{12}^2)^{1-d/2} \frac{ i^{d/2} 2^{d/2}  \Gamma \left( d/2 \right)}{ \left( k^+  \right)^{d/2} } \slashed{O}_q (z_{21}, m) \; , 
\end{gather}
where $\slashed{O}_q (z_{21}, m)$ is given by  
\begin{gather}
    \slashed{O}_q (z_{12}, m) = \frac{2^{1-d/2}}{ \displaystyle \Gamma \left( d/2 \right)} \bigg[ (m |\boldsymbol{z}_{12}|)^{d/2} K_{d/2} (m |\boldsymbol{z}_{12}|) - i m \slashed{z}_{12 \perp} (m |\boldsymbol{z}_{12}|)^{d/2-1} K_{d/2-1} (m |\boldsymbol{z}_{12}|) \bigg] \; ,
\end{gather}
and it encodes all finite-mass corrections. As expected, in the massless limit $m \rightarrow 0$, the operator $\slashed{O}_q(z_{12},m)$ reduces to the identity matrix in Dirac space, thereby recovering the massless kernel derived previously. The effect of the quark mass is therefore entirely encoded in $\slashed{O}_q (z_{21}, m)$, while the Wilson line structure of the evolution equation remains unchanged.    
Consequently, the evolution equation for the operator $Q_i(\boldsymbol{z})$ in the presence of a finite quark mass can be written as \footnote{Recall that the transverse coordinates have been relabeled according to $\boldsymbol{z}_1 \leftrightarrow \boldsymbol{z}_2$.}
\begin{align}
        \frac{\partial Q_i (\boldsymbol{z}_1)}{ \partial \eta} & = a_s \int d^d \boldsymbol{z}_2 \frac{ \slashed{O}_q (z_{21}, m)}{(\boldsymbol{z}_{12}^2)^{d-1}} \nonumber \\ 
        & \hspace{1cm} \times \frac{1}{2} \bigg \{  \left( {\rm Tr}[ \mathcal{U}_F (\boldsymbol{z}_1 ) \mathcal{U}_F^{\dagger} (\boldsymbol{z}_2 ) ] - \frac{1}{N_c} \right) \delta_{ij} - \frac{1}{N_c} \left( [ \mathcal{U}_F (\boldsymbol{z_1} ) \mathcal{U}_F^{\dagger} (\boldsymbol{z}_2 ) ]_{ij} - \delta_{ij} \right)  \bigg \} Q_j (\boldsymbol{z}_2) \; .
    \label{Eq:OneLoopEvoQuarkMass}
\end{align}

\subsection{Massive quark Regge trajectory from the coordinate space kernel}

From Eq.~(\ref{Eq:OneLoopEvoQuarkMass}), we observe that the effect of a finite quark mass is entirely encoded in the replacement 
\begin{gather}
  \frac{1}{(\boldsymbol{z}_{12}^2)^{d-1}} \longrightarrow  \frac{ \slashed{O}_q (z_{21}, m)}{(\boldsymbol{z}_{12}^2)^{d-1}} \; ,
\end{gather}
in the evolution kernel. Consequently, the one-loop massive quark Regge trajectory is naturally identified as
\begin{align}
    \delta^{(1)} (\slashed{p}_{\perp},m)  & = a_s C_F \int d^d \boldsymbol{z}_1 \frac{e^{-i \boldsymbol{p} \boldsymbol{z}_1}}{ (\boldsymbol{z}_{1}^{2})^{d - 1} } \nonumber \\ 
    & \hspace{1.8cm}\times \frac{2^{1-d/2}}{ \displaystyle \Gamma \left( d/2 \right)} \bigg[ (m |\boldsymbol{z}_{1}|)^{d/2} K_{d/2} (m |\boldsymbol{z}_{1}|) + i m \slashed{z}_{1\perp} (m |\boldsymbol{z}_{1}|)^{d/2-1} K_{d/2-1} (m |\boldsymbol{z}_{1}|) \bigg] \; .    \label{Eq:ReggeTrajMass}
\end{align}
This expression provides a coordinate-space representation of the one-loop Regge trajectory for massive quarks. 
To establish its equivalence with the momentum-space result as given in Eq.~(\ref{Eq:QuarkReggeTrajEpsilonExact}), we now evaluate the integral over $\boldsymbol{z}_1$ explicitly. We first consider the contribution proportional to $K_{d/2}$ for which we employ 
\begin{gather}
   K_{d/2} (m |\boldsymbol{z}_1|) = 2^{d -1} \Gamma (d/2) \left( \frac{m}{|\boldsymbol{z}_1|} \right)^{d/2}  \int \frac{d^d \boldsymbol{k}}{(2 \pi)^{d/2}} \frac{e^{i \boldsymbol{k} \cdot\boldsymbol{z}_1 }}{ (\boldsymbol{k}^2 + m^2)^{d} } \; . 
\end{gather}
Introducing
\begin{align}
    I_1  &= m^{d/2} \int d^d \boldsymbol{z}_1 \frac{e^{-i \boldsymbol{p} \boldsymbol{z}_1}}{ (\boldsymbol{z}_{1}^{2})^{3d/4 - 1} }    K_{d/2} (m |\boldsymbol{z}_{1}|) = 2^{d-1} \Gamma (d)  m^{d} \int \frac{d^d \boldsymbol{k}}{(2 \pi)^{d/2}} \frac{1}{ (\boldsymbol{k}^2 + m^2)^d }   \int d^d \boldsymbol{z}_1 \frac{ e^{-i(\boldsymbol p-\boldsymbol k)\cdot\boldsymbol z_1}. }{ (\boldsymbol{z}_{1}^{2})^{d - 1} } 
    \nonumber \\ 
    & = 2^{1-d/2} m^{d} \frac{\Gamma (d) \Gamma (1-d/2)}{\Gamma (d-1)} \int d^d \boldsymbol{k} \frac{1}{ (\boldsymbol{k}^2 + m^2)^d [(\boldsymbol{k}-\boldsymbol{p})^2 ]^{1-d/2} } \; ,  
\end{align}
and applying the Feynman parametrization, we obtain
\begin{align}
    I_1   & = 2^{1-d/2} m^{d} \frac{ \Gamma (1+d/2) }{\Gamma (d-1)}  \int_0^1 dx x^{-d/2} (1-x)^{d-1} \int d^d \boldsymbol{k}   \frac{1}{[ \boldsymbol{k}^2 + (1-x) (m^2 + x \boldsymbol{p}^2  ) ]^{1+d/2} } \nonumber \\
    & =  \frac{ 2^{1-d/2} m^{d-2} \pi^{d/2}  }{ \Gamma (d-1) }  \int_0^1 dx \; x^{1-d/2-1} (1-x)^{d-1-1}  \left( 1 + x \frac{\boldsymbol{p}^2}{m^2}  \right)^{-1} \nonumber \\ 
    & =  2^{1-d/2} m^{d-2} \pi^{d/2}    \frac{ \Gamma (1-d/2) 
    }{ \Gamma (d/2) } \; _2 F_1 \left( 1, 1-d/2; d/2 ; - \frac{\boldsymbol{p}^2}{m^2} \right) \; .
\end{align}
As expected, this is the only contribution that survives in the massless limit $m\to0$, and immediately reproduces the known massless Regge trajectory. Indeed, by using the asymptotic expansion of the Gauss hypergeometric function for $\boldsymbol{p}^2/m^2\rightarrow\infty$ (see Eq.~(\ref{Eq:Exp2F1})), we find
\begin{gather}
    \delta^{(1)} (\slashed{p}_{\perp},m=0) = 2^{2-d}  \pi^{d/2} a_s C_F      \frac{ \Gamma (1-d/2) 
    }{ \Gamma (d-1) } \;\left( \boldsymbol{p}^2 \right)^{d/2-1}   = - \frac{ g^2 C_F \Gamma (1 + \epsilon) }{(4 \pi)^{2-\epsilon}}         \frac{ \Gamma^2 \left( - \epsilon  \right)  
    }{  \Gamma (-2 \epsilon) } \;\left( \frac{\boldsymbol{p}^2}{\mu^2} \right)^{-\epsilon} = \delta^{(1)} (-p^2) \; .
\end{gather}
We next consider the second contribution in Eq.~(\ref{Eq:ReggeTrajMass}). Using
\begin{gather}
   K_{d/2-1} (m |\boldsymbol{z}_1|) = \left( \frac{m}{|\boldsymbol{z}_1|} \right)^{1-d/2}  \int \frac{d^d \boldsymbol{k}}{(2 \pi)^{d/2}} \frac{e^{i \boldsymbol{k} \cdot\boldsymbol{z}_1 }}{ (\boldsymbol{k}^2 + m^2) }  \; ,
\end{gather}
we obtain
\begin{align}
    I_2  & =  i m^{d/2} \int d^d \boldsymbol{z}_1 e^{-i \boldsymbol{p} \boldsymbol{z}_1} \frac{ \slashed{z}_{1\perp}}{ (\boldsymbol{z}_{1}^{2})^{3d/4 - 1/2} }     K_{d/2-1} (m |\boldsymbol{z}_{1}|) = m^{d/2} i \gamma^{\mu} 
    \int d^d \boldsymbol{z}_1 \frac{e^{-i \boldsymbol{p} \boldsymbol{z}_1} z_{1 \perp \mu}}{ (\boldsymbol{z}_{1}^{2})^{3d/4 - 1/2} }  K_{d/2-1} (m |\boldsymbol{z}_{1}|) 
    \nonumber \\ 
    & = 
    m i \gamma^{\mu}    \int \frac{d^d \boldsymbol{k}}{(2 \pi)^{d/2}} \frac{1}{ (\boldsymbol{k}^2 + m^2) }  \int d^d \boldsymbol{z}_1 \frac{e^{-i(\boldsymbol p-\boldsymbol k)\cdot\boldsymbol z_1}}{ (\boldsymbol{z}_{1}^{2})^{d/2} } z_{1 \perp \mu}  \; .
\end{align}
The remaining integral over $\boldsymbol{z}_1$ is readily evaluated as 
\begin{gather}
  I_{\mu} ( q_{\perp}) =  \int d^d \boldsymbol{z}_1 \frac{e^{-i \boldsymbol{q} \boldsymbol{z}_1}}{ (\boldsymbol{z}_{1}^{2})^{d/2} } i z_{1 \perp \mu} = \frac{ 2 \pi^{d/2} }{ \Gamma (d/2 ) } \frac{q_{ \perp \mu} }{ \boldsymbol{q}^{2}} \; .
\end{gather}
Substituting this result yields
\begin{align}
    I_2  & = \frac{ 2 \pi^{d/2} m }{ \Gamma (d/2 ) }
      \gamma^{\mu}    \int \frac{d^d \boldsymbol{k}}{(2 \pi)^{d/2}} \frac{ (p - k)_{\perp \mu} }{ (\boldsymbol{k}^2 + m^2) (\boldsymbol{k}-\boldsymbol{p})^2 }  = \frac{ 2^{1-d/2} m }{ \Gamma (d/2 ) }
      \slashed{p}_{\perp} \int_0^1 dx (1-x) \int d^d \boldsymbol{k} \frac{1  }{[ \boldsymbol{k}^2 + (1-x) (m^2 + x \boldsymbol{p}^2) ]^2} 
      \nonumber \\ 
      & =  \frac{ 2^{1-d/2} \pi^{d/2} }{ \Gamma (d/2 ) } \Gamma (2-d/2)
       m (m^2)^{d/2-2} \slashed{p}_{\perp} \int_0^1  dx \; x^{1-1} (1-x)^{d/2-1} \left(1 + x \frac{ \boldsymbol{p}^2 }{m^2} \right)^{d/2-2} 
       \nonumber \\ 
       & =   2^{1-d/2} \pi^{d/2} \frac{\Gamma (2-d/2)}{\Gamma (1+d/2)} 
        (m^2)^{d/2-1} \frac{\slashed{p}_{\perp}}{m} \; _2 F_1 \left(2-d/2, 1; 1+d/2; -\frac{ \boldsymbol{p}^2 }{m^2} \right) \; .
\end{align}
Combining the two contributions, we arrive at
\begin{align}
    \delta^{(1)} (\slashed{p}_{\perp},m)  & = a_s C_F \frac{2^{2-d} \pi^{d/2}}{ \displaystyle \Gamma \left( d/2 \right)} (m^2)^{d/2-1} \bigg \{       \frac{ \Gamma (1-d/2) 
    }{ \Gamma (d/2) } \; _2 F_1 \left( 1, 1-d/2; d/2 ; - \frac{\boldsymbol{p}^2}{m^2} \right) 
    \nonumber \\ 
    & + \,    \frac{\Gamma (2-d/2)}{\Gamma (1+d/2)} 
         \frac{\slashed{p}_{\perp}}{m} \; _2 F_1 \left(2-d/2, 1; 1+d/2; -\frac{ \boldsymbol{p}^2 }{m^2} \right) \bigg \} \; .
\end{align}
Expressing the result in terms of $\epsilon$ and extracting the overall prefactor gives
\begin{align}
    \delta^{(1)} (\slashed{p}_{\perp},m)  & = -\frac{g^2 \Gamma(1+\epsilon) }{ (4 \pi)^{2-\epsilon}}  2 C_F    \left(\frac{m^2}{\mu^2} \right)^{-\epsilon}  \bigg \{      - \frac{ \Gamma (\epsilon) 
    }{ \Gamma (1+\epsilon) } \; _2 F_1 \left( 1, \epsilon; 1-\epsilon ; - \frac{\boldsymbol{p}^2}{m^2} \right) \nonumber \\
    & - \,     \frac{\Gamma (1-\epsilon)}{\Gamma (2-\epsilon)} 
         \frac{\slashed{p}_{\perp}}{m} \; _2 F_1 \left(1 + \epsilon, 1; 2-\epsilon; -\frac{ \boldsymbol{p}^2 }{m^2} \right) \bigg \} \; .
\end{align}
To make contact with the representation of the massive Regge quark trajectory in (\ref{Eq:QuarkReggeTrajEpsilonExact}), it is convenient to add and subtract a contribution, so that the Regge trajectory can be rewritten as
\begin{align}
    \delta^{(1)} (\slashed{p}_{\perp},m) &  = -\frac{g^2 \Gamma(1+\epsilon) }{ (4 \pi)^{2-\epsilon}}  2 C_F    \left(\frac{m^2}{\mu^2} \right)^{-\epsilon}  \bigg \{      - \frac{ 1 
    }{ \epsilon } \; _2 F_1 \left( 1, \epsilon; 1-\epsilon ; - \frac{\boldsymbol{p}^2}{m^2} \right) - \frac{1}{ (1-\epsilon)} \; _2 F_1 \left(1 + \epsilon, 1; 2-\epsilon; -\frac{ \boldsymbol{p}^2 }{m^2} \right) \bigg \}\nonumber \\     
      & + \left(1- \frac{\slashed{p}_{\perp}}{m} \right) \frac{\Gamma (1-\epsilon)}{\Gamma (2-\epsilon)} \; _2 F_1 \left(1 + \epsilon, 1; 2-\epsilon; -\frac{ \boldsymbol{p}^2 }{m^2} \right) \bigg \} \; .
\end{align}
The combination of hypergeometric functions appearing in the first line can be related to that in Eq.~(\ref{Eq:QuarkReggeTrajEpsilonExact}) by means of Gauss' contiguous relations. One thus obtains
\begin{align}
    \delta^{(1)} (\slashed{p}_{\perp},m)  & = - \frac{g^2 \Gamma (1+\epsilon)}{(4 \pi)^{2-\epsilon}} 2 C_F  \left( \frac{m^2}{\mu^2} \right)^{-\epsilon} \left \{  \left( 1 + \frac{\boldsymbol{p}^2}{m^2} \right) \frac{\Gamma(-\epsilon)}{\Gamma(2-\epsilon)} \; _2 F_1 \left( 1+\epsilon, 2; 2-\epsilon; - \frac{\boldsymbol{p}^2}{m^2} \right) \right. 
    \nonumber \\ 
     & \hspace{4.9cm}\left. + \left( 1 - \frac{\slashed{p}_{\perp}}{m} \right) \frac{\Gamma(1-\epsilon)}{\Gamma(2-\epsilon)} \; _2 F_1 \left( 1+\epsilon, 1; 2-\epsilon; - \frac{\boldsymbol{p}^2}{m^2} \right) \right \} \; ,
\end{align}
which coincides exactly with the momentum-space result in Eq.~(\ref{Eq:QuarkReggeTrajEpsilonExact}), thereby completing the derivation of the massive quark Regge trajectory from the coordinate-space evolution kernel.

\section{Summary and Outlook}
\label{sec:Summary}

In this work, we have developed a Wilson-line formulation of quark
Reggeization within the shockwave formalism. Building on the correspondence
between Wilson-line evolution and gluon Reggeization established in
Ref.~\cite{Caron-Huot:2013fea}, we have shown that quark exchange can be
incorporated into the same operator framework by extending the shockwave
description beyond the strict eikonal approximation. The relevant
identity-changing operators contain an insertion of the target quark field
dressed by semi-infinite Wilson lines and naturally arise at next-to-eikonal
accuracy. Starting from the operator describing a photon-to-quark transition, we
derived its nonlinear one-loop rapidity evolution using the background-field
method. In the dilute regime, the evolution of the resulting RG-linearized quark operator is analyzed. At finite $N_c$, this operator does not
possess definite signature and consequently does not form an autonomous
Regge-pole sector within NLLA. In particular, its evolution contains a
Regge-pole-breaking contribution that couples the direct and $s \leftrightarrow u$-crossed
configurations through an additional Reggeized-gluon field. \\

Resolving the evolution into sectors of definite signature removes this
ambiguity. In the positive-signature sector, the Regge-pole-breaking
contributions cancel, and the resulting operator evolves autonomously with
the quark Regge trajectory. This identifies the
Reggeized-quark interpolating operator as
\begin{equation}
R_{Q,i}^{(+)}(\boldsymbol z)
=
\int dz^+\,
\left\{
\delta_{ij}
+
\frac{i f^{abc}t^a_{ij}}{2N_c}
\left[
\ln\!\left(
\mathcal U_A(\infty^+,z^+,\boldsymbol z)
\mathcal U_A^\dagger(z^+,-\infty^+,\boldsymbol z)
\right)
\right]^{bc}
\right\}
\Psi_j^{(-)}(z^+,\boldsymbol z)
+\mathcal O(g^2R^2).
\label{Eq:RQIoperatorSummary}
\end{equation}
The logarithm appearing in this expression isolates the Reggeized-gluon
content of the two complementary semi-infinite Wilson lines. The operator
therefore provides a Wilson-line realization of the Reggeized quark and extends the operator interpretation of parton Reggeization from gluon to quark. \\

The negative-signature sector has a qualitatively different structure. Its
leading component already contains an elementary quark accompanied by a
Reggeized gluon, and its rapidity evolution is coupled to the corresponding Reggeized-quark-gluon composite sector. This provides an operator-level realization
of the Regge-cut structure known from conventional analyses of
negative-signature fermion-exchange amplitudes. Importantly, rapidity
evolution preserves total signature: the nontrivial mixing occurs between
operators carrying the same overall signature, rather than between the
positive- and negative-signature sectors themselves. \\

The planar approximation provides a particularly simple limit of the construction. The nonplanar color structures responsible for the Regge-pole-breaking term are suppressed. In the large $N_c$, the non-linear evolution is a convolution of a dipole and the Born-level quark operator. In the dilute one-Reggeized-gluon sector, the latter reduces to a quark loop multiplying the Born-level quark operator, allowing the non-signaturized quark operator to evolve with the large-$N_c$ limit of the Regge trajectory within the NLLA. This gives a direct Wilson-line interpretation of the expected degeneracy between the two signature channels in the planar theory. \\

We have also illustrated the universality of the Reggeized-quark operator
by considering the more general gluon-to-quark transition. Unlike the
photon-initiated channel, this process admits several possible color
representations in the $t$-channel. Nevertheless, projecting onto
positive-signature exchange in the fundamental color channel yields
precisely the same interpolating operator as in the photon-initiated
process. The Reggeized quark is therefore not tied to a particular
nonlinear next-to-eikonal operator: it emerges as the universal dilute
degree of freedom governing positive-signature quark exchange. An interesting future direction is to explore the Regge structure of amplitudes involving $\bar{\mathbf{6}}$ and $\mathbf{15}$ color exchange within this framework. \\

The construction applies equally to massless and massive quarks. The
operator structure and the separation into signature sectors remain
unchanged, while the dependence on the quark mass is encoded in the
rapidity-evolution kernel. We derived the corresponding coordinate-space
kernel and showed that its Fourier transform reproduces the conventional
massive-quark Regge trajectory, including its nontrivial spinor structure
and its exact dependence on the dimensional-regularization parameter. The
massless result is recovered smoothly from the resulting expression. \\

The structure of the evolution also provides useful information about the
extension beyond LLA. In the positive-signature sector, signature-conservation forbids
mixing with a composite Reggeized-quark-Reggeized-gluon operator of
total negative signature. This is the fermionic analogue of the absence of
mixing between one- and two-Reggeized-gluon states in the negative-signature
gluon sector. The first allowed composite contribution instead involves a
Reggeized gluon multiplying a negative-signature quark operator. Its field
content and coupling suppression imply that the corresponding
off-diagonal transition starts parametrically later than the diagonal evolution. The explicit calculation identifies one representative operator belonging
to this positive-signature composite sector. Other operators with the same
quantum numbers may occur at the same order, but their multiplicity does
not modify the coupling power counting. Assuming the expected Hermiticity
of the rapidity Hamiltonian in an appropriately normalized Reggeon basis,
the reverse transition has the same parametric suppression. This strongly
suggests that multi-Reggeon mixing cannot modify the Reggeized-quark
eigenvalue through NLLA.  \\

\begin{figure}[t]
    \centering
    \includegraphics[width=0.25\linewidth]{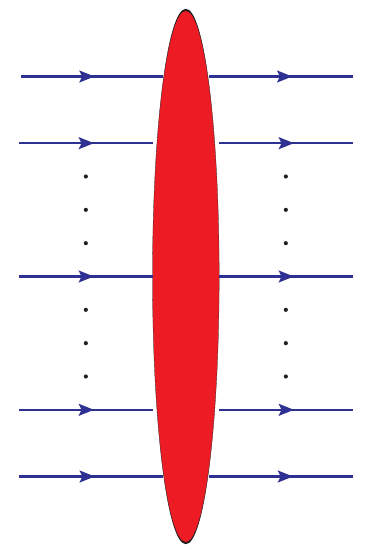}
    \hspace{3cm}
    \includegraphics[width=0.25\linewidth]{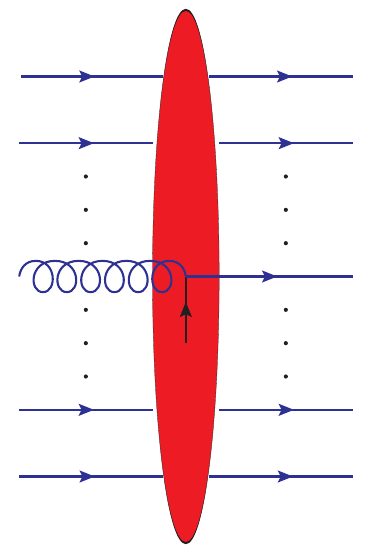}
    \\
    (a) \hspace{7cm} (b)
    \caption{Schematic comparison between the eikonal and next-to-eikonal
    evolution hierarchies. (a) A projectile composed of an arbitrary number
    of Wilson lines in the fundamental representation, whose evolution is
    described by the Balitsky--JIMWLK hierarchy. (b) A projectile in which
    one of its constituents changes its partonic identity through a
    next-to-eikonal interaction with the target quark background field,
    converting an adjoint Wilson line into a fundamental one.}
    \label{figNeikBalitHiera}
\end{figure}

The most immediate continuation of this work is the completion of the
Wilson-line proof of quark Reggeization at NLLA. This requires determining
the two-loop diagonal evolution of the Reggeized-quark operator and
completing the analysis of its mixing with the allowed positive-signature
composite sector. In particular, the Fourier transform of the diagonal
kernel must reproduce the two-loop quark Regge trajectory extracted from
the scattering-amplitude calculation of Ref.~\cite{Bogdan:2002sr}, while
the power counting of the off-diagonal transitions must confirm that
multi-Reggeon mixing does not modify the Reggeized-quark eigenvalue at this
accuracy. Completing these two steps would have a significance that goes beyond an
independent reproduction of the known two-loop trajectory. The existing
two-loop calculation provides fixed-order evidence for quark Reggeization
at NLLA. By contrast, establishing that the positive-signature
Reggeized-quark operator remains an eigenstate of the rapidity Hamiltonian,
with its eigenvalue given by the two-loop trajectory, would yield an
operator-level proof of quark Reggeization to all orders in the NLLA
resummation. A closely related direction is the extension of the massive-quark construction to NLLA, including the determination of the massive-quark Regge trajectory at two loops, which remains unknown. Its derivation will require computing the relevant two-loop rapidity evolution within the shockwave formalism, following the methods developed in Refs.~\cite{Balitsky:2013fea,Kovner:2013ona,Grabovsky:2013mba}; see also Ref.~\cite{Brunello:2025rhh} for recent progress at three loops. \\

Looking further ahead, the identity-changing operators studied here suggest
an extension of the Balitsky--JIMWLK hierarchy beyond the eikonal
approximation. Whereas the ordinary hierarchy describes the nonlinear
evolution of products of Wilson lines whose partonic identity is preserved,
a genuinely next-to-eikonal hierarchy would also contain quark-field
insertions and would allow transitions between different partonic degrees
of freedom. It would consequently couple operators carrying different
color and spin structures. This distinction is illustrated
schematically in Fig.~\ref{figNeikBalitHiera}. The linearization of such a hierarchy would provide systematic access to
multi-Reggeon states carrying fermion quantum numbers. It could therefore
clarify how Regge-pole-breaking contributions and Regge cuts arise from
nonlinear Wilson-line evolution in the quark sector, in close analogy with
the role played by the Balitsky--JIMWLK framework in gluon exchange~\cite{Caron-Huot:2013fea}. More generally, the present construction shows that the shockwave formalism,
once extended beyond the eikonal approximation, provides a unified operator
language for both gluon and quark Reggeization. It thus emerges as a systematic framework for uncovering the structure of scattering amplitudes in the Regge limit.

\section{Acknowledgments}

We are grateful to R. Boussarie, F. Cougoulic, V. S. Fadin, G. Falcioni, M. A. Nefedov, A. Papa, L. Szymanowski, R. Venugopalan, and S. Wallon for insightful discussions. TA and JF are supported in part by the National Science Centre (Poland) under the research Grant No. 2023/50/E/ST2/00133 (SONATA BIS 13). GB is supported in part by the National Science Centre (Poland) under the research Grant No. 2020/38/E/ST2/00122 (SONATA BIS 10). The work of MF is supported by the ULAM fellowship program of NAWA No. BNI/ULM/2024/1/00065 “Color glass condensate effective theory beyond the eikonal approximation”.

\bibliographystyle{apsrev}
\bibliography{mybib_New}

@article{Chachamis:2012gh,
    author = "Chachamis, G. and Hentschinski, M. and Madrigal Martinez, J. D. and Sabio Vera, A.",
    title = "{Quark contribution to the gluon Regge trajectory at NLO from the high energy effective action}",
    eprint = "1202.0649",
    archivePrefix = "arXiv",
    primaryClass = "hep-ph",
    doi = "10.1016/j.nuclphysb.2012.03.015",
    journal = "Nucl. Phys. B",
    volume = "861",
    pages = "133--144",
    year = "2012"
}

@article{Hentschinski:2011xg,
    author = "Hentschinski, Martin",
    title = "{Pole prescription of higher order induced vertices in Lipatov's QCD effective action}",
    eprint = "1112.4509",
    archivePrefix = "arXiv",
    primaryClass = "hep-ph",
    reportNumber = "IFT-UAM-CSIC-11-105, LPN11-95",
    doi = "10.1016/j.nuclphysb.2012.02.001",
    journal = "Nucl. Phys. B",
    volume = "859",
    pages = "129--142",
    year = "2012"
}

@article{Nefedov:2019mrg,
    author = "Nefedov, Maxim A.",
    title = "{Computing one-loop corrections to effective vertices with two scales in the EFT for Multi-Regge processes in QCD}",
    eprint = "1902.11030",
    archivePrefix = "arXiv",
    primaryClass = "hep-ph",
    doi = "10.1016/j.nuclphysb.2019.114715",
    journal = "Nucl. Phys. B",
    volume = "946",
    pages = "114715",
    year = "2019"
}

@article{Kotsky:2002aq,
    author = "Kotsky, M. I. and Lipatov, L. N. and Principe, A. and Vyazovsky, M. I.",
    title = "{Radiative corrections to the quark gluon Reggeized quark vertex in QCD}",
    eprint = "hep-ph/0207169",
    archivePrefix = "arXiv",
    reportNumber = "BUDKER-INP-2002-32",
    doi = "10.1016/S0550-3213(02)00967-7",
    journal = "Nucl. Phys. B",
    volume = "648",
    pages = "277--292",
    year = "2003"
}

@article{Bogdan:2004cg,
    author = "Bogdan, A. V. and Fadin, V. S.",
    title = "{Quark Regge trajectory in two loops from unitarity relations}",
    eprint = "hep-ph/0408127",
    archivePrefix = "arXiv",
    reportNumber = "BUDKER-INP-2004-1",
    doi = "10.1134/1.2053342",
    journal = "Phys. Atom. Nucl.",
    volume = "68",
    pages = "1599--1615",
    year = "2005"
}

@article{Bogdan:2007qj,
    author = "Bogdan, A. V. and Grabovsky, A. V.",
    title = "{Radiative corrections to the Reggeized quark - Reggeized quark - gluon effective vertex}",
    eprint = "hep-ph/0701144",
    archivePrefix = "arXiv",
    doi = "10.1016/j.nuclphysb.2007.03.017",
    journal = "Nucl. Phys. B",
    volume = "773",
    pages = "65--83",
    year = "2007"
}

@article{Bogdan:2006wq,
    author = "Bogdan, A. V. and Grabovsky, A. V.",
    title = "{Verification of bootstrap conditions for amplitudes with quark exchanges in QMRK}",
    eprint = "hep-ph/0606132",
    archivePrefix = "arXiv",
    doi = "10.1016/j.nuclphysb.2006.09.004",
    journal = "Nucl. Phys. B",
    volume = "757",
    pages = "211--232",
    year = "2006"
}

@article{Jalilian-Marian:2000pwi,
    author = "Jalilian-Marian, Jamal and Jeon, Sangyong and Venugopalan, Raju",
    title = "{Wong's equations and the small x effective action in QCD}",
    eprint = "hep-ph/0003070",
    archivePrefix = "arXiv",
    reportNumber = "BNL-NT-00-5, LBNL-45248, LBL-45248",
    doi = "10.1103/PhysRevD.63.036004",
    journal = "Phys. Rev. D",
    volume = "63",
    pages = "036004",
    year = "2001"
}

@phdthesis{Li:2023ihv,
    author = "Li, Emilie",
    title = "{Probing gluon saturation in semi-hard {\ensuremath{\gamma}}(*)+p/A processes}",
    reportNumber = "tel-04310201, 2023UPASP131",
    school = "Laboratoire de Physique des 2 Infinis Ir{\`e}ne Joliot-Curie, France, U. Paris-Saclay",
    year = "2023"
}

@article{Boussarie:2024pax,
    author = "Boussarie, Renaud and Fucilla, Michael and Szymanowski, Lech and Wallon, Samuel",
    title = "{Probing Gluonic Saturation in Deeply Virtual Meson Production beyond Leading Power}",
    eprint = "2407.18203",
    archivePrefix = "arXiv",
    primaryClass = "hep-ph",
    doi = "10.1103/PhysRevLett.134.041901",
    journal = "Phys. Rev. Lett.",
    volume = "134",
    number = "4",
    pages = "041901",
    year = "2025"
}

@article{Lipatov:2000se,
    author = "Lipatov, L. N. and Vyazovsky, M. I.",
    title = "{QuasimultiRegge processes with a quark exchange in the t channel}",
    eprint = "hep-ph/0009340",
    archivePrefix = "arXiv",
    reportNumber = "SPBU-IP-00-14",
    doi = "10.1016/S0550-3213(00)00709-4",
    journal = "Nucl. Phys. B",
    volume = "597",
    pages = "399--409",
    year = "2001"
}

@article{Brunello:2025rhh,
    author = "Brunello, Giacomo and Caron-Huot, Simon and Crisanti, Giulio and Giroux, Mathieu and Smith, Sid",
    title = "{High-energy evolution in planar QCD to three loops: the non-conformal contribution}",
    eprint = "2508.03794",
    archivePrefix = "arXiv",
    primaryClass = "hep-ph",
    doi = "10.1007/JHEP11(2025)055",
    journal = "JHEP",
    volume = "11",
    pages = "055",
    year = "2025"
}

@article{Grabovsky:2013mba,
    author = "Grabovsky, A. V.",
    title = "{Connected contribution to the kernel of the evolution equation for 3-quark Wilson loop operator}",
    eprint = "1307.5414",
    archivePrefix = "arXiv",
    primaryClass = "hep-ph",
    doi = "10.1007/JHEP09(2013)141",
    journal = "JHEP",
    volume = "09",
    pages = "141",
    year = "2013"
}

@article{Kovner:2013ona,
    author = "Kovner, Alex and Lublinsky, Michael and Mulian, Yair",
    title = "{Jalilian-Marian, Iancu, McLerran, Weigert, Leonidov, Kovner evolution at next to leading order}",
    eprint = "1310.0378",
    archivePrefix = "arXiv",
    primaryClass = "hep-ph",
    doi = "10.1103/PhysRevD.89.061704",
    journal = "Phys. Rev. D",
    volume = "89",
    number = "6",
    pages = "061704",
    year = "2014"
}

@article{Balitsky:2013fea,
    author = "Balitsky, Ian and Chirilli, Giovanni A.",
    title = "{Rapidity evolution of Wilson lines at the next-to-leading order}",
    eprint = "1309.7644",
    archivePrefix = "arXiv",
    primaryClass = "hep-ph",
    reportNumber = "JLAB-THY-13-1806",
    doi = "10.1103/PhysRevD.88.111501",
    journal = "Phys. Rev. D",
    volume = "88",
    pages = "111501",
    year = "2013"
}

@article{Falcioni:2021buo,
    author = "Falcioni, Giulio and Gardi, Einan and Maher, Niamh and Milloy, Calum and Vernazza, Leonardo",
    title = "{Scattering amplitudes in the Regge limit and the soft anomalous dimension through four loops}",
    eprint = "2111.10664",
    archivePrefix = "arXiv",
    primaryClass = "hep-ph",
    reportNumber = "CERN-TH-2021-200",
    doi = "10.1007/JHEP03(2022)053",
    journal = "JHEP",
    volume = "03",
    pages = "053",
    year = "2022"
}

@article{Falcioni:2020lvv,
    author = "Falcioni, Giulio and Gardi, Einan and Milloy, Calum and Vernazza, Leonardo",
    title = "{Climbing three-Reggeon ladders: four-loop amplitudes in the high-energy limit in full colour}",
    eprint = "2012.00613",
    archivePrefix = "arXiv",
    primaryClass = "hep-ph",
    doi = "10.1103/PhysRevD.103.L111501",
    journal = "Phys. Rev. D",
    volume = "103",
    pages = "L111501",
    year = "2021"
}

@article{Grisaru:1973ku,
    author = "Grisaru, Marcus T. and Schnitzer, Howard J. and Tsao, Hung-Sheng",
    title = "{THE REGGEIZATION OF ELEMENTARY PARTICLES IN RENORMALIZABLE GAUGE THEORIES: SCALARS}",
    reportNumber = "Print-74-0203 (BRANDEIS)",
    doi = "10.1103/PhysRevD.9.2864",
    journal = "Phys. Rev. D",
    volume = "9",
    pages = "2864",
    year = "1974"
}

@article{Grisaru:1973wbb,
    author = "Grisaru, Marcus T. and Schnitzer, H. J. and Tsao, Hung-Sheng",
    title = "{Reggeization of elementary particles in renormalizable gauge theories - vectors and spinors}",
    doi = "10.1103/PhysRevD.8.4498",
    journal = "Phys. Rev. D",
    volume = "8",
    pages = "4498--4509",
    year = "1973"
}

@article{Grisaru:1973vw,
    author = "Grisaru, Marcus T. and Schnitzer, H. J. and Tsao, Hung-Sheng",
    title = "{Reggeization of yang-mills gauge mesons in theories with a spontaneously broken symmetry}",
    doi = "10.1103/PhysRevLett.30.811",
    journal = "Phys. Rev. Lett.",
    volume = "30",
    pages = "811--814",
    year = "1973"
}

@article{Gribov:1983ivg,
    author = "Gribov, L. V. and Levin, E. M. and Ryskin, M. G.",
    title = "{Semihard Processes in QCD}",
    doi = "10.1016/0370-1573(83)90022-4",
    journal = "Phys. Rept.",
    volume = "100",
    pages = "1--150",
    year = "1983"
}

@article{Caron-Huot:2017fxr,
    author = "Caron-Huot, Simon and Gardi, Einan and Vernazza, Leonardo",
    title = "{Two-parton scattering in the high-energy limit}",
    eprint = "1701.05241",
    archivePrefix = "arXiv",
    primaryClass = "hep-ph",
    reportNumber = "EDINBURGH-2017-01",
    doi = "10.1007/JHEP06(2017)016",
    journal = "JHEP",
    volume = "06",
    pages = "016",
    year = "2017"
}

@article{Caron-Huot:2017zfo,
    author = "Caron-Huot, Simon and Gardi, Einan and Reichel, Joscha and Vernazza, Leonardo",
    title = "{Infrared singularities of QCD scattering amplitudes in the Regge limit to all orders}",
    eprint = "1711.04850",
    archivePrefix = "arXiv",
    primaryClass = "hep-ph",
    reportNumber = "EDINBURGH-2017-24, NIKHEF-2017-060, Edinburgh 2017/24, NIKHEF/2017-060",
    doi = "10.1007/JHEP03(2018)098",
    journal = "JHEP",
    volume = "03",
    pages = "098",
    year = "2018"
}

@article{Caron-Huot:2020grv,
    author = "Caron-Huot, Simon and Gardi, Einan and Reichel, Joscha and Vernazza, Leonardo",
    title = "{Two-parton scattering amplitudes in the Regge limit to high loop orders}",
    eprint = "2006.01267",
    archivePrefix = "arXiv",
    primaryClass = "hep-ph",
    doi = "10.1007/JHEP08(2020)116",
    journal = "JHEP",
    volume = "08",
    pages = "116",
    year = "2020"
}

@article{DelDuca:2001gu,
    author = "Del Duca, Vittorio and Glover, E. W. Nigel",
    title = "{The High-energy limit of QCD at two loops}",
    eprint = "hep-ph/0109028",
    archivePrefix = "arXiv",
    reportNumber = "DCPT-01-66, IPPP-01-33, DFTT-25-2001",
    doi = "10.1088/1126-6708/2001/10/035",
    journal = "JHEP",
    volume = "10",
    pages = "035",
    year = "2001"
}

@article{DelDuca:2014cya,
    author = "Del Duca, Vittorio and Falcioni, Giulio and Magnea, Lorenzo and Vernazza, Leonardo",
    title = "{Analyzing high-energy factorization beyond next-to-leading logarithmic accuracy}",
    eprint = "1409.8330",
    archivePrefix = "arXiv",
    primaryClass = "hep-ph",
    doi = "10.1007/JHEP02(2015)029",
    journal = "JHEP",
    volume = "02",
    pages = "029",
    year = "2015"
}

@article{DelDuca:2017twk,
    author = "Del Duca, V. and Laenen, E. and Magnea, L. and Vernazza, L. and White, C. D.",
    title = "{Universality of next-to-leading power threshold effects for colourless final states in hadronic collisions}",
    eprint = "1706.04018",
    archivePrefix = "arXiv",
    primaryClass = "hep-ph",
    reportNumber = "NIKHEF-2017-25, EDINBURGH-2017-10, QMUL-PH-17-07, ARC-17-03",
    doi = "10.1007/JHEP11(2017)057",
    journal = "JHEP",
    volume = "11",
    pages = "057",
    year = "2017"
}

@article{DelDuca:2013ara,
    author = "Del Duca, Vittorio and Falcioni, Giulio and Magnea, Lorenzo and Vernazza, Leonardo",
    title = "{High-energy QCD amplitudes at two loops and beyond}",
    eprint = "1311.0304",
    archivePrefix = "arXiv",
    primaryClass = "hep-ph",
    reportNumber = "LPN13-086",
    doi = "10.1016/j.physletb.2014.03.033",
    journal = "Phys. Lett. B",
    volume = "732",
    pages = "233--240",
    year = "2014"
}

@article{Chachamis:2013hma,
    author = "Chachamis, G. and Hentschinski, M. and Madrigal Martinez, J. D. and Sabio Vera, A.",
    title = "{Gluon Regge trajectory at two loops from Lipatov's high energy effective action}",
    eprint = "1307.2591",
    archivePrefix = "arXiv",
    primaryClass = "hep-ph",
    reportNumber = "LPN13-045",
    doi = "10.1016/j.nuclphysb.2013.08.013",
    journal = "Nucl. Phys. B",
    volume = "876",
    pages = "453--472",
    year = "2013"
}

@article{Braun:1999uz,
    author = "Braun, Mikhail and Vacca, Gian Paolo",
    title = "{The Bootstrap for impact factors and the gluon wave function}",
    eprint = "hep-ph/9910432",
    archivePrefix = "arXiv",
    doi = "10.1016/S0370-2693(00)00196-9",
    journal = "Phys. Lett. B",
    volume = "477",
    pages = "156--162",
    year = "2000"
}

@article{Fadin:2002hz,
    author = "Fadin, V. S. and Papa, A.",
    title = "{A Proof of fulfillment of the strong bootstrap condition}",
    eprint = "hep-ph/0206079",
    archivePrefix = "arXiv",
    reportNumber = "BUDKER-INP-2002-38, DESY-02-074, DFCAL-TH-02-2",
    doi = "10.1016/S0550-3213(02)00579-5",
    journal = "Nucl. Phys. B",
    volume = "640",
    pages = "309--330",
    year = "2002"
}

@article{Fadin:2000ww,
    author = "Fadin, Victor S. and Fiore, R. and Kotsky, M. I. and Papa, A.",
    title = "{Strong bootstrap conditions}",
    eprint = "hep-ph/0008057",
    archivePrefix = "arXiv",
    reportNumber = "BUDKER-INP-2000-64, UNICAL-TH-00-6",
    doi = "10.1016/S0370-2693(00)01260-0",
    journal = "Phys. Lett. B",
    volume = "495",
    pages = "329--337",
    year = "2000"
}

@article{Balitsky:1978ic,
    author = "Balitsky, I. I. and Lipatov, L. N.",
    title = "{The Pomeranchuk Singularity in Quantum Chromodynamics}",
    journal = "Sov. J. Nucl. Phys.",
    volume = "28",
    pages = "822--829",
    year = "1978"
}

@article{Kuraev:1977fs,
    author = "Kuraev, E. A. and Lipatov, L. N. and Fadin, Victor S.",
    title = "{The Pomeranchuk singularity in nonabelian gauge theories}",
    journal = "Sov. Phys. JETP",
    volume = "45",
    pages = "199--204",
    year = "1977"
}

@article{Kuraev:1976ge,
    author = "Kuraev, E. A. and Lipatov, L. N. and Fadin, Victor S.",
    title = "{Multiregge processes in the Yang-Mills theory}",
    journal = "Sov. Phys. JETP",
    volume = "44",
    number = "3",
    pages = "443--451",
    year = "1976"
}

@article{Fadin:1975cb,
    author = "Fadin, Victor S. and Kuraev, E. A. and Lipatov, L. N.",
    title = "{On the Pomeranchuk Singularity in Asymptotically Free Theories}",
    doi = "10.1016/0370-2693(75)90524-9",
    journal = "Phys. Lett. B",
    volume = "60",
    pages = "50--52",
    year = "1975"
}

@article{Abreu:2024xoh,
    author = "Abreu, Samuel and De Laurentis, Giuseppe and Falcioni, Giulio and Gardi, Einan and Milloy, Calum and Vernazza, Leonardo",
    title = "{The two-loop Lipatov vertex in QCD}",
    eprint = "2412.20578",
    archivePrefix = "arXiv",
    primaryClass = "hep-ph",
    reportNumber = "CERN-TH-2024-226, ZU-TH 68/24",
    doi = "10.1007/JHEP04(2025)161",
    journal = "JHEP",
    volume = "04",
    pages = "161",
    year = "2025"
}

@article{Buccioni:2024gzo,
    author = "Buccioni, Federico and Caola, Fabrizio and Devoto, Federica and Gambuti, Giulio",
    title = "{Investigating the universality of five-point QCD scattering amplitudes at high energy}",
    eprint = "2411.14050",
    archivePrefix = "arXiv",
    primaryClass = "hep-ph",
    reportNumber = "OUTP-24-06P, SLAC-PUB-241120, TUM-HEP-1537/24",
    doi = "10.1007/JHEP03(2025)129",
    journal = "JHEP",
    volume = "03",
    pages = "129",
    year = "2025"
}

@article{Fadin:2023roz,
    author = "Fadin, Victor S. and Fucilla, Michael and Papa, Alessandro",
    title = "{One-loop Lipatov vertex in QCD with higher {\ensuremath{\epsilon}}-accuracy}",
    eprint = "2302.09868",
    archivePrefix = "arXiv",
    primaryClass = "hep-ph",
    doi = "10.1007/JHEP04(2023)137",
    journal = "JHEP",
    volume = "04",
    pages = "137",
    year = "2023"
}

@article{Byrne:2022wzk,
    author = "Byrne, Emmet P. and Del Duca, Vittorio and Dixon, Lance J. and Gardi, Einan and Smillie, Jennifer M.",
    title = "{One-loop central-emission vertex for two gluons in $ \mathcal{N} $ = 4 super Yang-Mills theory}",
    eprint = "2204.12459",
    archivePrefix = "arXiv",
    primaryClass = "hep-ph",
    reportNumber = "SLAC-PUB-17654",
    doi = "10.1007/JHEP08(2022)271",
    journal = "JHEP",
    volume = "08",
    pages = "271",
    year = "2022"
}

@article{Falcioni:2021dgr,
    author = "Falcioni, Giulio and Gardi, Einan and Maher, Niamh and Milloy, Calum and Vernazza, Leonardo",
    title = "{Disentangling the Regge Cut and Regge Pole in Perturbative QCD}",
    eprint = "2112.11098",
    archivePrefix = "arXiv",
    primaryClass = "hep-ph",
    reportNumber = "CERN-TH-2021-225",
    doi = "10.1103/PhysRevLett.128.132001",
    journal = "Phys. Rev. Lett.",
    volume = "128",
    number = "13",
    pages = "132001",
    year = "2022"
}

@article{Caola:2021izf,
    author = "Caola, Fabrizio and Chakraborty, Amlan and Gambuti, Giulio and von Manteuffel, Andreas and Tancredi, Lorenzo",
    title = "{Three-Loop Gluon Scattering in QCD and the Gluon Regge Trajectory}",
    eprint = "2112.11097",
    archivePrefix = "arXiv",
    primaryClass = "hep-ph",
    reportNumber = "OUTP-21-28P, MSUHEP-21-035, TUM-HEP-1382/21",
    doi = "10.1103/PhysRevLett.128.212001",
    journal = "Phys. Rev. Lett.",
    volume = "128",
    number = "21",
    pages = "212001",
    year = "2022"
}

@article{DelDuca:2021vjq,
    author = "Del Duca, Vittorio and Marzucca, Robin and Verbeek, Bram",
    title = "{The gluon Regge trajectory at three loops from planar Yang-Mills theory}",
    eprint = "2111.14265",
    archivePrefix = "arXiv",
    primaryClass = "hep-ph",
    reportNumber = "UUIPT-59/21",
    doi = "10.1007/JHEP01(2022)149",
    journal = "JHEP",
    volume = "01",
    pages = "149",
    year = "2022"
}

@article{Fadin:2024eyf,
    author = "Fadin, V. S.",
    title = "{Peculiarities of Regge Cuts in QCD}",
    eprint = "2409.01698",
    archivePrefix = "arXiv",
    primaryClass = "hep-ph",
    doi = "10.1134/S1547477124701942",
    journal = "Phys. Part. Nucl. Lett.",
    volume = "22",
    number = "1",
    pages = "117--125",
    year = "2025"
}

@article{Fadin:2017nka,
    author = "Fadin, V. S. and Lipatov, L. N.",
    title = "{Reggeon cuts in QCD amplitudes with negative signature}",
    eprint = "1712.09805",
    archivePrefix = "arXiv",
    primaryClass = "hep-ph",
    reportNumber = "BUDKER-INP-2017-15",
    doi = "10.1140/epjc/s10052-018-5910-1",
    journal = "Eur. Phys. J. C",
    volume = "78",
    number = "6",
    pages = "439",
    year = "2018"
}

@article{Fadin:2023aen,
    author = "Fadin, V. S.",
    title = "{Regge Cuts in QCD}",
    doi = "10.1134/S1547477123030275",
    journal = "Phys. Part. Nucl. Lett.",
    volume = "20",
    number = "3",
    pages = "341--346",
    year = "2023"
}

@article{Boussarie:2024bdo,
    author = "Boussarie, Renaud and Fucilla, Michael and Szymanowski, Lech and Wallon, Samuel",
    title = "{Twist corrections to exclusive vector meson production in a saturation framework}",
    eprint = "2407.18115",
    archivePrefix = "arXiv",
    primaryClass = "hep-ph",
    doi = "10.1103/PhysRevD.111.014032",
    journal = "Phys. Rev. D",
    volume = "111",
    number = "1",
    pages = "014032",
    year = "2025"
}

@article{Kirschner:1994gd,
    author = "Kirschner, R. and Lipatov, L. N. and Szymanowski, L.",
    title = "{Effective action for multi - Regge processes in QCD}",
    eprint = "hep-th/9402010",
    archivePrefix = "arXiv",
    reportNumber = "SI-94-1",
    doi = "10.1016/0550-3213(94)90288-7",
    journal = "Nucl. Phys. B",
    volume = "425",
    pages = "579--594",
    year = "1994"
}

@article{Kirschner:1994xi,
    author = "Kirschner, R. and Lipatov, L. N. and Szymanowski, L.",
    title = "{Symmetry properties of the effective action for high-energy scattering in QCD}",
    eprint = "hep-th/9403082",
    archivePrefix = "arXiv",
    reportNumber = "DESY-94-064",
    doi = "10.1103/PhysRevD.51.838",
    journal = "Phys. Rev. D",
    volume = "51",
    pages = "838--855",
    year = "1995"
}

@article{Lipatov:1991nf,
    author = "Lipatov, L. N.",
    title = "{High-energy scattering in QCD and in quantum gravity and two-dimensional field theories}",
    reportNumber = "IPNO-TH-91-21",
    doi = "10.1016/0550-3213(91)90512-V",
    journal = "Nucl. Phys. B",
    volume = "365",
    pages = "614--632",
    year = "1991"
}

@article{Bogdan:2002sr,
    author = "Bogdan, A. V. and Del Duca, V. and Fadin, Victor S. and Glover, E. W. Nigel",
    title = "{The Quark Regge trajectory at two loops}",
    eprint = "hep-ph/0201240",
    archivePrefix = "arXiv",
    reportNumber = "DCPT-02-08, IPPP-02-04, DFTT-03-2002, BUDKER-INP-2002-4",
    doi = "10.1088/1126-6708/2002/03/032",
    journal = "JHEP",
    volume = "03",
    pages = "032",
    year = "2002"
}

@article{Bogdan:2006af,
    author = "Bogdan, A. V. and Fadin, V. S.",
    title = "{A Proof of the reggeized form of amplitudes with quark exchanges}",
    eprint = "hep-ph/0601117",
    archivePrefix = "arXiv",
    doi = "10.1016/j.nuclphysb.2006.01.033",
    journal = "Nucl. Phys. B",
    volume = "740",
    pages = "36--57",
    year = "2006"
}

@article{Fadin:1976nw,
    author = "Fadin, Victor S. and Sherman, V. E.",
    title = "{Fermion Reggeization in Nonabelian Calibration Theories}",
    journal = "Pisma Zh. Eksp. Teor. Fiz.",
    volume = "23",
    pages = "599--602",
    year = "1976"
}

@article{Fadin:1977jr,
    author = "Fadin, Victor S. and Sherman, V. E.",
    title = "{Processes with fermion exchange in nonabelian gauge theories}",
    journal = "Sov. Phys. JETP",
    volume = "45",
    pages = "861--870",
    year = "1977"
}

@article{Nefedov:2017qzc,
    author = "Nefedov, Maxim and Saleev, Vladimir",
    title = "{On the one-loop calculations with Reggeized quarks}",
    eprint = "1709.06246",
    archivePrefix = "arXiv",
    primaryClass = "hep-th",
    doi = "10.1142/S0217732317502078",
    journal = "Mod. Phys. Lett. A",
    volume = "32",
    number = "40",
    pages = "1750207",
    year = "2017"
}

@article{Gao:2024fyz,
    author = "Gao, Anjie and Moult, Ian and Raman, Sanjay and Ridgway, Gregory and Stewart, Iain W.",
    title = "{Reggeization in Color}",
    eprint = "2411.09692",
    archivePrefix = "arXiv",
    primaryClass = "hep-ph",
    reportNumber = "MIT-CTP 5808, UWThPh 2024-22",
    month = "11",
    year = "2024"
}

@article{Moult:2022lfy,
    author = "Moult, Ian and Raman, Sanjay and Ridgway, Gregory and Stewart, Iain W.",
    title = "{Anomalous dimensions from soft Regge constants}",
    eprint = "2207.02859",
    archivePrefix = "arXiv",
    primaryClass = "hep-ph",
    reportNumber = "MIT-CTP 5448",
    doi = "10.1007/JHEP05(2023)025",
    journal = "JHEP",
    volume = "05",
    pages = "025",
    year = "2023"
}

@article{Gao:2024qsg,
    author = "Gao, Anjie and Moult, Ian and Raman, Sanjay and Ridgway, Gregory and Stewart, Iain W.",
    title = "{A collinear perspective on the Regge limit}",
    eprint = "2401.00931",
    archivePrefix = "arXiv",
    primaryClass = "hep-ph",
    reportNumber = "MIT-CTP 5628",
    doi = "10.1007/JHEP05(2024)328",
    journal = "JHEP",
    volume = "05",
    pages = "328",
    year = "2024"
}

@article{Rothstein:2016bsq,
    author = "Rothstein, Ira Z. and Stewart, Iain W.",
    title = "{An Effective Field Theory for Forward Scattering and Factorization Violation}",
    eprint = "1601.04695",
    archivePrefix = "arXiv",
    primaryClass = "hep-ph",
    reportNumber = "MIT-CTP-4655, MIT-CTP 4655",
    doi = "10.1007/JHEP08(2016)025",
    journal = "JHEP",
    volume = "08",
    pages = "025",
    year = "2016"
}

@article{Moult:2019vou,
    author = "Moult, Ian and Vita, Gherardo and Yan, Kai",
    title = "{Subleading power resummation of rapidity logarithms: the energy-energy correlator in $ \mathcal{N} $ = 4 SYM}",
    eprint = "1912.02188",
    archivePrefix = "arXiv",
    primaryClass = "hep-ph",
    reportNumber = "MIT-CTP 5162, MPP-2019-243",
    doi = "10.1007/JHEP07(2020)005",
    journal = "JHEP",
    volume = "07",
    pages = "005",
    year = "2020"
}

@article{Moult:2017xpp,
    author = "Moult, Ian and Solon, Mikhail P. and Stewart, Iain W. and Vita, Gherardo",
    title = "{Fermionic Glauber Operators and Quark Reggeization}",
    eprint = "1709.09174",
    archivePrefix = "arXiv",
    primaryClass = "hep-ph",
    reportNumber = "MIT-CTP-4933, CALT-TH-2017-055",
    doi = "10.1007/JHEP02(2018)134",
    journal = "JHEP",
    volume = "02",
    pages = "134",
    year = "2018"
}

@article{Caron-Huot:2013fea,
    author = "Caron-Huot, Simon",
    title = "{When does the gluon reggeize?}",
    eprint = "1309.6521",
    archivePrefix = "arXiv",
    primaryClass = "hep-th",
    doi = "10.1007/JHEP05(2015)093",
    journal = "JHEP",
    volume = "05",
    pages = "093",
    year = "2015"
}

@article{Fadin:2001dc,
    author = "Fadin, Victor S. and Fiore, R.",
    title = "{Calculation of Reggeon vertices in QCD}",
    eprint = "hep-ph/0107010",
    archivePrefix = "arXiv",
    reportNumber = "BUDKER-INP-2001-32, DFCAL-TH-01-3",
    doi = "10.1103/PhysRevD.64.114012",
    journal = "Phys. Rev. D",
    volume = "64",
    pages = "114012",
    year = "2001"
}

@article{Lipatov:1976zz,
    author = "Lipatov, L. N.",
    title = "{Reggeization of the Vector Meson and the Vacuum Singularity in Nonabelian Gauge Theories}",
    journal = "Sov. J. Nucl. Phys.",
    volume = "23",
    pages = "338--345",
    year = "1976"
}

@article{Altinoluk:2014oxa,
    author = "Altinoluk, Tolga and Armesto, N\'estor and Beuf, Guillaume and Mart\'\i{}nez, Mauricio and Salgado, Carlos A.",
    title = "{Next-to-eikonal corrections in the CGC: gluon production and spin asymmetries in pA collisions}",
    eprint = "1404.2219",
    archivePrefix = "arXiv",
    primaryClass = "hep-ph",
    doi = "10.1007/JHEP07(2014)068",
    journal = "JHEP",
    volume = "07",
    pages = "068",
    year = "2014"
}

@article{Altinoluk:2015gia,
    author = "Altinoluk, Tolga and Armesto, N\'estor and Beuf, Guillaume and Moscoso, Alexis",
    title = "{Next-to-next-to-eikonal corrections in the CGC}",
    eprint = "1505.01400",
    archivePrefix = "arXiv",
    primaryClass = "hep-ph",
    doi = "10.1007/JHEP01(2016)114",
    journal = "JHEP",
    volume = "01",
    pages = "114",
    year = "2016"
}

@article{Altinoluk:2015xuy,
    author = "Altinoluk, Tolga and Dumitru, Adrian",
    title = "{Particle production in high-energy collisions beyond the shockwave limit}",
    eprint = "1512.00279",
    archivePrefix = "arXiv",
    primaryClass = "hep-ph",
    doi = "10.1103/PhysRevD.94.074032",
    journal = "Phys. Rev. D",
    volume = "94",
    number = "7",
    pages = "074032",
    year = "2016"
}

@article{Agostini:2019avp,
    author = "Agostini, Pedro and Altinoluk, Tolga and Armesto, N\'estor",
    title = "{Non-eikonal corrections to multi-particle production in the Color Glass Condensate}",
    eprint = "1902.04483",
    archivePrefix = "arXiv",
    primaryClass = "hep-ph",
    doi = "10.1140/epjc/s10052-019-7097-5",
    journal = "Eur. Phys. J. C",
    volume = "79",
    number = "7",
    pages = "600",
    year = "2019"
}

@article{Agostini:2019hkj,
    author = "Agostini, Pedro and Altinoluk, Tolga and Armesto, N\'estor",
    title = "{Effect of non-eikonal corrections on azimuthal asymmetries in the Color Glass Condensate}",
    eprint = "1907.03668",
    archivePrefix = "arXiv",
    primaryClass = "hep-ph",
    reportNumber = "CERN-TH-2019-102",
    doi = "10.1140/epjc/s10052-019-7315-1",
    journal = "Eur. Phys. J. C",
    volume = "79",
    number = "9",
    pages = "790",
    year = "2019"
}

@article{Altinoluk:2020oyd,
    author = "Altinoluk, Tolga and Beuf, Guillaume and Czajka, Alina and Tymowska, Arantxa",
    title = "{Quarks at next-to-eikonal accuracy in the CGC: Forward quark-nucleus scattering}",
    eprint = "2012.03886",
    archivePrefix = "arXiv",
    primaryClass = "hep-ph",
    doi = "10.1103/PhysRevD.104.014019",
    journal = "Phys. Rev. D",
    volume = "104",
    number = "1",
    pages = "014019",
    year = "2021"
}

@article{Altinoluk:2021lvu,
    author = "Altinoluk, Tolga and Beuf, Guillaume",
    title = "{Quark and scalar propagators at next-to-eikonal accuracy in the CGC through a dynamical background gluon field}",
    eprint = "2109.01620",
    archivePrefix = "arXiv",
    primaryClass = "hep-ph",
    doi = "10.1103/PhysRevD.105.074026",
    journal = "Phys. Rev. D",
    volume = "105",
    number = "7",
    pages = "074026",
    year = "2022"
}

@article{Agostini:2022ctk,
    author = "Agostini, Pedro and Altinoluk, Tolga and Armesto, N\'estor and Dominguez, Fabio and Milhano, Jos\'e Guilherme",
    title = "{Multiparticle production in proton\textendash{}nucleus collisions beyond eikonal accuracy}",
    eprint = "2207.10472",
    archivePrefix = "arXiv",
    primaryClass = "hep-ph",
    doi = "10.1140/epjc/s10052-022-10962-1",
    journal = "Eur. Phys. J. C",
    volume = "82",
    number = "11",
    pages = "1001",
    year = "2022"
}

@article{Agostini:2022oge,
    author = "Agostini, Pedro and Altinoluk, Tolga and Armesto, N\'estor",
    title = "{Finite width effects on the azimuthal asymmetry in proton-nucleus collisions in the Color Glass Condensate}",
    eprint = "2212.03633",
    archivePrefix = "arXiv",
    primaryClass = "hep-ph",
    doi = "10.1016/j.physletb.2023.137892",
    journal = "Phys. Lett. B",
    volume = "840",
    pages = "137892",
    year = "2023"
}

@article{Altinoluk:2022jkk,
    author = "Altinoluk, Tolga and Beuf, Guillaume and Czajka, Alina and Tymowska, Arantxa",
    title = "{DIS dijet production at next-to-eikonal accuracy in the CGC}",
    eprint = "2212.10484",
    archivePrefix = "arXiv",
    primaryClass = "hep-ph",
    doi = "10.1103/PhysRevD.107.074016",
    journal = "Phys. Rev. D",
    volume = "107",
    number = "7",
    pages = "074016",
    year = "2023"
}

@article{Altinoluk:2023qfr,
    author = "Altinoluk, Tolga and Armesto, Nestor and Beuf, Guillaume",
    title = "{Probing quark transverse momentum distributions in the color glass condensate: Quark-gluon dijets in deep inelastic scattering at next-to-eikonal accuracy}",
    eprint = "2303.12691",
    archivePrefix = "arXiv",
    primaryClass = "hep-ph",
    doi = "10.1103/PhysRevD.108.074023",
    journal = "Phys. Rev. D",
    volume = "108",
    number = "7",
    pages = "074023",
    year = "2023"
}

@article{Agostini:2023cvc,
    author = "Agostini, Pedro",
    title = "{Scalar propagator in a background gluon field beyond the eikonal approximation}",
    eprint = "2307.13573",
    archivePrefix = "arXiv",
    primaryClass = "hep-ph",
    doi = "10.1007/JHEP11(2023)099",
    journal = "JHEP",
    volume = "11",
    pages = "099",
    year = "2023"
}

@article{Agostini:2024xqs,
    author = "Agostini, Pedro and Altinoluk, Tolga and Armesto, N\'estor",
    title = "{Next-to-eikonal corrections to dijet production in Deep Inelastic Scattering in the dilute limit of the Color Glass Condensate}",
    eprint = "2403.04603",
    archivePrefix = "arXiv",
    primaryClass = "hep-ph",
    doi = "10.1007/JHEP07(2024)137",
    journal = "JHEP",
    volume = "07",
    pages = "137",
    year = "2024"
}

@article{Altinoluk:2024zom,
    author = "Altinoluk, Tolga and Beuf, Guillaume and Czajka, Alina and Marquet, Cyrille",
    title = "{Back-to-back dijet production in DIS at next-to-eikonal accuracy and twist-3 gluon TMDs}",
    eprint = "2410.00612",
    archivePrefix = "arXiv",
    primaryClass = "hep-ph",
    doi = "10.1103/PhysRevD.111.014010",
    journal = "Phys. Rev. D",
    volume = "111",
    number = "1",
    pages = "014010",
    year = "2025"
}

@article{Altinoluk:2024dba,
    author = "Altinoluk, Tolga and Beuf, Guillaume and Mulani, Swaleha",
    title = "{Forward parton-nucleus scattering at next-to-eikonal accuracy in the color glass condensate}",
    eprint = "2411.15047",
    archivePrefix = "arXiv",
    primaryClass = "hep-ph",
    doi = "10.1103/PhysRevD.111.034028",
    journal = "Phys. Rev. D",
    volume = "111",
    number = "3",
    pages = "034028",
    year = "2025"
}

@article{Altinoluk:2024tyx,
    author = "Altinoluk, Tolga and Beuf, Guillaume and Blanco, Etienne and Mulani, Swaleha",
    title = "{Quark TMDs from back-to-back dijet production at forward rapidities in pA collisions beyond eikonal accuracy in the CGC}",
    eprint = "2412.08485",
    archivePrefix = "arXiv",
    primaryClass = "hep-ph",
    doi = "10.1007/JHEP06(2025)097",
    journal = "JHEP",
    volume = "06",
    pages = "097",
    year = "2025"
}

@article{Altinoluk:2025ang,
    author = "Altinoluk, Tolga and Beuf, Guillaume and Mulani, Swaleha",
    title = "{Parton model contributions as next-to-eikonal corrections to the dipole factorization of DIS and SIDIS at low xBj}",
    eprint = "2510.13571",
    archivePrefix = "arXiv",
    primaryClass = "hep-ph",
    doi = "10.1103/15qz-1krh",
    journal = "Phys. Rev. D",
    volume = "113",
    number = "3",
    pages = "034011",
    year = "2026"
}

@article{Agostini:2025vvx,
    author = "Agostini, Pedro and Altinoluk, Tolga and Armesto, N{\'e}stor and Beuf, Guillaume and Cougoulic, Florian and Mulani, Swaleha",
    title = "{Dijet production in DIS off a large nucleus at next-to-eikonal accuracy in a Gaussian model within the CGC framework}",
    eprint = "2512.17848",
    archivePrefix = "arXiv",
    primaryClass = "hep-ph",
    doi = "10.1103/fpy6-9s8c",
    journal = "Phys. Rev. D",
    volume = "113",
    number = "5",
    pages = "054035",
    year = "2026"
}

@article{Altinoluk:2025ivn,
    author = "Altinoluk, Tolga and Beuf, Guillaume and Favrel, Jules and Fucilla, Michael",
    title = "{Next-to-leading order corrections to the next-to-eikonal DIS structure functions}",
    eprint = "2512.16788",
    archivePrefix = "arXiv",
    primaryClass = "hep-ph",
    doi = "10.1007/JHEP06(2026)124",
    journal = "JHEP",
    volume = "06",
    pages = "124",
    year = "2026"
}

@article{Kovchegov:2015pbl,
    author = "Kovchegov, Yuri V. and Pitonyak, Daniel and Sievert, Matthew D.",
    title = "{Helicity Evolution at Small-x}",
    eprint = "1511.06737",
    archivePrefix = "arXiv",
    primaryClass = "hep-ph",
    reportNumber = "RBRC-1159, BNL-111624-2015-JA",
    doi = "10.1007/JHEP01(2016)072",
    journal = "JHEP",
    volume = "01",
    pages = "072",
    year = "2016",
    note = "[Erratum: JHEP 10, 148 (2016)]"
}

@article{Kovchegov:2016zex,
    author = "Kovchegov, Yuri V. and Pitonyak, Daniel and Sievert, Matthew D.",
    title = "{Helicity Evolution at Small $x$: Flavor Singlet and Non-Singlet Observables}",
    eprint = "1610.06197",
    archivePrefix = "arXiv",
    primaryClass = "hep-ph",
    reportNumber = "LA-UR-16-27996",
    doi = "10.1103/PhysRevD.95.014033",
    journal = "Phys. Rev. D",
    volume = "95",
    number = "1",
    pages = "014033",
    year = "2017"
}

@article{Kovchegov:2016weo,
    author = "Kovchegov, Yuri V. and Pitonyak, Daniel and Sievert, Matthew D.",
    title = "{Small-$x$ asymptotics of the quark helicity distribution}",
    eprint = "1610.06188",
    archivePrefix = "arXiv",
    primaryClass = "hep-ph",
    reportNumber = "LA-UR-16-27995, RBRC-1207",
    doi = "10.1103/PhysRevLett.118.052001",
    journal = "Phys. Rev. Lett.",
    volume = "118",
    number = "5",
    pages = "052001",
    year = "2017"
}

@article{Kovchegov:2017jxc,
    author = "Kovchegov, Yuri V. and Pitonyak, Daniel and Sievert, Matthew D.",
    title = "{Small-$x$ Asymptotics of the Quark Helicity Distribution: Analytic Results}",
    eprint = "1703.05809",
    archivePrefix = "arXiv",
    primaryClass = "hep-ph",
    reportNumber = "LA-UR-17-22988",
    doi = "10.1016/j.physletb.2017.06.032",
    journal = "Phys. Lett. B",
    volume = "772",
    pages = "136--140",
    year = "2017"
}

@article{Kovchegov:2017lsr,
    author = "Kovchegov, Yuri V. and Pitonyak, Daniel and Sievert, Matthew D.",
    title = "{Small-$x$ Asymptotics of the Gluon Helicity Distribution}",
    eprint = "1706.04236",
    archivePrefix = "arXiv",
    primaryClass = "nucl-th",
    reportNumber = "LA-UR-16-27995",
    doi = "10.1007/JHEP10(2017)198",
    journal = "JHEP",
    volume = "10",
    pages = "198",
    year = "2017"
}

@article{Kovchegov:2018znm,
    author = "Kovchegov, Yuri V. and Sievert, Matthew D.",
    title = "{Small-$x$ Helicity Evolution: an Operator Treatment}",
    eprint = "1808.09010",
    archivePrefix = "arXiv",
    primaryClass = "hep-ph",
    doi = "10.1103/PhysRevD.99.054032",
    journal = "Phys. Rev. D",
    volume = "99",
    number = "5",
    pages = "054032",
    year = "2019"
}

@article{Kovchegov:2018zeq,
    author = "Kovchegov, Yuri V. and Sievert, Matthew D.",
    title = "{Valence Quark Transversity at Small $x$}",
    eprint = "1808.10354",
    archivePrefix = "arXiv",
    primaryClass = "hep-ph",
    doi = "10.1103/PhysRevD.99.054033",
    journal = "Phys. Rev. D",
    volume = "99",
    number = "5",
    pages = "054033",
    year = "2019"
}

@article{Cougoulic:2019aja,
    author = "Cougoulic, Florian and Kovchegov, Yuri V.",
    title = "{Helicity-dependent generalization of the JIMWLK evolution}",
    eprint = "1910.04268",
    archivePrefix = "arXiv",
    primaryClass = "hep-ph",
    doi = "10.1103/PhysRevD.100.114020",
    journal = "Phys. Rev. D",
    volume = "100",
    number = "11",
    pages = "114020",
    year = "2019"
}

@article{Kovchegov:2020kxg,
    author = "Kovchegov, Yuri V. and Santiago, M. Gabriel",
    title = "{Lensing mechanism meets small- $x$ physics: Single transverse spin asymmetry in $p^{\uparrow}+p$ and $p^{\uparrow}+A$ collisions}",
    eprint = "2003.12650",
    archivePrefix = "arXiv",
    primaryClass = "hep-ph",
    doi = "10.1103/PhysRevD.102.014022",
    journal = "Phys. Rev. D",
    volume = "102",
    number = "1",
    pages = "014022",
    year = "2020"
}

@article{Kovchegov:2020hgb,
    author = "Kovchegov, Yuri V. and Tawabutr, Yossathorn",
    title = "{Helicity at Small $x$: Oscillations Generated by Bringing Back the Quarks}",
    eprint = "2005.07285",
    archivePrefix = "arXiv",
    primaryClass = "hep-ph",
    doi = "10.1007/JHEP08(2020)014",
    journal = "JHEP",
    volume = "08",
    pages = "014",
    year = "2020"
}

@article{Cougoulic:2020tbc,
    author = "Cougoulic, Florian and Kovchegov, Yuri V.",
    title = "{Helicity-dependent extension of the McLerran\textendash{}Venugopalan model}",
    eprint = "2005.14688",
    archivePrefix = "arXiv",
    primaryClass = "hep-ph",
    doi = "10.1016/j.nuclphysa.2020.122051",
    journal = "Nucl. Phys. A",
    volume = "1004",
    pages = "122051",
    year = "2020"
}

@article{Adamiak:2021ppq,
    author = "Adamiak, Daniel and Kovchegov, Yuri V. and Melnitchouk, W. and Pitonyak, Daniel and Sato, Nobuo and Sievert, Matthew D.",
    collaboration = "Jefferson Lab Angular Momentum",
    title = "{First analysis of world polarized DIS data with small-x helicity evolution}",
    eprint = "2102.06159",
    archivePrefix = "arXiv",
    primaryClass = "hep-ph",
    reportNumber = "JLAB-THY-21-3318",
    doi = "10.1103/PhysRevD.104.L031501",
    journal = "Phys. Rev. D",
    volume = "104",
    number = "3",
    pages = "L031501",
    year = "2021"
}

@article{Kovchegov:2021lvz,
    author = "Kovchegov, Yuri V. and Tarasov, Andrey and Tawabutr, Yossathorn",
    title = "{Helicity evolution at small x: the single-logarithmic contribution}",
    eprint = "2104.11765",
    archivePrefix = "arXiv",
    primaryClass = "hep-ph",
    doi = "10.1007/JHEP03(2022)184",
    journal = "JHEP",
    volume = "03",
    pages = "184",
    year = "2022"
}

@article{Kovchegov:2021iyc,
    author = "Kovchegov, Yuri V. and Santiago, M. Gabriel",
    title = "{Quark sivers function at small $x$: spin-dependent odderon and the sub-eikonal evolution}",
    eprint = "2108.03667",
    archivePrefix = "arXiv",
    primaryClass = "hep-ph",
    doi = "10.1007/JHEP11(2021)200",
    journal = "JHEP",
    volume = "11",
    pages = "200",
    year = "2021",
    note = "[Erratum: JHEP 09, 186 (2022)]"
}

@article{Cougoulic:2022gbk,
    author = "Cougoulic, Florian and Kovchegov, Yuri V. and Tarasov, Andrey and Tawabutr, Yossathorn",
    title = "{Quark and gluon helicity evolution at small x: revised and updated}",
    eprint = "2204.11898",
    archivePrefix = "arXiv",
    primaryClass = "hep-ph",
    doi = "10.1007/JHEP07(2022)095",
    journal = "JHEP",
    volume = "07",
    pages = "095",
    year = "2022"
}

@article{Kovchegov:2022kyy,
    author = "Kovchegov, Yuri V. and Santiago, M. Gabriel",
    title = "{T-odd leading-twist quark TMDs at small x}",
    eprint = "2209.03538",
    archivePrefix = "arXiv",
    primaryClass = "hep-ph",
    doi = "10.1007/JHEP11(2022)098",
    journal = "JHEP",
    volume = "11",
    pages = "098",
    year = "2022"
}

@article{Borden:2023ugd,
    author = "Borden, Jeremy and Kovchegov, Yuri V.",
    title = "{Analytic solution for the revised helicity evolution at small x and large Nc: New resummed gluon-gluon polarized anomalous dimension and intercept}",
    eprint = "2304.06161",
    archivePrefix = "arXiv",
    primaryClass = "hep-ph",
    doi = "10.1103/PhysRevD.108.014001",
    journal = "Phys. Rev. D",
    volume = "108",
    number = "1",
    pages = "014001",
    year = "2023"
}

@article{Kovchegov:2024aus,
    author = "Kovchegov, Yuri V. and Li, Ming",
    title = "{Gluon double-spin asymmetry in the longitudinally polarized p + p collisions}",
    eprint = "2403.06959",
    archivePrefix = "arXiv",
    primaryClass = "hep-ph",
    doi = "10.1007/JHEP05(2024)177",
    journal = "JHEP",
    volume = "05",
    pages = "177",
    year = "2024"
}

@article{Borden:2024bxa,
    author = "Borden, Jeremy and Kovchegov, Yuri V. and Li, Ming",
    title = "{Helicity Evolution at Small $x$: Quark to Gluon and Gluon to Quark Transition Operators}",
    eprint = "2406.11647",
    archivePrefix = "arXiv",
    primaryClass = "hep-ph",
    month = "6",
    year = "2024"
}

@article{Adamiak:2025dpw,
    author = "Adamiak, Daniel and Baldonado, Nicholas and Kovchegov, Yuri V. and Li, Ming and Melnitchouk, W. and Pitonyak, Daniel and Sato, Nobuo and Sievert, Matthew D. and Tarasov, Andrey and Tawabutr, Yossathorn",
    collaboration = "JAM Collaboration (Small-x Analysis Group)",
    title = "{First study of polarized proton-proton scattering with small-x helicity evolution}",
    eprint = "2503.21006",
    archivePrefix = "arXiv",
    primaryClass = "hep-ph",
    reportNumber = "JLAB-THY-25-4253",
    doi = "10.1103/9gnx-ycs4",
    journal = "Phys. Rev. D",
    volume = "112",
    number = "9",
    pages = "094032",
    year = "2025"
}

@article{Kovchegov:2025gcg,
    author = "Kovchegov, Yuri V. and Li, Ming",
    title = {{Weizs{\"a}cker-Williams gluon helicity distribution and inclusive dijet production in longitudinally polarized electron-proton collisions}},
    eprint = "2504.12979",
    archivePrefix = "arXiv",
    primaryClass = "hep-ph",
    doi = "10.1007/JHEP08(2025)206",
    journal = "JHEP",
    volume = "08",
    pages = "206",
    year = "2025"
}

@article{Borden:2025ehe,
    author = "Borden, Jeremy and Kovchegov, Yuri V.",
    title = "{Analytic Solution for the Helicity Evolution Equations at Small $x$ and Large $N_c\&N_f$}",
    eprint = "2508.00195",
    archivePrefix = "arXiv",
    primaryClass = "hep-ph",
    month = "7",
    year = "2025"
}

@article{Balitsky:2015qba,
    author = "Balitsky, I. and Tarasov, A.",
    title = "{Rapidity evolution of gluon TMD from low to moderate x}",
    eprint = "1505.02151",
    archivePrefix = "arXiv",
    primaryClass = "hep-ph",
    reportNumber = "JLAB-THY-15-2040",
    doi = "10.1007/JHEP10(2015)017",
    journal = "JHEP",
    volume = "10",
    pages = "017",
    year = "2015"
}

@article{Balitsky:2016dgz,
    author = "Balitsky, I. and Tarasov, A.",
    title = "{Gluon TMD in particle production from low to moderate x}",
    eprint = "1603.06548",
    archivePrefix = "arXiv",
    primaryClass = "hep-ph",
    reportNumber = "JLAB-THY-16-2229",
    doi = "10.1007/JHEP06(2016)164",
    journal = "JHEP",
    volume = "06",
    pages = "164",
    year = "2016"
}

@article{Balitsky:2017flc,
    author = "Balitsky, I. and Tarasov, A.",
    title = "{Higher-twist corrections to gluon TMD factorization}",
    eprint = "1706.01415",
    archivePrefix = "arXiv",
    primaryClass = "hep-ph",
    reportNumber = "BNL-113982-2017-JA, JLAB-THY-17-2484",
    doi = "10.1007/JHEP07(2017)095",
    journal = "JHEP",
    volume = "07",
    pages = "095",
    year = "2017"
}

@article{Chirilli:2018kkw,
    author = "Chirilli, Giovanni Antonio",
    title = "{Sub-eikonal corrections to scattering amplitudes at high energy}",
    eprint = "1807.11435",
    archivePrefix = "arXiv",
    primaryClass = "hep-ph",
    doi = "10.1007/JHEP01(2019)118",
    journal = "JHEP",
    volume = "01",
    pages = "118",
    year = "2019"
}

@article{Chirilli:2021lif,
    author = "Chirilli, Giovanni Antonio",
    title = "{High-energy operator product expansion at sub-eikonal level}",
    eprint = "2101.12744",
    archivePrefix = "arXiv",
    primaryClass = "hep-ph",
    doi = "10.1007/JHEP06(2021)096",
    journal = "JHEP",
    volume = "06",
    pages = "096",
    year = "2021"
}

@article{Chirilli:2026pkv,
    author = "Chirilli, Giovanni Antonio",
    title = "{From Sub-eikonal DIS to Quark Distributions and their High-Energy Evolution}",
    eprint = "2603.30000",
    archivePrefix = "arXiv",
    primaryClass = "hep-ph",
    month = "3",
    year = "2026"
}

@article{Jalilian-Marian:2017ttv,
    author = "Jalilian-Marian, Jamal",
    title = "{Elastic scattering of a quark from a color field: longitudinal momentum exchange}",
    eprint = "1708.07533",
    archivePrefix = "arXiv",
    primaryClass = "hep-ph",
    doi = "10.1103/PhysRevD.96.074020",
    journal = "Phys. Rev. D",
    volume = "96",
    number = "7",
    pages = "074020",
    year = "2017"
}

@article{Jalilian-Marian:2018iui,
    author = "Jalilian-Marian, Jamal",
    title = "{Quark jets scattering from a gluon field: from saturation to high $p_t$}",
    eprint = "1809.04625",
    archivePrefix = "arXiv",
    primaryClass = "hep-ph",
    doi = "10.1103/PhysRevD.99.014043",
    journal = "Phys. Rev. D",
    volume = "99",
    number = "1",
    pages = "014043",
    year = "2019"
}

@article{Jalilian-Marian:2019kaf,
    author = "Jalilian-Marian, Jamal",
    title = "{Rapidity loss, spin, and angular asymmetries in the scattering of a quark from the color field of a proton or nucleus}",
    eprint = "1912.08878",
    archivePrefix = "arXiv",
    primaryClass = "hep-ph",
    doi = "10.1103/PhysRevD.102.014008",
    journal = "Phys. Rev. D",
    volume = "102",
    number = "1",
    pages = "014008",
    year = "2020"
}

@article{Boussarie:2023xun,
    author = "Boussarie, Renaud and Mehtar-Tani, Yacine",
    title = "{Low and moderate x gluon contribution to exclusive Compton scattering processes}",
    eprint = "2309.16576",
    archivePrefix = "arXiv",
    primaryClass = "hep-ph",
    doi = "10.1007/JHEP10(2024)056",
    journal = "JHEP",
    volume = "10",
    pages = "056",
    year = "2024"
}

@article{Li:2023tlw,
    author = "Li, Ming",
    title = "{Small x physics beyond eikonal approximation: an effective Hamiltonian approach}",
    eprint = "2304.12842",
    archivePrefix = "arXiv",
    primaryClass = "hep-ph",
    doi = "10.1007/JHEP07(2023)158",
    journal = "JHEP",
    volume = "07",
    pages = "158",
    year = "2023"
}

@article{Li:2024fdb,
    author = "Li, Ming",
    title = "{Quasiclassical Gluon Fields and Low\textquoteright{}s Soft Theorem at Small Momentum-Fraction x}",
    eprint = "2402.17568",
    archivePrefix = "arXiv",
    primaryClass = "hep-ph",
    doi = "10.1103/PhysRevLett.133.021902",
    journal = "Phys. Rev. Lett.",
    volume = "133",
    number = "2",
    pages = "021902",
    year = "2024"
}

@article{Li:2024xra,
    author = "Li, Ming",
    title = "{Quasiclassical evaluation of gluon saturation induced helicity effects}",
    eprint = "2411.13431",
    archivePrefix = "arXiv",
    primaryClass = "hep-ph",
    doi = "10.1103/PhysRevD.111.034027",
    journal = "Phys. Rev. D",
    volume = "111",
    number = "3",
    pages = "034027",
    year = "2025"
}

@article{Li:2026azt,
    author = "Li, Ming",
    title = "{High Energy Evolution of Dipole Gluon Distribution Beyond Eikonal Approximation}",
    eprint = "2607.12554",
    archivePrefix = "arXiv",
    primaryClass = "hep-ph",
    month = "7",
    year = "2026"
}

@article{Hatta:2016aoc,
    author = "Hatta, Yoshitaka and Nakagawa, Yuya and Yuan, Feng and Zhao, Yong and Xiao, Bowen",
    title = "{Gluon orbital angular momentum at small-$x$}",
    eprint = "1612.02445",
    archivePrefix = "arXiv",
    primaryClass = "hep-ph",
    reportNumber = "YITP-16-133",
    doi = "10.1103/PhysRevD.95.114032",
    journal = "Phys. Rev. D",
    volume = "95",
    number = "11",
    pages = "114032",
    year = "2017"
}

@article{Kovchegov:2019rrz,
    author = "Kovchegov, Yuri V.",
    title = "{Orbital Angular Momentum at Small $x$}",
    eprint = "1901.07453",
    archivePrefix = "arXiv",
    primaryClass = "hep-ph",
    reportNumber = "INT pre-print number INT-PUB-18-065",
    doi = "10.1007/JHEP03(2019)174",
    journal = "JHEP",
    volume = "03",
    pages = "174",
    year = "2019"
}

@article{Boussarie:2019icw,
    author = "Boussarie, Renaud and Hatta, Yoshitaka and Yuan, Feng",
    title = "{Proton Spin Structure at Small-$x$}",
    eprint = "1904.02693",
    archivePrefix = "arXiv",
    primaryClass = "hep-ph",
    doi = "10.1016/j.physletb.2019.134817",
    journal = "Phys. Lett. B",
    volume = "797",
    pages = "134817",
    year = "2019"
}

@article{Kovchegov:2023yzd,
    author = "Kovchegov, Yuri V. and Manley, Brandon",
    title = "{Orbital angular momentum at small x revisited}",
    eprint = "2310.18404",
    archivePrefix = "arXiv",
    primaryClass = "hep-ph",
    doi = "10.1007/JHEP02(2024)060",
    journal = "JHEP",
    volume = "02",
    pages = "060",
    year = "2024"
}

@article{Kovchegov:2024wjs,
    author = "Kovchegov, Yuri V. and Manley, Brandon",
    title = "{Elastic Dijet Production in Electron Scattering on a Longitudinally Polarized Proton at Small $x$: A Portal to Orbital Angular Momentum Distributions}",
    eprint = "2410.21260",
    archivePrefix = "arXiv",
    primaryClass = "hep-ph",
    month = "10",
    year = "2024"
}

@article{Mukherjee:2026cte,
    author = "Mukherjee, Swagato and Skokov, Vladimir V. and Tarasov, Andrey and Tiwari, Shaswat and Yao, Fei",
    title = "{Back-to-back dijet production in DIS at arbitrary Bjorken-x: TMD gluon distributions to twist-3 accuracy}",
    eprint = "2602.15137",
    archivePrefix = "arXiv",
    primaryClass = "hep-ph",
    month = "2",
    year = "2026"
}

@article{Kar:2026vzk,
    author = "Kar, Tiyasa and Tarasov, Andrey and Skokov, Vladimir V.",
    title = "{DIS dijet production in Background Field Approach: General formalism and methods}",
    eprint = "2603.08805",
    archivePrefix = "arXiv",
    primaryClass = "hep-ph",
    month = "3",
    year = "2026"
}

@article{Mukherjee:2026six,
    author = "Mukherjee, Swagato and Skokov, Vladimir. V. and Tarasov, Andrey and Tiwari, Shaswat and Yao, Fei",
    title = "{Back-to-back dijet production in DIS at arbitrary Bjorken-x: TMD quark distributions to twist-3 accuracy}",
    eprint = "2607.12268",
    archivePrefix = "arXiv",
    primaryClass = "hep-ph",
    month = "7",
    year = "2026"
}

@article{Gelis:2010nm,
    author = "Gelis, Francois and Iancu, Edmond and Jalilian-Marian, Jamal and Venugopalan, Raju",
    title = "{The Color Glass Condensate}",
    eprint = "1002.0333",
    archivePrefix = "arXiv",
    primaryClass = "hep-ph",
    doi = "10.1146/annurev.nucl.010909.083629",
    journal = "Ann. Rev. Nucl. Part. Sci.",
    volume = "60",
    pages = "463--489",
    year = "2010"
}

@article{McLerran:1994vd,
    author = "McLerran, Larry D. and Venugopalan, Raju",
    title = "{Green's functions in the color field of a large nucleus}",
    eprint = "hep-ph/9402335",
    archivePrefix = "arXiv",
    reportNumber = "TPI-MINN-94-7-T, NUC-MINN-94-2-T, HEP-MINN-94-1242-T",
    doi = "10.1103/PhysRevD.50.2225",
    journal = "Phys. Rev. D",
    volume = "50",
    pages = "2225--2233",
    year = "1994"
}

@article{Balitsky:1995ub,
    author = "Balitsky, I.",
    title = "{Operator expansion for high-energy scattering}",
    eprint = "hep-ph/9509348",
    archivePrefix = "arXiv",
    reportNumber = "MIT-CTP-2470",
    doi = "10.1016/0550-3213(95)00638-9",
    journal = "Nucl. Phys. B",
    volume = "463",
    pages = "99--160",
    year = "1996"
}

@article{Kovchegov:1999yj,
    author = "Kovchegov, Yuri V.",
    title = "{Small x F(2) structure function of a nucleus including multiple pomeron exchanges}",
    eprint = "hep-ph/9901281",
    archivePrefix = "arXiv",
    reportNumber = "NUC-MN-99-1-T, TPI-MINN-99-05",
    doi = "10.1103/PhysRevD.60.034008",
    journal = "Phys. Rev. D",
    volume = "60",
    pages = "034008",
    year = "1999"
}

@article{Kovchegov:1999ua,
    author = "Kovchegov, Yuri V.",
    title = "{Unitarization of the BFKL pomeron on a nucleus}",
    eprint = "hep-ph/9905214",
    archivePrefix = "arXiv",
    reportNumber = "NUC-MN-99-8-T, TPI-MINN-99-26",
    doi = "10.1103/PhysRevD.61.074018",
    journal = "Phys. Rev. D",
    volume = "61",
    pages = "074018",
    year = "2000"
}

@article{Jalilian-Marian:1996mkd,
    author = "Jalilian-Marian, Jamal and Kovner, Alex and McLerran, Larry D. and Weigert, Heribert",
    title = "{The Intrinsic glue distribution at very small x}",
    eprint = "hep-ph/9606337",
    archivePrefix = "arXiv",
    reportNumber = "TPI-MINN-96-1429, NUC-MINN-96-10-T, TPI-MINN-96-08",
    doi = "10.1103/PhysRevD.55.5414",
    journal = "Phys. Rev. D",
    volume = "55",
    pages = "5414--5428",
    year = "1997"
}

@article{Jalilian-Marian:1997qno,
    author = "Jalilian-Marian, Jamal and Kovner, Alex and Leonidov, Andrei and Weigert, Heribert",
    title = "{The BFKL equation from the Wilson renormalization group}",
    eprint = "hep-ph/9701284",
    archivePrefix = "arXiv",
    reportNumber = "TPI-MINN-96-28-T, NUC-MINN-96-22-T, HEP-MINN-96-1524",
    doi = "10.1016/S0550-3213(97)00440-9",
    journal = "Nucl. Phys. B",
    volume = "504",
    pages = "415--431",
    year = "1997"
}

@article{Jalilian-Marian:1997jhx,
    author = "Jalilian-Marian, Jamal and Kovner, Alex and Leonidov, Andrei and Weigert, Heribert",
    title = "{The Wilson renormalization group for low x physics: Towards the high density regime}",
    eprint = "hep-ph/9706377",
    archivePrefix = "arXiv",
    reportNumber = "TPI-MINN-97-20-T, NUC-MINN-97-6-T, HEP-MINN-97-1546, CAVENDISH-HEP-97-09",
    doi = "10.1103/PhysRevD.59.014014",
    journal = "Phys. Rev. D",
    volume = "59",
    pages = "014014",
    year = "1998"
}

@article{Jalilian-Marian:1997ubg,
    author = "Jalilian-Marian, Jamal and Kovner, Alex and Weigert, Heribert",
    title = "{The Wilson renormalization group for low x physics: Gluon evolution at finite parton density}",
    eprint = "hep-ph/9709432",
    archivePrefix = "arXiv",
    reportNumber = "TPI-MINN-97-26, NUC-MINN-97-11-T, HEP-MINN-1607, OUTP-97-45-P, CAVENDISH-HEP-97-15",
    doi = "10.1103/PhysRevD.59.014015",
    journal = "Phys. Rev. D",
    volume = "59",
    pages = "014015",
    year = "1998"
}

@article{Kovner:2000pt,
    author = "Kovner, Alex and Milhano, J. Guilherme and Weigert, Heribert",
    title = "{Relating different approaches to nonlinear QCD evolution at finite gluon density}",
    eprint = "hep-ph/0004014",
    archivePrefix = "arXiv",
    reportNumber = "OUTP-00-10-P, NORDITA-2000-14-HE",
    doi = "10.1103/PhysRevD.62.114005",
    journal = "Phys. Rev. D",
    volume = "62",
    pages = "114005",
    year = "2000"
}

@article{Weigert:2000gi,
    author = "Weigert, Heribert",
    title = "{Unitarity at small Bjorken x}",
    eprint = "hep-ph/0004044",
    archivePrefix = "arXiv",
    reportNumber = "NORDITA-2000-34-HE",
    doi = "10.1016/S0375-9474(01)01668-2",
    journal = "Nucl. Phys. A",
    volume = "703",
    pages = "823--860",
    year = "2002"
}

@article{Iancu:2000hn,
    author = "Iancu, Edmond and Leonidov, Andrei and McLerran, Larry D.",
    title = "{Nonlinear gluon evolution in the color glass condensate. 1.}",
    eprint = "hep-ph/0011241",
    archivePrefix = "arXiv",
    reportNumber = "SACLAY-T00-166, BNL-NT-00-24",
    doi = "10.1016/S0375-9474(01)00642-X",
    journal = "Nucl. Phys. A",
    volume = "692",
    pages = "583--645",
    year = "2001"
}

@article{Iancu:2001ad,
    author = "Iancu, Edmond and Leonidov, Andrei and McLerran, Larry D.",
    title = "{The Renormalization group equation for the color glass condensate}",
    eprint = "hep-ph/0102009",
    archivePrefix = "arXiv",
    reportNumber = "BNL-NT-01-3",
    doi = "10.1016/S0370-2693(01)00524-X",
    journal = "Phys. Lett. B",
    volume = "510",
    pages = "133--144",
    year = "2001"
}

@article{Ferreiro:2001qy,
    author = "Ferreiro, Elena and Iancu, Edmond and Leonidov, Andrei and McLerran, Larry",
    title = "{Nonlinear gluon evolution in the color glass condensate. 2.}",
    eprint = "hep-ph/0109115",
    archivePrefix = "arXiv",
    reportNumber = "SACLAY-T01-085, BNL-NT-01-21",
    doi = "10.1016/S0375-9474(01)01329-X",
    journal = "Nucl. Phys. A",
    volume = "703",
    pages = "489--538",
    year = "2002"
}

@article{Boussarie:2020fpb,
    author = "Boussarie, Renaud and Mehtar-Tani, Yacine",
    title = "{A novel formulation of the unintegrated gluon distribution for DIS}",
    eprint = "2006.14569",
    archivePrefix = "arXiv",
    primaryClass = "hep-ph",
    doi = "10.1016/j.physletb.2022.137125",
    journal = "Phys. Lett. B",
    volume = "831",
    pages = "137125",
    year = "2022"
}

@article{Boussarie:2021wkn, 
    author = "Boussarie, Renaud and Mehtar-Tani, Yacine",
    title = "{Gluon-mediated inclusive Deep Inelastic Scattering from Regge to Bjorken kinematics}",
    eprint = "2112.01412",
    archivePrefix = "arXiv",
    primaryClass = "hep-ph",
    doi = "10.1007/JHEP07(2022)080",
    journal = "JHEP",
    volume = "07",
    pages = "080",
    year = "2022"
}

\end{document}